\documentclass[resetfootnote]{aastex701}

\newcommand{\rmm}{\mathrm{rad/m^2}}

\begin{document}

\title{Correcting Ionospheric Faraday Rotation for the VLA and MeerKAT}

\author[orcid=0000-0001-7097-8360,gname=Richard,sname=Perley]{Richard A. Perley}
\affiliation{National Radio Astronomy Observatory, P.O. Box 'O',
  Socorro, NM, 87801, USA}
\email[show]{rperley@nrao.edu}

\author[orcid=0000-0002-5344-820X,gname=Bryan,sname=Butler]{Bryan J. Butler}
\affiliation{National Radio Astronomy Observatory, P.O. Box 'O',
  Socorro NM, 87801, USA}
\email[show]{bbutler@nrao.edu}

\author[orcid=0009-0004-7873-604X,gname=Greisen,sname=Eric]{Eric W. Greisen}
\affiliation{National Radio Astronomy Observatory, P.O. Box 'O',
  Socorro NM, 87801, USA}
\email[show]{egreisen@nrao.edu}

\author[orcid=0000-0002-2933-9134,gname=Benjamin,sname=Hugo]{Benjamin V. Hugo}
\affiliation{South African Radio Astronomy Observatory, Liesbeek
  House, River Park Liesbeek Parkway, Mowbray, Cape Town, 7705 South Africa}
\affiliation{Rhodes University, Artillery Road, Grahamstown, South Africa}
\email[show]{bhugo@sarao.ac.za}

\author[orcid=0000-0002-4039-6703,gname=Evangelina,sname=Tremou]{Evangelia Tremou}
\affiliation{National Radio Astronomy Observatory, P.O. Box 'O',
  Socorro NM, 87801, USA}
\email[show]{etremou@nrao.edu}

\author[orcid=0000-0002-2173-6151,gname=Anthony,sname=Willis]{A. G.  Willis}
\affiliation{Dominion Radio Astrophysical Observatory, P.O. Box 248,
  Penticton, B.C. V2A 6J9, Canada}
\email[show]{tony.willis.research@gmail.com}

\begin{abstract}

We report here on studies to determine the accuracy of estimated
corrections of Ionospheric Faraday Rotation Measure (IFRM) using
observations of the Moon with the Very Large Array (VLA) and MeerKAT
telescopes.  To estimate the IFRM requires an estimate of the total
electron content along the line-of-sight to the observed sources (the
so-called Slant Total Electron Content, or STEC).  Estimating the STEC
requires an estimate of the global 2-D map of Vertical Total Electron
Content (VTEC) along with the ray path from the telescope to the
source.  Traditionally, these global VTEC maps have been utilized
along with an assumption that the electrons are in a thin shell at a
given altitude to provide an estimate of the IFRM as a function of
time.  We find that this traditional technique significantly
overestimates the IFRM -- typically by $\sim$ 0.5 to 1.1 $\rmm$ for
the VLA, and $\sim$ $-0.3$ $\rmm$ for MeerKAT.  Alternatively, the
software package ALBUS utilizes raw data from nearby Global Navigation
Satellite System (GNSS) stations, to generate a local estimate of the
IFRM as a function of time.  ALBUS provides considerably better
estimates of the IFRM -- accurate to $\sim 0.1\, \rmm$ for both the
VLA and MeerKAT, provided the stations utilized have known receiver
bias values.  A byproduct of our study is the establishment of the
intrinsic electric vector position angle (EVPA) of the standard
polarized calibrators 3C286 and 3C138 from 500 MHz to 50 GHz, using
additional VLA observations of the Moon, Venus, and Mars.

\end{abstract}

\keywords{\uat{Radio interferometry}{1346} ---
  \uat{Polarimetry}{1278}--- \uat{Calibration}{2179}--- \uat{Earth ionosphere}{860}} 
\section{Introduction}

Modern interferometric arrays such as the Very Large Array (VLA) and
MeerKAT produce full polarimetric visibility data, permitting imaging in
all four Stokes parameters I, Q, U, and V.  Accurate polarimetry
requires establishing the orientation of the electric vector position
angle (EVPA) of polarized emission to high accuracy -- ideally five
degrees or better.  Attaining this accuracy requires corrections for a
number of instrumental and environmental factors.  A major environmental
factor at low frequencies is the effect of the Earth's ionosphere.

As an electromagnetic wave passes through the ionosphere, the
polarization angle of the wave undergoes a rotation due to the
combined effect of the electrons and the magnetic field.  The
magnitude of this rotation becomes significant at longer wavelengths
(the effect is proportional to wavelength squared), i.e., at long
radio wavelengths.  This effect is important at frequencies below
$\sim$ 4 GHz and is the dominant correction term in observations below
$\sim$ 1 GHz.

In order to calibrate polarization observations at these long
wavelengths, an estimate of the Ionospheric Faraday Rotation Measure
(IFRM) along the line-of-sight to the source being observed as a
function of time must be determined.  Time-variable global maps of
Vertical Total Electron Content (VTEC), produced by a number of
organizations, along with an assumption of the vertical distribution of
the electrons, the known direction from a telescope to a source, and a
model of the Earth's magnetic field, can be used to estimate the Slant
Total Electron Content (STEC) along the path and hence the IFRM.
Accurate polarimetry requires the determination of the IFRM with an
accuracy of at least 0.1 $\rmm$ for $\lambda \sim 1$m on timescales of
an hour, or less.

In this paper we describe and compare methods for correcting for the
IFRM based on both globally and locally-derived estimates of the 
STEC, with the goal of determining which of these methods is most
accurate and reliable.  As part of this work, we have accurately
measured the EVPA of the polarization calibrators 3C286 and 3C138 from
0.5 to 50 GHz, using data from the VLA and MeerKAT.

\section{Ionospheric Faraday Rotation} \label{sec:IFRM}

Passage of a linearly polarized electromagnetic wave through a
magnetized ionized medium results in a rotation of the plane of
polarization given by
\begin{equation}
  \Delta\chi = RM\lambda^2 \quad\mathrm{rad}
\end{equation}
where $RM$ is the rotation measure, given by
\begin{equation}
  RM = 2.62\times10^{-17} \int n_e\mathbf{B}\bullet d\mathbf{l} \qquad\rmm
\end{equation}
with the electron density $n_e$ in $\mathrm{m^{-3}}$, the magnetic
field $\mathbf{B}$ in Gauss, and the propagation path $\mathbf{l}$ and
wavelength $\lambda$ in meters\footnote{This simple relation holds
  when there is neither polarized emission nor absorption in the
  rotating medium and when the frequency of observation is much higher
  than the plasma frequency (typically 12 MHz) and the gyro frequency
  (typically 1.4 MHz) -- conditions which apply to the Earth's
  ionosphere and the observations reported in this paper.}.

Solar UV and X-ray emissions cause partial ionization ($\sim$ 1\%) of
the Earth's atmosphere at heights above $\sim$ 100 km, resulting in
daytime electron densities of up to $10^{12.5}\mathrm{m^{-3}}$ at
heights of $\sim$ 350 km during solar maximum, and about an order of
magnitude less during solar minimum.  Typical night-time densities are
about $10^{11}\mathrm{m^{-3}}$.  Combined with the Earth's
approximately dipolar magnetic field of $\sim$ 1 Gauss, the resulting
typical night-time IFRMs at the VLA and MeerKAT sites have a magnitude
of $\sim 0.5\,\rmm$, with daytime values between 2 and 6 $\rmm$,
(positive for the VLA, negative for MeerKAT) depending on solar
activity and source direction.  In addition to the regular day--night
variations, significant variations on hourly, or shorter, timescales
are common, especially during periods of significant solar activity.

\section{Estimating the IFRM} \label{sec:Models}

\subsection{Using Global VTEC Maps}

Correction of the ionospheric Faraday rotation requires knowing the
electron density and magnetic field distributions along the
propagation path from the source to the observer at the time of
observation -- a difficult prospect.  Because of this, simpler
approximations utilizing available global vertical total electron
content (VTEC) maps defined as a function of longitude, latitude and
time, a model of the magnetic field, and a thin-shell approximation
for the distribution of the ionospheric electrons have been developed.
These global VTEC map files are developed by various Ionospheric
Associated Analysis Centers (IAACs) of the International GNSS (Global
Navigation Satellite System) Service (IGS), using timing data from up
to four GNSS constellations (GPS, Galileo, BeiDou, and GLONASS) and a
network of globally distributed stations.  The timing referred to here
is the additional delay caused by the ionosphere to signals broadcast
by the satellites.  For signals propagating through an ionized medium,
this additional, dispersive group-path delay is proportional to the
product of the electron column density and the wavelength squared:
$\Delta T \propto TEC\times\lambda^2$.  For two frequencies, analysis
\citep{L64} shows the column density (called the `tau-TEC')
is related to the difference in propagation times by
\begin{equation}
  TEC_{TU} = 6.75\times10^{-2}\frac{\Delta T}{\lambda_1^2 - \lambda_2^2}
\end{equation}
where the time difference $\Delta T$ is in nsec, the wavelength
$\lambda$ is in meters, and the units of TEC are in TUs, defined as
$1\, TU = 10^{16}$ electrons/$\mathrm{m^2}$ \citep{E01}. For the GPS
system, $\lambda_1$ = 0.1904 meters, and $\lambda_2$ = 0.2444 meters.
With these values, the relation between the dispersive delay and the
path TEC in TU units is $TEC_{TU} = 2.853\Delta T$.

A much more accurate estimation of the TEC is obtained from the
relative phases of the carrier frequencies, $\Delta\phi$.  In this
case, the column density (called the `phase-TEC') is related to this
phase by
\begin{equation}
  TEC_{TU} = 6.28\times10^{-4}\frac{\Delta\phi}{\lambda_1 - \lambda_2}
\end{equation}
where the phase difference is in degrees, the wavelength in meters and
TEC in TEC units.  For the GPS system, the relation between the
carrier phase difference and the path TEC is $TEC_{TU} =
1.16\times10^{-2}\Delta\phi$.  Although this estimate is subject to
full cycle ambiguities, current software (including the ALBUS program
discussed later in this paper) utilizes the tau-TEC estimates to
detect and remove these ambiguities.

Many institutions generate VTEC maps using different combinations of
the four available global satellite constellations and ground
stations.  These global maps are typically produced on a 2.5 by 5
degree grid (latitude, longitude) on timescales of 30 minutes to 2
hours.  Table~\ref{tab:GNSS} gives typical characteristics of seven
available VTEC maps.  These maps are then interpolated in time and
space to generate the desired STEC values in the direction of interest
needed for the calculation of the IFRM.
\begin{deluxetable}{ccccccc}
\tablewidth{0pt} \tablecaption{Global VTEC Map Characteristics}
\tablehead{ \colhead{Name} & \colhead{Organization} &
  \colhead{ElCutoff} & \colhead{\#Satellites} &\colhead{SatCon} &
  \colhead{\#Stations}&\colhead{TimeStep}}
\startdata jplg&Jet Propulsion Laboratory &10& 32&G &170&2hr\\
codg&Center for Orbit Determination in Europe&10& 79&G,E &235&1\\
esag&European Space Operations Center &10& 78&G,E &248&2\\
igsg&Geodynamics Research Lab,GRL/UWN & 0& 31&G &297&2\\
upcg&Polytechnic Univrsity of Catalonia & 0& 54&G,E &235&2\\
casg&Chinese Academy of Science & 0&108&G,C,E,R& ? &0.5\\
emrg&Space Weather Canada &10& 52&G,R &270&1\\
\enddata
\tablecomments{Explanation of Satellite Constellations: G: US Global
  Positioning System (GPS, 32 satellites); E: European Union Galileo (24); C: China
  BeiDou Navigation Satellite System (24); R: Russian Global
  Navigation Satellite System (GLONASS, 24)}
\label{tab:GNSS}
\end{deluxetable}

Efforts to estimate IFRM corrections for the VLA began in the 1990s with
the implementation of the P-band (330 MHz) receivers and the public
availability of the VTEC maps.  \cite{E01} described VLA observations of
the polarized pulsar PSR 1932+109 taken in the late 1990s.  They
compared the observed pulsar EVPA to predictions derived by utilizing
timing data from a GPS receiver temporarily placed at the center of the
VLA, concluding that IFRM corrections accurate to 0.2 $\rmm$ were
possible.  As a result of this investigative effort, the program {\tt
TECOR} in the software package {\tt AIPS} \citep{G03} was implemented in
1998 to generate IFRM estimates utilizing published global maps of the
vertical total electron content available from the CDDIS (Crustal
Dynamics Data Interchange System) and a model of the Earth's magnetic
field \citep{C40,A21}, assuming the electrons are all in a thin shell at
a height of 450 km.  This height is assumed because it is roughly the
appropriate height for a thin-shell approximation over a wide variety of
conditions and in mid-latitudes \citep{Z19}.

There are two alternative programs known to us which provide IFRM
estimates -- {\tt ionFR} \citep{S13} and {\tt RMextract} \citep{M18}.
Both of these programs work similarly to TECOR, i.e., using the same
model of the Earth's magnetic field, thin-shell approximation, and
global VTEC maps.
ionFR\footnote{\href{https://github.com/csobey/ionFR}{https://github.com/csobey/ionFR}}
is no longer maintained.  RMextract has recently been superseded by {\tt
spinifex}\footnote{\href{https://git.astron.nl/RD/spinifex}{https://git.astron.nl/RD/spinifex}}\citep{spinifex2025},
which is under development for use with LOFAR and SKA-LO observations,
both of which operate at wavelengths longer than 1 meter.  In
Appendix~\ref{sec:RMExtractVsTECOR} we show comparisons of RMExtract,
TECOR, and spinifex for a particular observation.

\subsection{Using Locally-Derived Estimates of the STEC}

An alternative to utilization of global VTEC maps is a regional model
provided by the {\tt ALBUS} program (Advanced Long Baseline User
Software)
\footnote{\href{https://github.com/twillis449/ALBUS_ionosphere/blob/master/twillis_ALBUS_paper.pdf}{\url{https://github.com/twillis449/ALBUS_ionosphere/blob/master/twillis_ALBUS_paper.pdf}}}
\citep{W22}, written to estimate dispersive delays for VLBI phase
corrections, and subsequently modified to provide estimation of the STEC
for a given location and direction as a function of time.  ALBUS offers
ten models, labelled `G00' through 'G09', including both `2-D'
(thin-shell) and `3-D' electron density distributions.

\begin{figure}[t]
  \plottwo{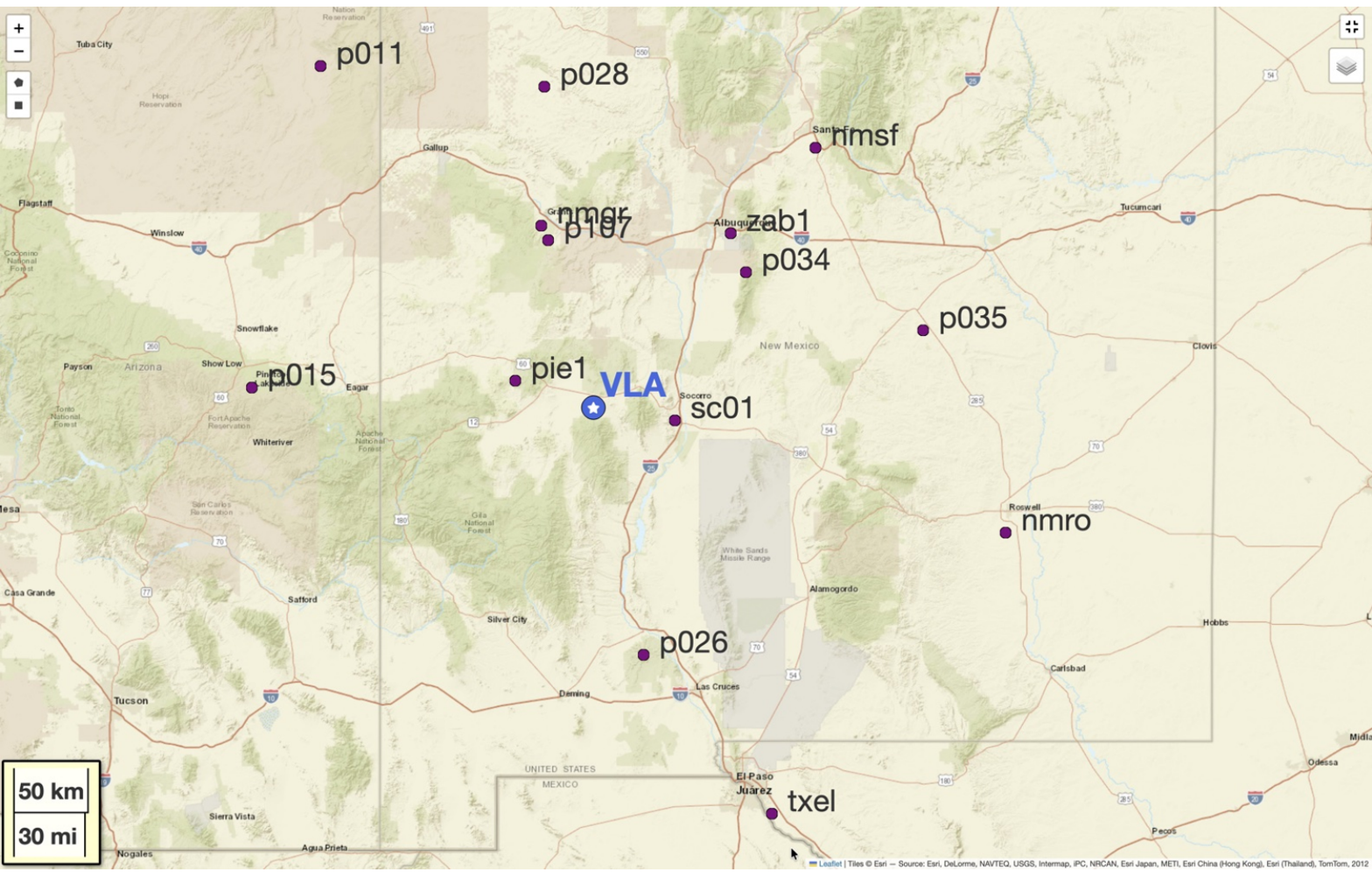}{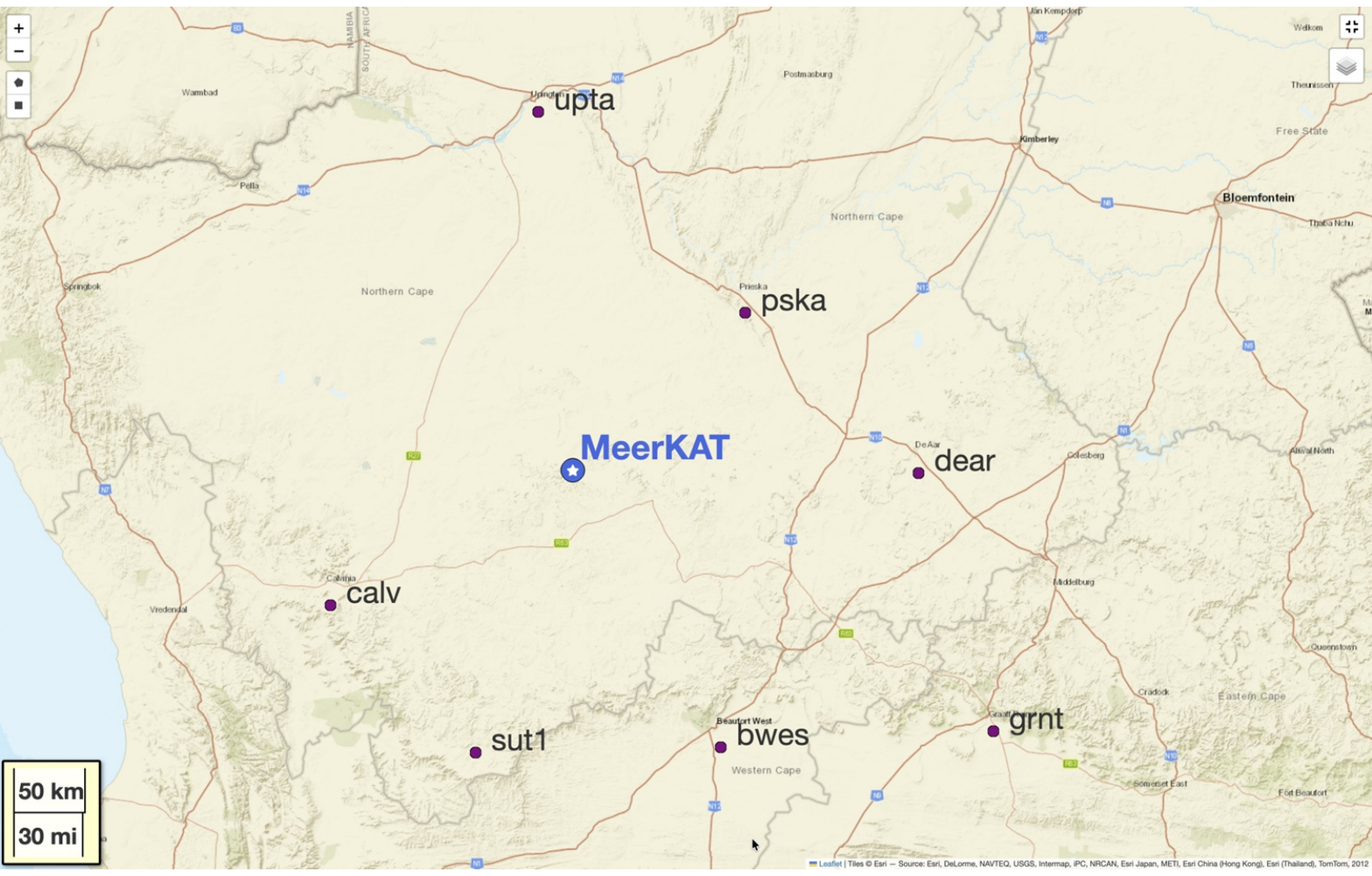}
  \caption{Locations of the GNSS stations for the VLA (left) and
    MeerKAT (right) potentially utilized by ALBUS are shown by the
    small dots.  The larger blue dots show the locations of the two
    arrays.}
  \label{fig:GNSS}
\end{figure}

The ALBUS program calculates IFRM estimates by utilizing GNSS timing
data from ground stations within a specified distance of a given
location combined with, for the 3-D models, a parametrically generated
ionospheric vertical distribution of electrons.  For both the VLA and
MeerKAT, there are typically up to $\sim 20$ potential GNSS stations
within a 300 km radius whose data can be used to generate an estimate of
the STEC. The locations of these stations utilized for this paper are
shown in Fig~\ref{fig:GNSS} for both the VLA and MeerKAT arrays.  The
ALBUS program is available stand-alone, or can be executed within {\tt
AIPS} through the use of a container.
Although ALBUS knows of three satellite constellations, at present it
utilizes only the GPS system data.

We have extensively tested the results of these models on our data,
the details of which are in Appendix~\ref{sec:ALBUS}.  The summary
conclusion is that the 2-dimensional `G01' model, when utilizing the
closest GNSS station for which the timing bias value is known,
provides the best estimate of the IFRM for both the VLA and MeerKAT.
For the rest of this paper, when utilizing the imaging results
including IFRM corrections, we have used the G01 model.

\section{Solar System Bodies as EVPA Calibrators}\label{sec:Planets}

A proper assessment of the accuracy of an estimation of the IFRM
requires observations of a source of {\it a priori} known EVPA.  This
is required because, as will be shown in subsequent sections, although
all IFRM estimates based on the global VTEC maps are effective in
removing the day to night change in the IFRM, they also generate (at
least for the VLA and MeerKAT) a significant overestimation of the
VTEC, resulting in an overcorrection of the IFRM rotation.

The standard set of strongly polarized calibrator sources (such as
3C286 and 3C138) are not sufficient for the purpose of assessing the
accuracy of these corrections for two reasons.  First, their intrinsic
EVPA values are a function of frequency, and are not known at the
frequencies where ionospheric Faraday rotation becomes important, so
that one cannot determine if an IFRM-corrected image provides the
correct EVPA.  Secondly, they are completely depolarized at
frequencies below 500 MHz, where the ionopheric Faraday rotations
become especially important. An additional point is that, with the
exception of 3C286, compact polarized sources are time-variable in
both total and polarized flux density on timescales of weeks to
years. Similarly, although observations of pulsars have been
successfully used to judge diurnal IFRM correction accuracy
\citep[see,][]{Porayko19,Porayko23}
  they cannot be used to
  judge the accuracy of the corrected EVPA, since their intrinsic EVPA
  values are not known.

\begin{figure}{}
  \gridline{\fig{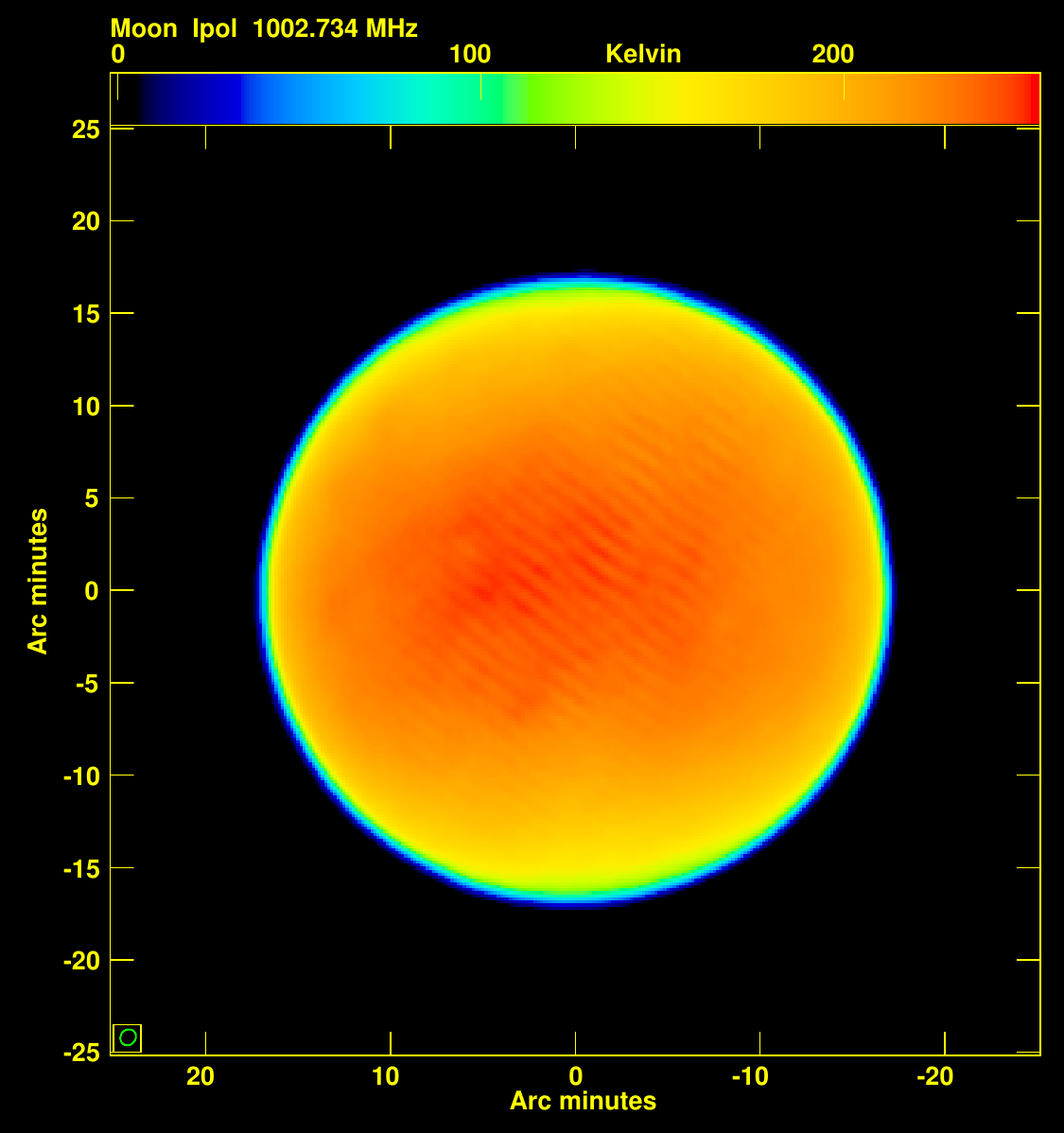}{0.33\textwidth}{Brightness Temperature}
            \fig{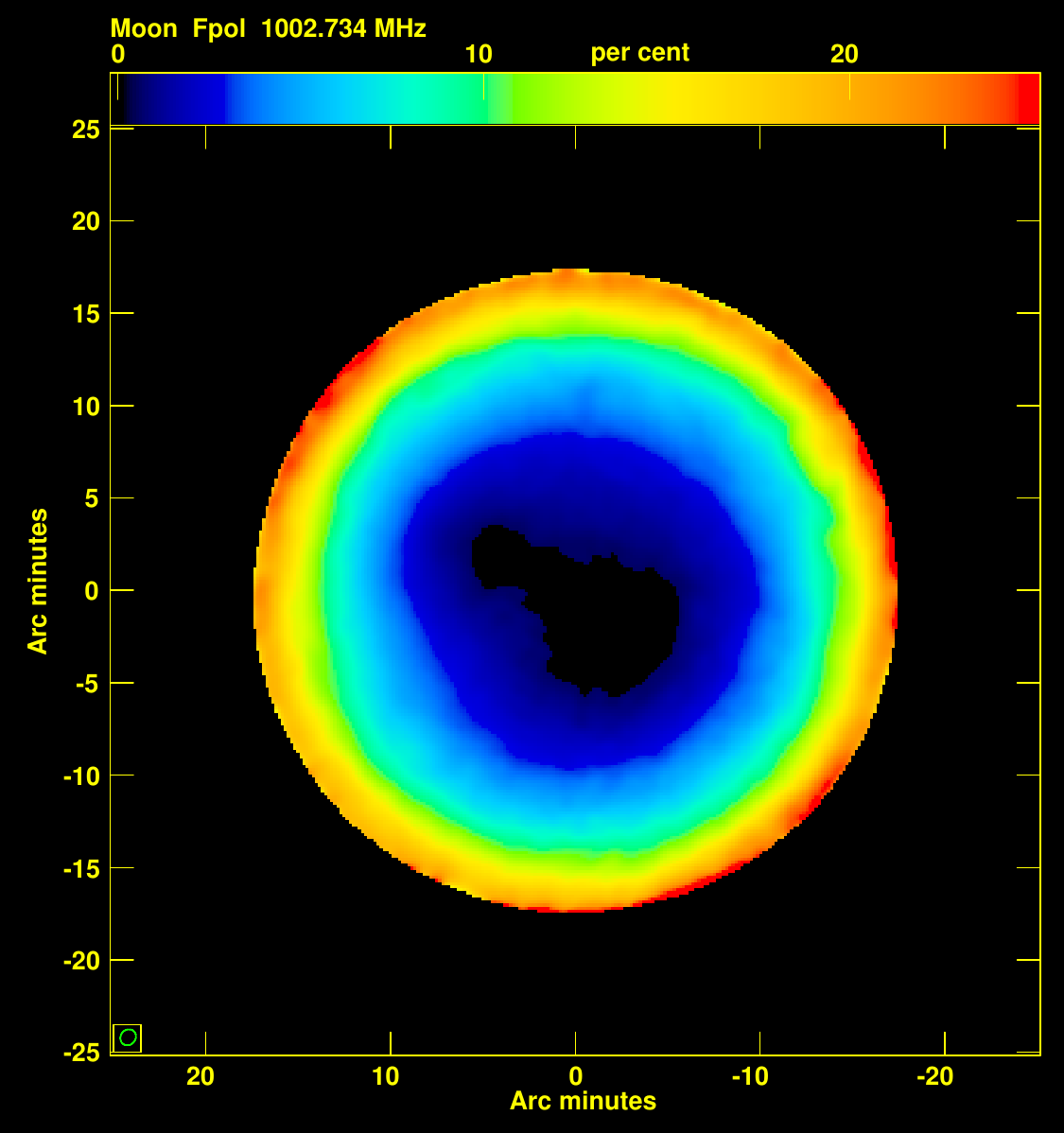}{0.33\textwidth}{Fractional Polarization}
            \fig{Moon-1002-EVPA.pdf}{0.33\textwidth}{E-Vector Orientation}}
  \caption{Color-coded MeerKAT images of the Moon at 1002 MHz. The
    Stokes I brightness (left) is nearly featureless while the
    fractional polarized brightness (center) is strongly peaked toward
    the limb, with fractional polarization reaching 30\% near the
    lunar limb.  The EVPA vectors (right) are nearly radial. The
    slight ripple seen in the Stokes I image is from a deconvolution
    instability.}
  \label{fig:MoonImages}
\end{figure}

To our knowledge, sources of wideband polarized radio emission with
{\em a priori} known EVPA distributions that are useful (have enough
flux density, are the proper size, etc.) are limited to the Moon and
the planets Mercury, Venus, and Mars.  For these objects, the
originally unpolarized thermal radio emission originates from 10 -- 20
$\lambda$ below the surface, and becomes linearly polarized due to
differential refraction of the orthogonal planes of polarization upon
passage through the planetary surface. The physical principles which
cause this polarization are discussed by \citet{Heiles63} and
\citet{PB13}. The resulting observed emission is highly polarized near
the limb of the body with a nearly perfect radial orientation.  A good
example of lunar polarimetric images using data taken with MeerKAT is
shown in Fig.~\ref{fig:MoonImages}.

Imaging of the polarized thermal emission of these extended objects is
difficult, especially at frequencies below 2 GHz, as their surface
brightnesses and flux densities decline with the square of the
frequency.  The observational requirements for effective planetary
polarimetry are that the Stokes I surface brightness be sufficiently
high that the detected flux density within the array resolution beam
has an SNR of at least 20 on a timescale comparable to any ionospheric
variations, and that there be at least three, and preferably more than
five resolution elements across the source.  The result of these two
requirements for VLA observations is that only the Moon can be used at
frequencies below 1 GHz, and only when the VLA is in its compact D or
C configurations. Additionally, the 25-meter diameter of the VLA
antennas limits lunar polarimetry to frequencies below 1.5 GHz.  For
higher frequency work, Mars can be utilized for polarimetry at
frequencies above 4 GHz, and Venus for frequencies between 1 and 10
GHz (the upper limit being due to absorption in the Venusian
atmosphere).

For MeerKAT, only the Moon can be utilized for polarimetric studies,
as the array resolution is too low for the other planets.  The
high central concentration of MeerKAT stations, and the 13.5 meter
antenna diameter, make this array ideal for lunar polarimetric
observations in its lowest two frequency bands.

The VLA's sensitivity in P-band is significantly degraded by strong
mutual coupling between the antennas on baselines less than $\sim$ 500
m, and by a poor RFI environment due to various satellite and local
transmissions.  Nearly 50\% of the bandwidth is lost for observations
in the compact C and D configurations due to RFI.  The result of
these factors is that on-source integrations of at least 10 minutes
are required to obtain sufficiently good polarimetric sensitivity for
lunar observing at 300 MHz.  These issues are not a problem for VLA
observations at higher frequencies thanks to interferometer
fringe-winding attenuating the RFI and mutual coupling, nor for the
MeerKAT array, for which careful design has virtually eliminated
detectable signal coupling between antennas.  The isolated location of
the MeerKAT array combined with tight control of local sources of RFI
result in less than 5\% of the bandwidth being lost to RFI in the UHF
band.

\section{Observations} \label{sec:Observations}

Multiple observations at multiple frequency bands have been taken with
the VLA and MeerKAT for this program.  We utilize `band codes' to
simplify presentation and discussion of the results.
Table~\ref{tab:Bands} gives the band codes and frequencies spanned
by the VLA and MeerKAT receivers used in this study.
\begin{deluxetable}{ccc|ccc}
	\label{tab:Bands}
    \tablewidth{0pt}
    \tablecaption{VLA and MeerKAT Band Codes}
\tablehead{ \colhead{Array} & \colhead{BandCode} &
  \colhead{Freq. Range (GHz)} &
            \colhead{Array} & \colhead{BandCode} &
            \colhead{Freq. Range (GHz)}}
\startdata
VLA & P& 0.23 --0.45 & VLA    & K   & 18 -- 28\\
VLA & L& 1.0 -- 2.0  & VLA    & A   & 28 -- 40\\
VLA & S& 2.0 -- 4.0  & VLA    & Q   & 40 -- 50\\
VLA & C& 4.0 -- 8.0  & MeerKAT& UHF & 0.54 -- 1.10\\
VLA & X& 8.0 -- 12.0 & MeerKAT& L   & 0.85 -- 1.70\\
VLA & Ku&12.0 -- 18.0 & MeerKAT& S0  & 1.75 -- 2.62\\
\enddata
\end{deluxetable}

Accurate assessment of the IFRM corrections requires multiple extended
observations of the planetary bodies, especially at the lower
frequencies where the emission is weakest and IFRM rotation
corrections are large. We have thus observed the Moon with the VLA on
10 occasions, and with MeerKAT on six, as listed in
Table~\ref{tab:MoonData}.  In this table, column \#4 gives an
observation code which is extensively used in the following sections
to concisely identify which observation is being discussed.

VLA observations of the planets Venus and Mars, used to establish the
high-frequency EVPA for 3C286 and 3C138, are summarized in
Table~\ref{tab:PlanetData}.
\begin{deluxetable}{lccclccc}
\tablewidth{0pt} \tablecaption{Polarimetric Lunar Observations with
  the VLA and MeerKAT} \tablehead{\colhead{Date} &
  \colhead{Array/Config} & \colhead{Band} &\colhead{Code}&
  \colhead{Polarized Sources} & \colhead{Separation} & \colhead{IFRM
    Range} & \colhead{ToS} \\ \colhead{} & \colhead{} & \colhead{} &
  \colhead{} & \colhead{} &
  \colhead{Degrees}&\colhead{$\mathrm{rad/m^2}$} & \colhead{min}}
\startdata 10 May 2017 & VLA-D & P & D1 & DA240, 3C345 & T171 & 0.4 --
0.7 & 9x15 \\
06 Aug 2017 & VLA-C & P & C1 & 3C345 & T162 & 0.6 -- 1.4& 56x5\\
30 Dec 2018 & VLA-C & P & C2 & DA240, 3C303, 3C345 & L75 &0.3 -- 1.5 & 58x8\\
31 May 2021 & VLA-D & L & L1 & 3C286, 3C138 & L112& 0.3 -- 2.4 & 18x20\\
22 Jun 2021 &MeerKAT& UHF&U1 & 3C286 & T150&$-0.10$ -- $-0.04$ & 17x10\\
26 Jun 2021 &MeerKAT& S0& MS & 3C286 &L156 &$-0.10$ -- $-0.05$ & 16x10\\
18 Aug 2021 &MeerKAT& L & ML &3C286 & T128 &$-0.25$ -- $-0.15$ & 19x10\\
20 Sep 2022 & VLA-D & L & L2 & 3C286, 3C138 & L59 & 1.4 -- 2.3 & 30x10\\
13 Dec 2022 & VLA-C & P & C3 & DA240, 3C303, 3C345 & L122 & 0.3 -- 1.2 & 32x10\\
04 Nov 2023 &VLA-D  & P & D2 & DA240 & L100 & 0.4 -- 2.4 & 14x5\\
20 Jan 2024 &VLA-D$\Rightarrow$C&P&DC&DA240 & T125 & 0.5 -- 2.5 & 22x10\\
09 Mar 2025 & VLA-D & P & D3 & DA240, 3C303, 3C345 & T120 & 0.5 -- 2.7 &13x1.5\\
09 Mar 2025 & VLA-D & L & L3 & DA240, 3C303, 3C345 & T120 &0.5 -- 2.7 & 13x3\\
18 Aug 2025 &MeerKAT& UHF&U2 & 3C138 & L64 &$-2.5$ -- $-0.4$ & 23x8\\
19 Aug 2025 &MeerKAT& UHF&U3 & 3C138 & L51 &$-3.0$ -- $-0.3$ & 23x8\\
19 Aug 2025 & VLA-C$\Rightarrow$B&P&CB&DA240& L46 & 0.3 -- 2.4 &19x14\\
30 Oct 2025 &MeerKAT& UHF&U4 & 3C286 &T100 & $-0.7$ -- $-2.1$ & 18x8\\
\enddata
\tablecomments{{\bf Separation:} The angular offset between the Moon and the Sun.  `T'
  means the Moon trails the Sun, `L' means the Moon leads the Sun.
  {\bf ToS:} Time on Source, given as the number of observations times
  the duration of each in minutes.  Proposal codes for the MeerKAT
  data are EXT-20220902-BH-01 for the 2021 and 30 Oct 2025
  observations, and EXT-20220615-BH-01 for the August 2025
  observations.}
\label{tab:MoonData}
\end{deluxetable}
\begin{deluxetable}{cclccccc}
\tablewidth{0pt} \tablecaption{Planetary Observations with the VLA}
\tablehead{ \colhead{Date} & \colhead{Config} & \colhead{Bands Used} &
  \colhead{Planet} &\colhead{Solar Separation} & \colhead{IFRM
    Range}&\colhead{Ang.Size}\\
  \colhead{} & \colhead{} & \colhead{} &
  \colhead{} & \colhead{Degrees} & \colhead{$\rmm$}&\colhead{arcsec}}
\startdata
03 Mar 2022 & A & L, S      & Venus & 45 & 1.2 -- 3.0& 30 \\
14 Apr 2022 & A & C, X      & Mars  & 35 & 0.6 -- 4.5& 5.4 \\
31 Jan 2025 & A & L, S      & Venus & 45 & 2.0 -- 4.5& 32 \\
31 Jan 2025 & A & C, X, U   & Mars  & 158& 0.5 -- 1.3& 13.\\
09 Mar 2025 & D & S, C, X   & Venus & 21 & 3.4 -- 5.1& 55 \\
09 Mar 2025 & D & U, K, A, Q& Mars  & 118& 0.5 -- 3.0&10.0\\
\enddata
\tablecomments{The 2025 observations included all
  bands, but only those shown in the table could be used for this work
  due to array resolution or sensitivity limits.  Venus cannot be
  used above 10 GHz due to absorption in its atmosphere.}
\label{tab:PlanetData}
\end{deluxetable}

Editing and calibration of all data were done using the {\tt AIPS}
software package using well-established procedures for calibration
utilizing the parallel-hand correlations.  Polarimetry requires
calibration of the cross-hand correlations, the details of which
differ between linearly and circularly polarized systems.  The VLA's
high frequency receivers utilize circularly polarized receivers, while
MeerKAT and the VLA's P-band receivers utilize linearly polarized
receivers.  In the following paragraphs, we detail the procedures.

For both linear and circular systems, calibration of the antenna
cross-polarization (so-called D-terms) was done using known unpolarized
sources, for which the procedure is the same.  We utilized 3C84 for
the VLA high-frequency observations, 3C147 (J0542+4951) for the VLA's
P-band obsrvations, and J0408-6545 or J1939-6342 for the MeerKAT
observations.

The critical difference between circularly-polarized and
linearly-polarized receivers is the effect of a cross-hand phase
offset between the parallel-hand channels which causes rotation
between the Stokes parameters.  Standard parallel-hand calibration
references the opposite hand phases (R and L, or H and V) to those of
the chosen reference antenna.  The phase difference between the
parallel-hand receivers of this antenna is imprinted on all cross-hand
correlations.  For circularly-polarized system, a residual phase
offset of $\delta\phi$ is manifested by a rotation between Stokes Q
and U by an angle $\delta\chi = 0.5 \delta\phi$.  For linear systems
(after correction for the antenna parallactic angle), the rotation is
between Stokes U and Stokes V\footnote{These rotations are best
  understood as rotations in the Poincar\'e sphere about the V axis
  for a circular system, and about the Q axis for a linear system.}.

In both cases, correction of these phase offsets requires either an
on-board cross-hand calibration system, or observations of a strongly
polarized source.  The VLA does not have an on-board cross-hand
calibration system, so an observation of a strongly polarized source
of known EVPA is required for determination of this cross-hand phase
offset for its circular-based receivers.  The stable highly polarized
source 3C286 is most commonly utilized for this purpose.

For linear systems, we take advantage of the fact that these polarized
sources have negligible circularly polarized emission, and utilize the
V=0 method detailed in \citet[][EVLA Memo 219]{PGH2022}.  The sources
utilized were DA240, 3C303, or 3C345 for the VLA's P-band
observations, and either 3C286 or 3C138, both of which remain
sufficiently polarized down to 544 MHz, for the MeerKAT observations.

Conversion of the linearly-polarized MeerKAT data to a circular basis
was done as described by \citet[][EVLA Memo 229]{PG24}, in order to
enable better calibration using linearly polarized calibrator
sources\footnote{This conversion was not needed for the
  linear-polarized VLA P-band data, since all calibrators utilized for
  these observations are known to be unpolarized.}.

The ALBUS G01 model, using data from the nearest well-calibrated
GNSS station (`pie1' for the VLA, and `suth' for MeerKAT) was used to
estimate the IFRM for all frequencies up to 4 GHz.

\begin{figure}{}
  \gridline{\rotatefig{-90}{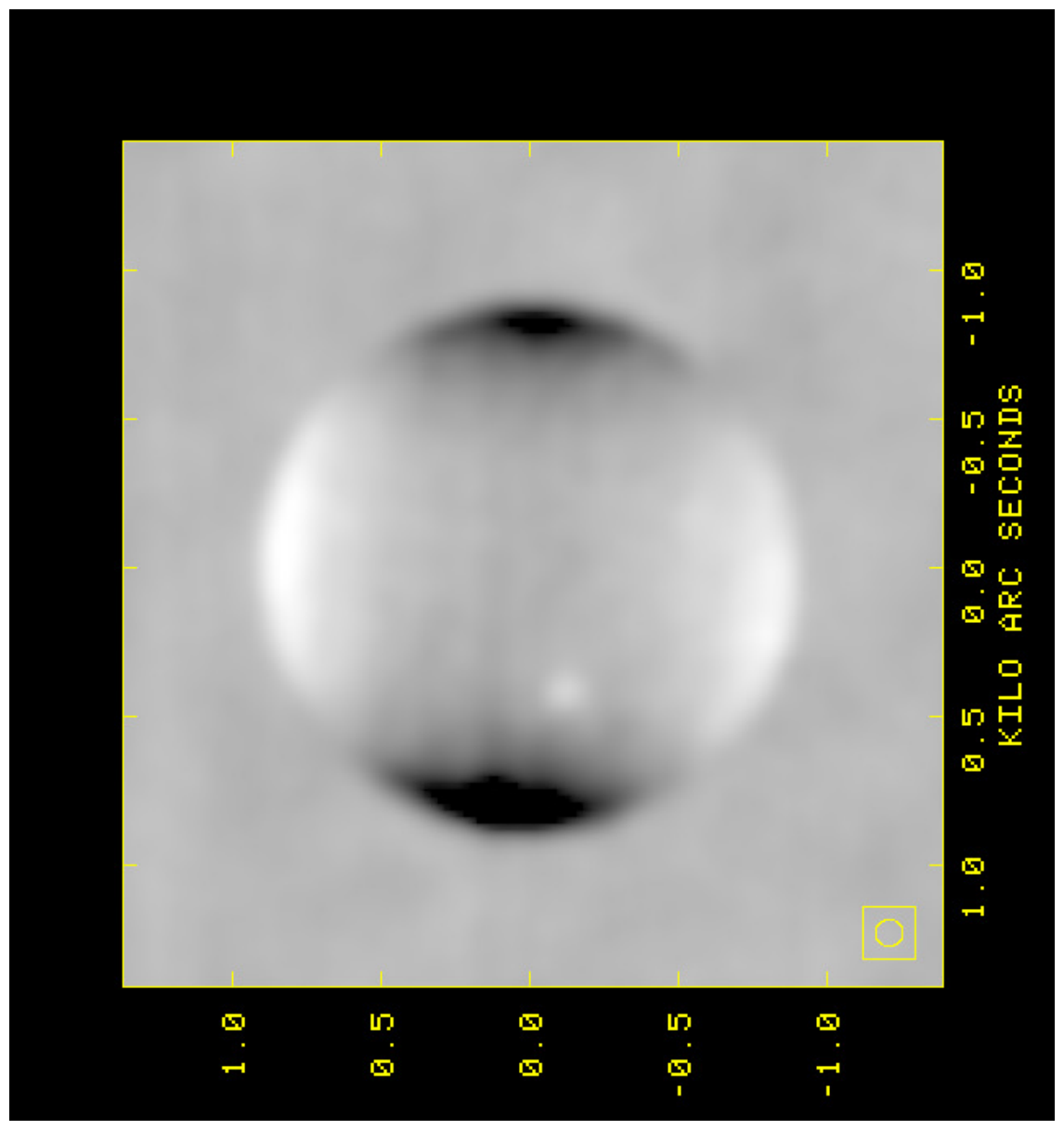}{0.25\textwidth}{Stokes Q}
    \rotatefig{-90}{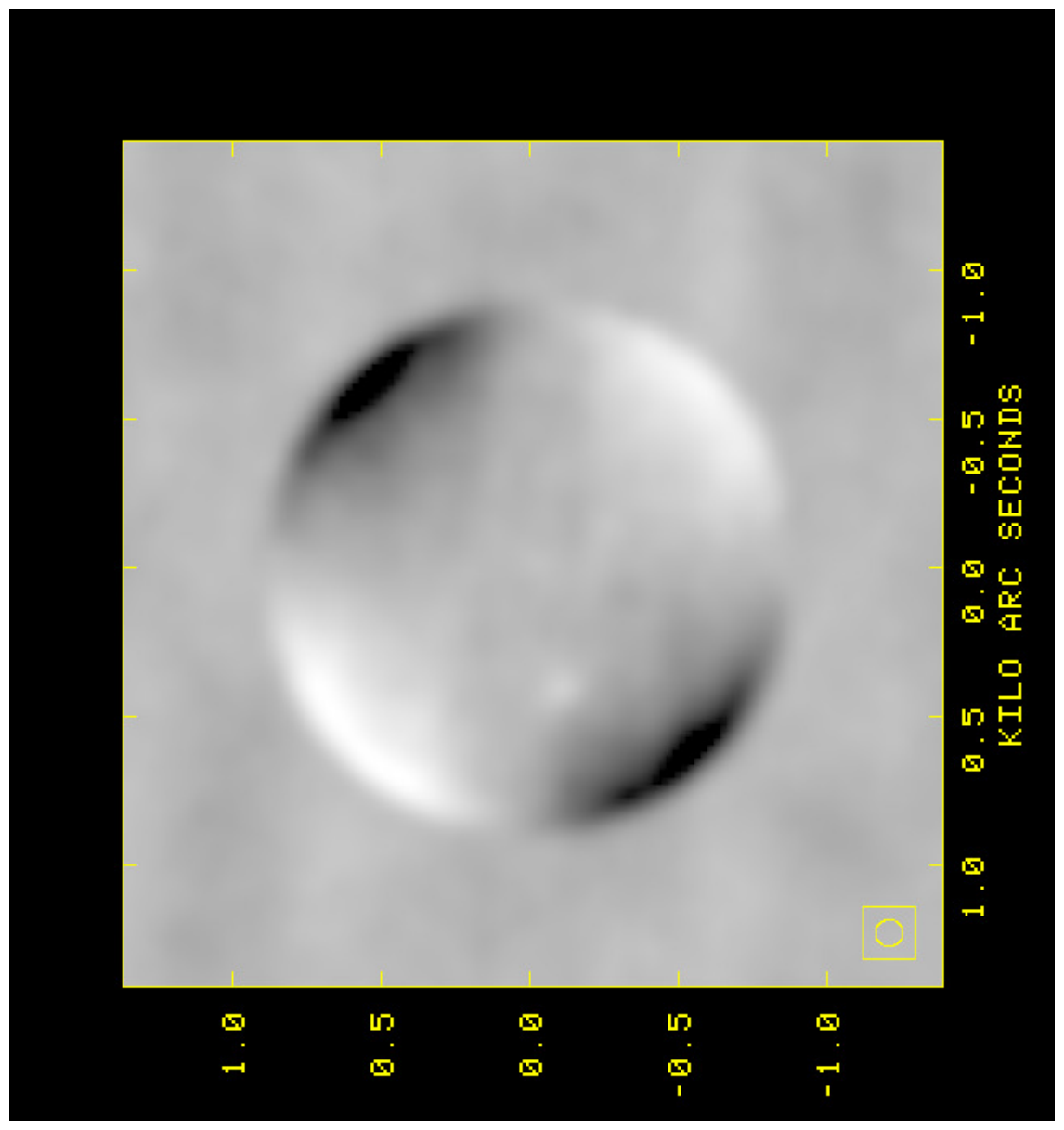}{0.25\textwidth}{Stokes U}
    \rotatefig{-90}{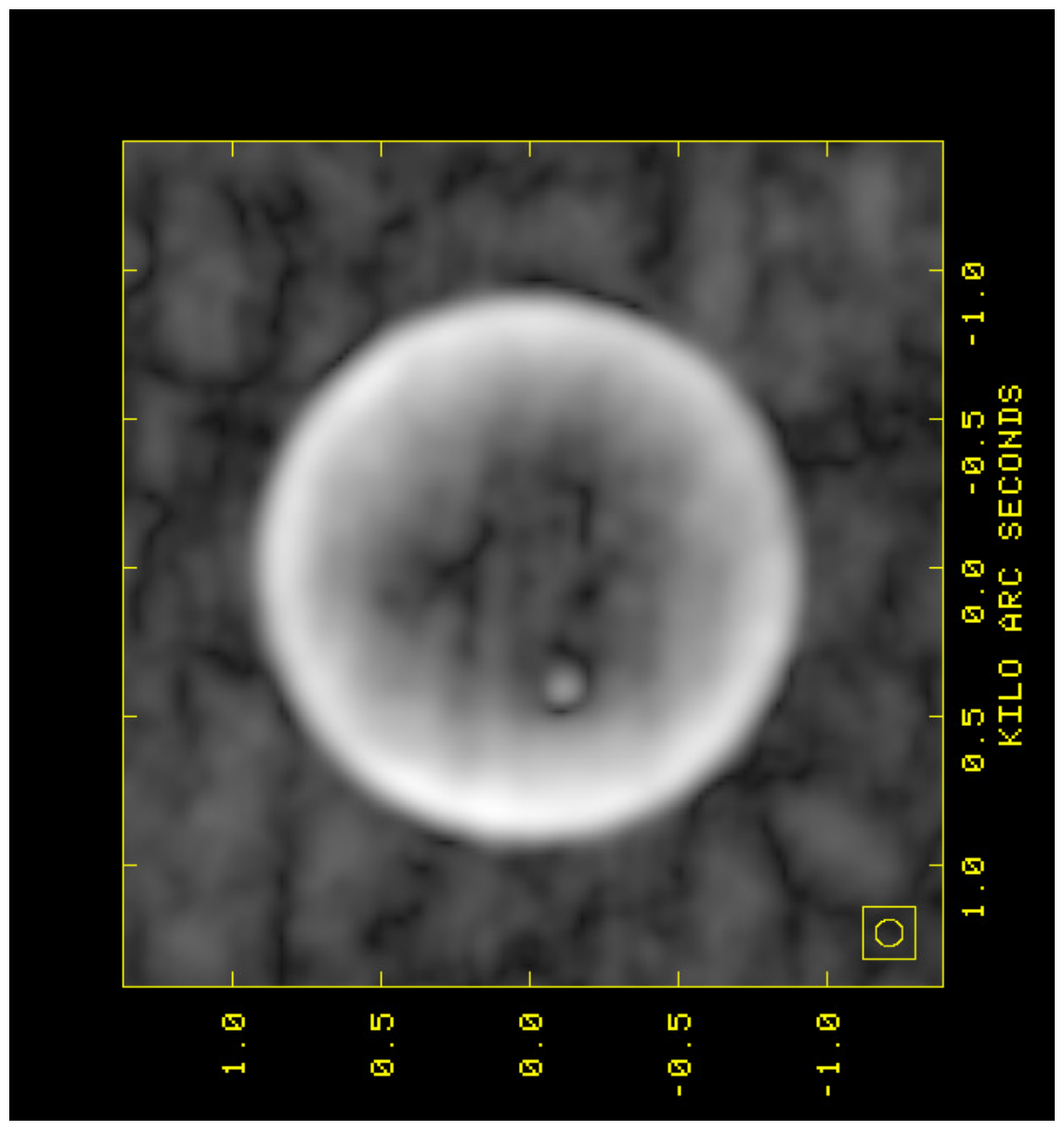}{0.25\textwidth}{Polarization Intensity}
  \fig{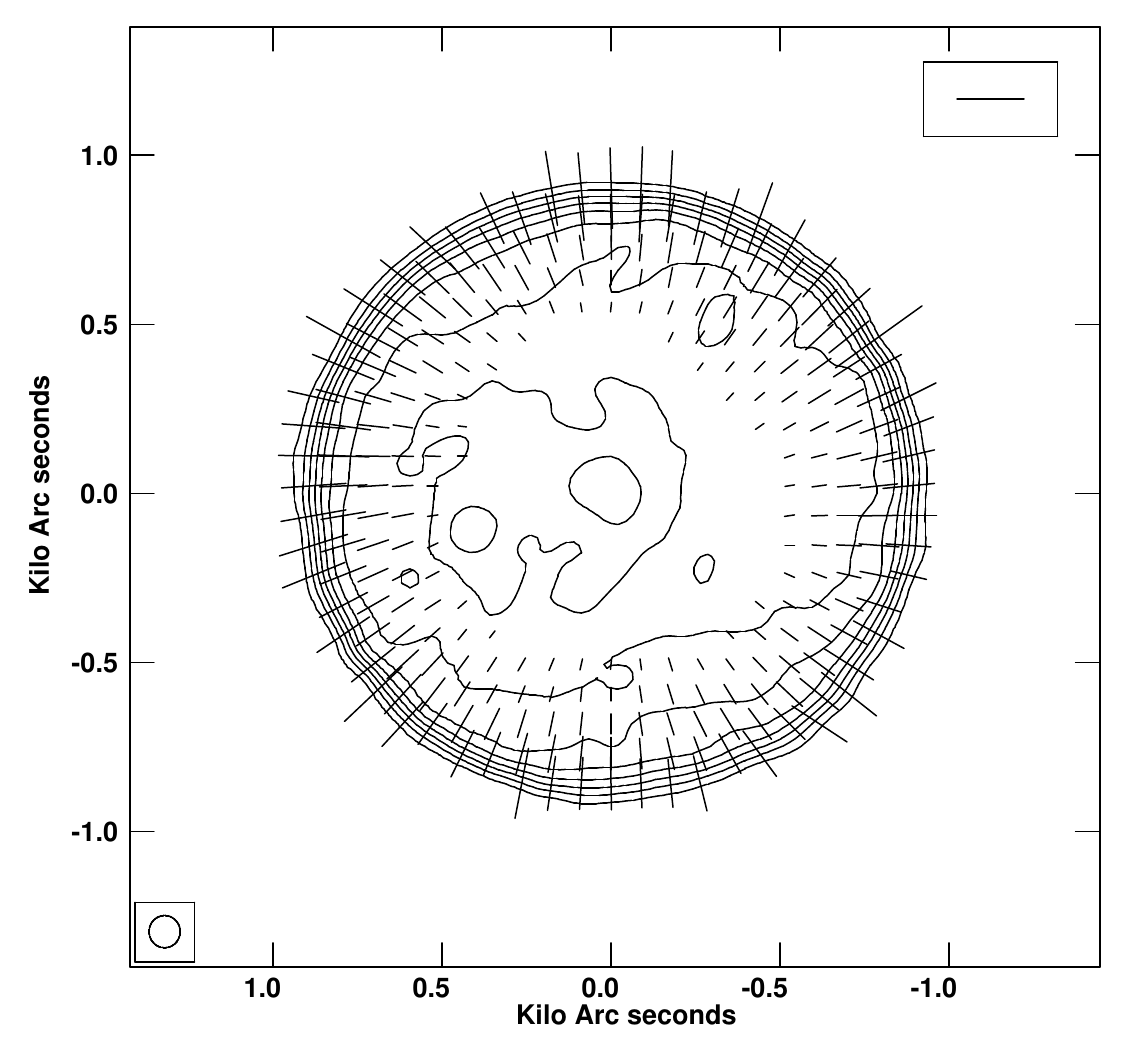}{0.25\textwidth}{EVPA Image}}
  \caption{Showing the image products from the C3 observation.  The
    bright spot in the three leftmost panels is due to reflection of
    solar emission.}
  \label{fig:C3}
\end{figure}

Imaging was done utilizing the {\tt AIPS} task {\tt IMAGR}, applying
the various estimates of the IFRM on timescales appropriate to
tracking the changes in the EVPA.  An example of the lunar images from
the VLA `C3' observation is shown in Fig.~\ref{fig:C3}.  The prominent
emission spot seen in the Q, U, and polarized intensity image is due
to reflected solar emission, which is orthogonally polarized to the
lunar thermal emission.

As noted in Section~\ref{sec:Planets}, the instrinsic EVPA
distribution for the moon and planets is known to be radial --
providing the image resolution is lower than the scale size of
planetary features.  An incorrect removal of the ionospheric rotation
-- or an error in the data calibration or image processing -- will
show up as an offset of the observed EVPA from radial around the limb
of the polarization image.  To obtain a quantitative measure of this
rotation, the {\tt AIPS} program {\tt MARSP} was used to determine the
mean and standard deviation of the EVPA offset from radial for
circularly symmetric objects.  The program permits specification of
the minimum brightness between the inner and outer radii of an annulus
within which the data are to be included for processing.

\section{VLA Results} \label{sec:Results}

The procedure for determining the accuracy of the various IFRM
estimates was to generate both uncorrected and IFRM-corrected lunar
polarization images as a function of time and frequency, applying
different estimates of the IFRM.  These images were then analysed for
the mean offset from the known intrinsic radial EVPA.

\begin{figure}{hbt}
  \gridline{\fig{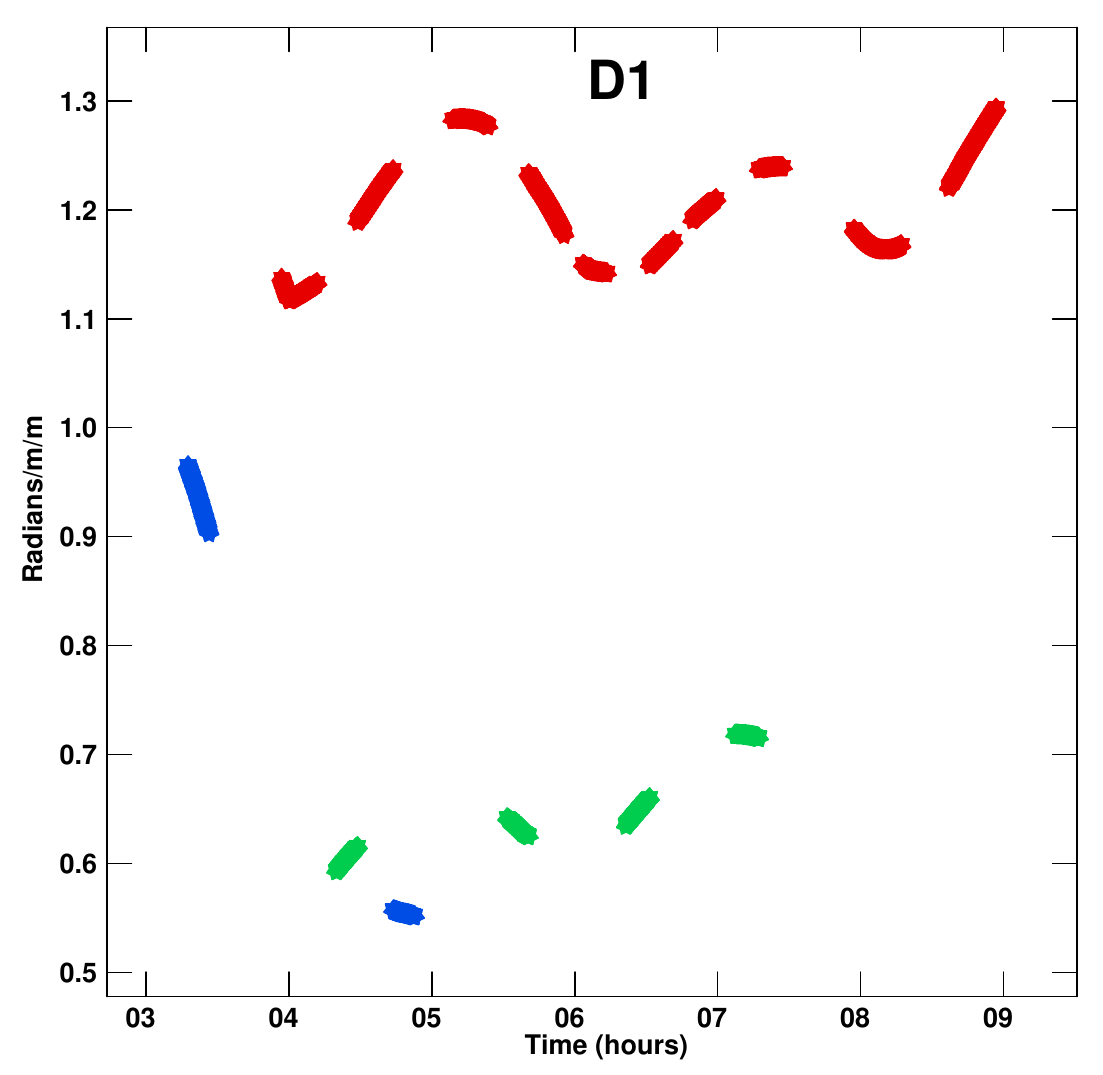}{0.33\textwidth}{10May2017}
            \fig{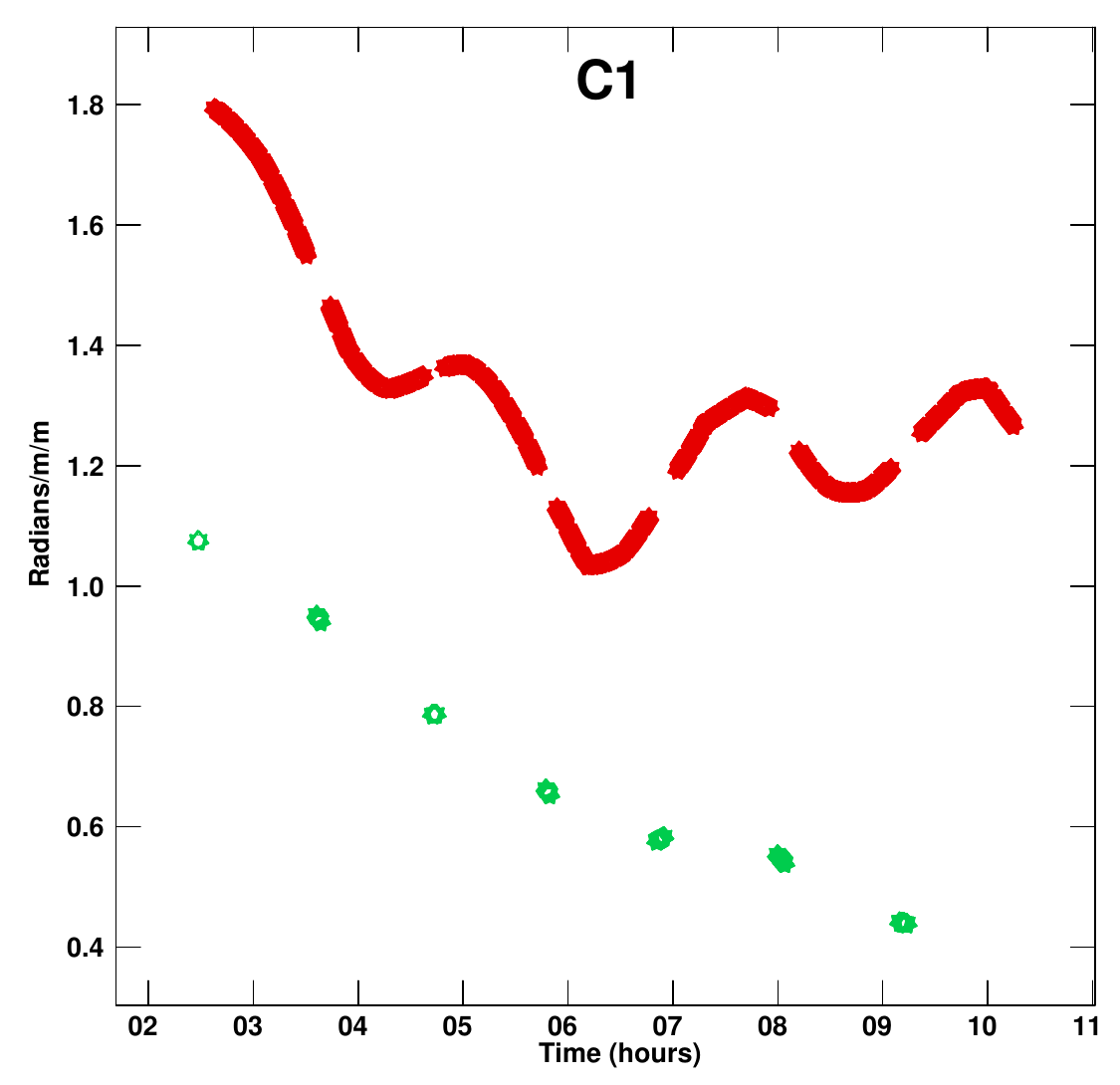}{0.33\textwidth}{06Aug2017}
            \fig{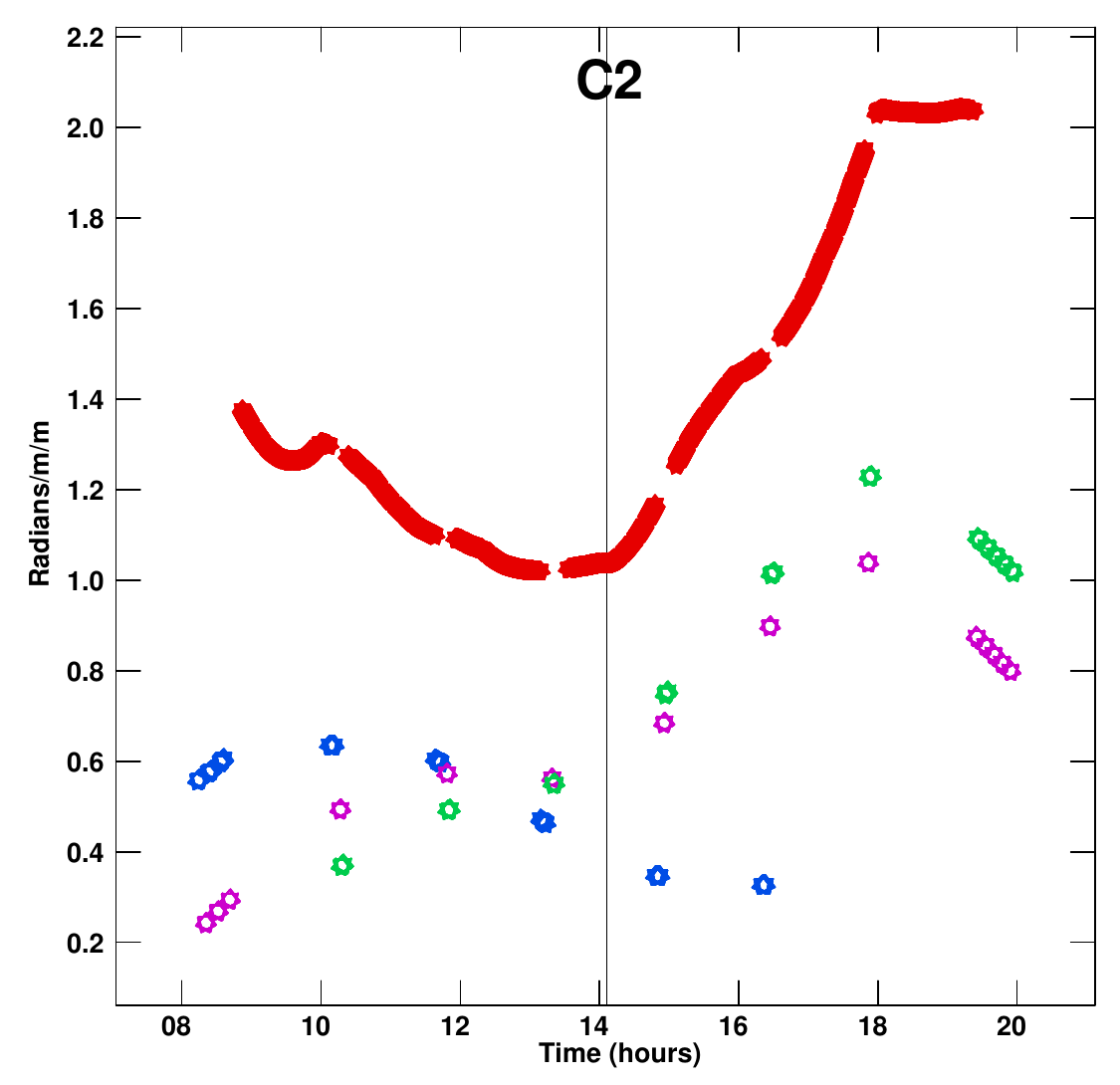}{0.33\textwidth}{30Dec2018}}
  \gridline{\fig{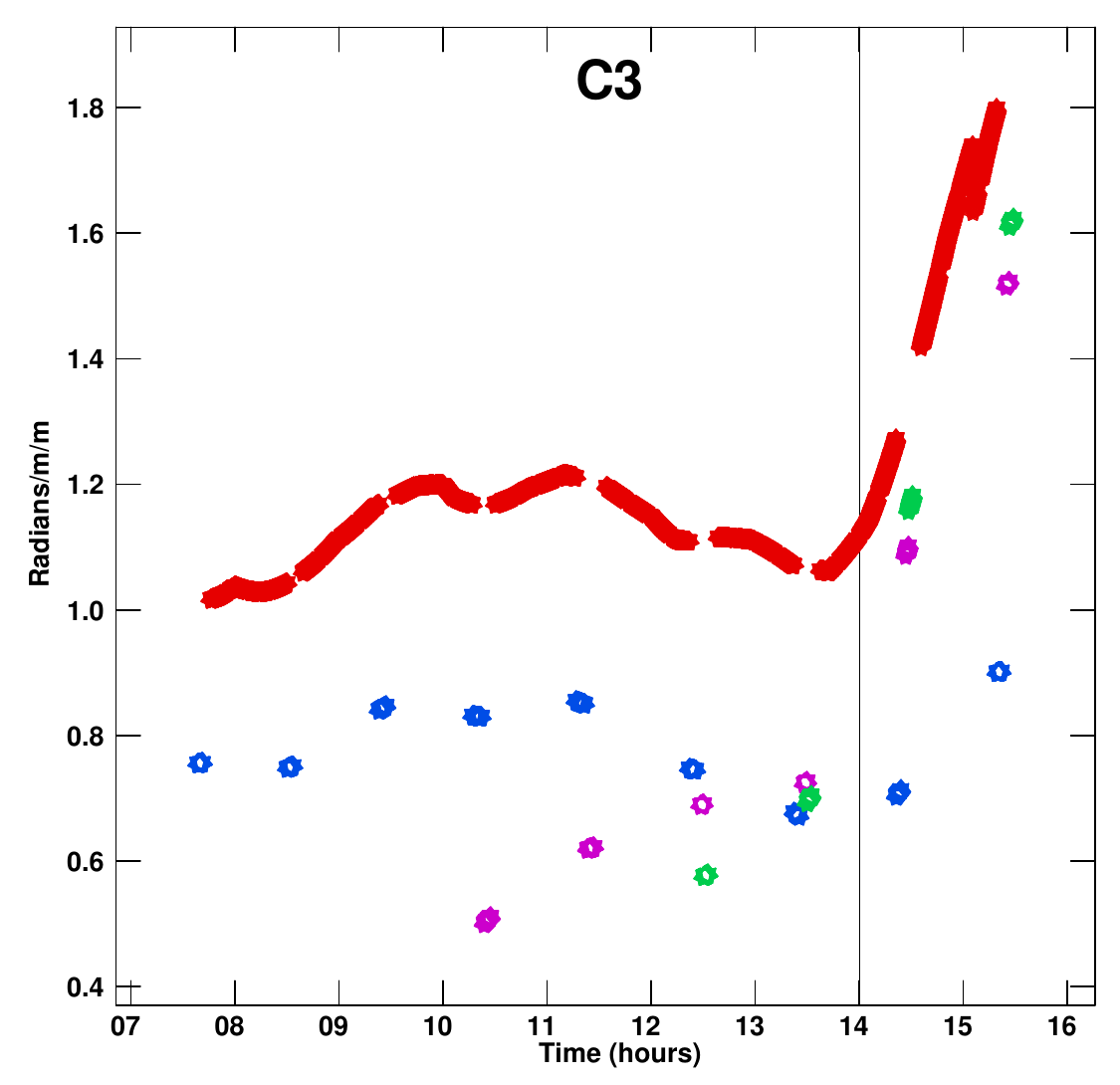}{0.33\textwidth}{13Dec2022}
            \fig{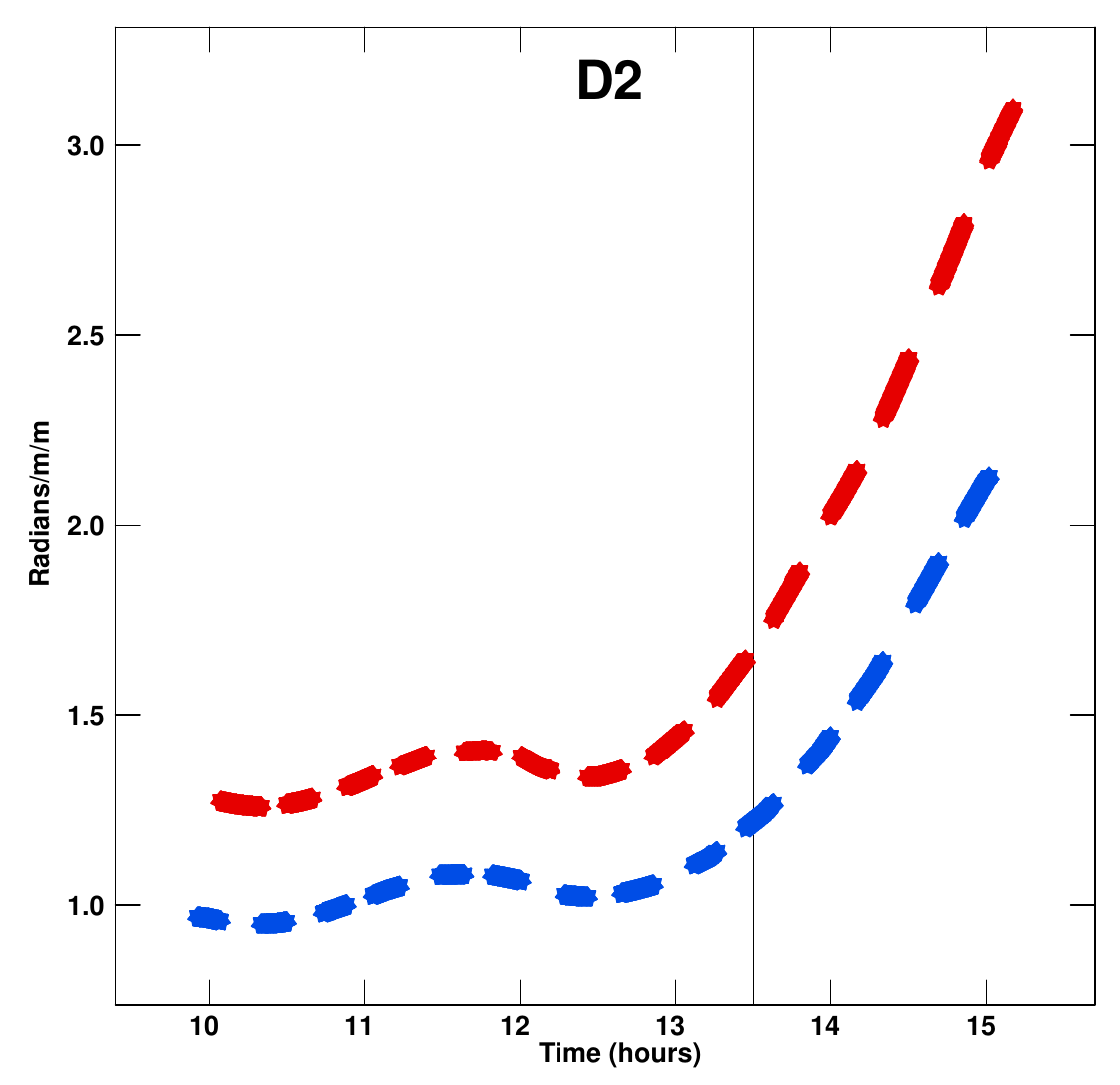}{0.33\textwidth}{04Nov2023}
            \fig{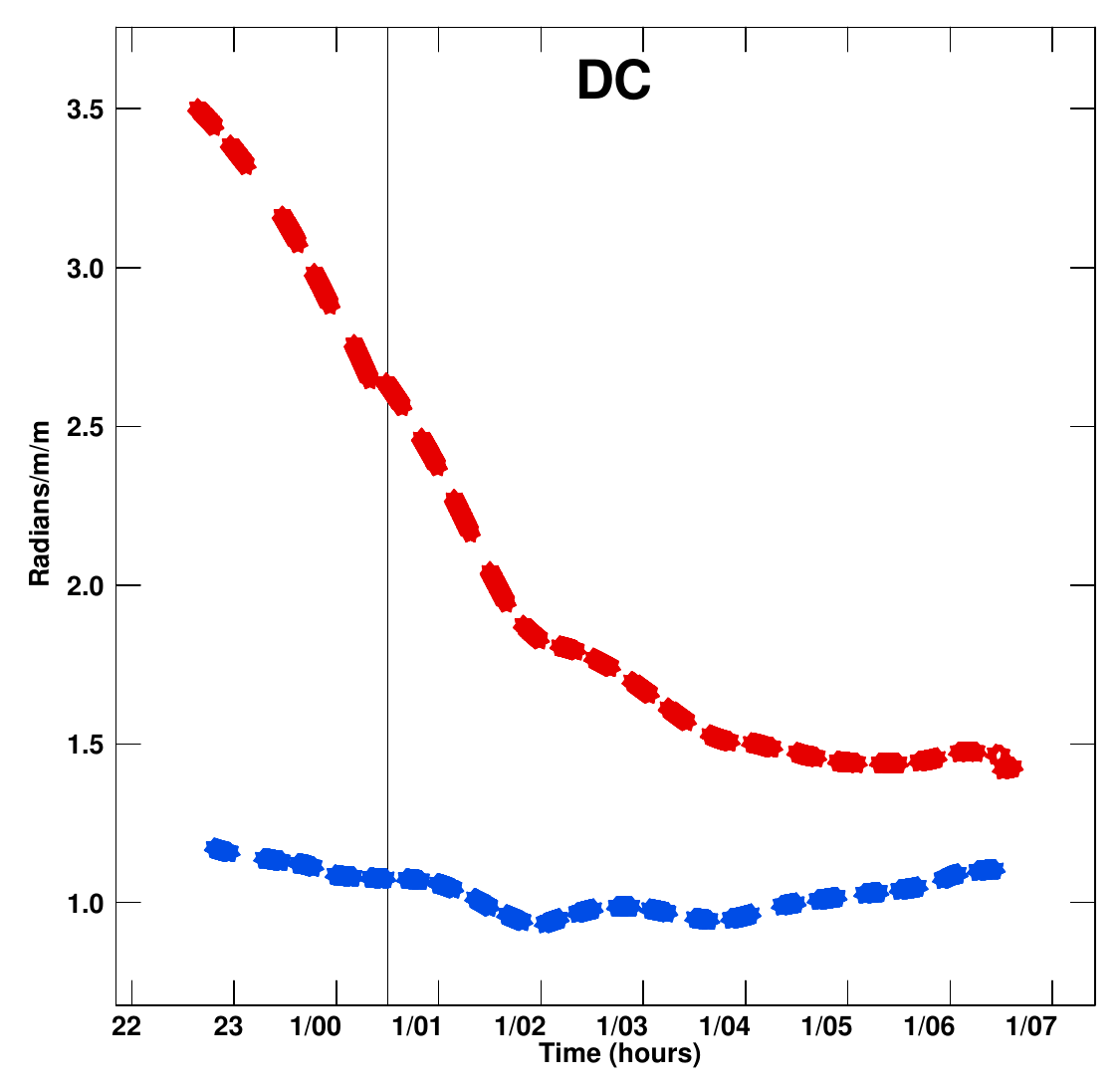}{0.33\textwidth}{20Jan2024}}
  \caption{Examples of the estimated lunar IFRMs by TECOR, using the
    `jplg' gobal VTEC maps.  Time is in UTC hours -- subtract seven for
    local MST.  Sources are color-coded: red for the Moon, blue for
    DA240, purple for 3C303, and green for 3C345.  The vertical line
    indicates time of local sunrise or (for the DC observation)
    sunset.}
  \label{fig:jplg}
\end{figure}

Examples of the IFRM estimates generated by the `jplg' VTEC maps are
shown in Fig~\ref{fig:jplg} for six of the P-band VLA lunar
observations.
The red traces in each panel show the IFRM estimates for the Moon.
The polarized calibrator IFRM estimates are shown in color -- blue for
DA240, purple for 3C303, and green for 3C345.  Of these six
observations, the last two -- `D2' and `DC' -- proved the most useful
for our study, as these observations spanned sunrise or sunset, and
the changes in IFRM during the observation period were larger and
smoother than in the other observations.

\subsection{IFRM Estimation} \label{sec:IFRM Removal}

\begin{figure*}[hbt]
 \epsscale{1.1}
\plottwo{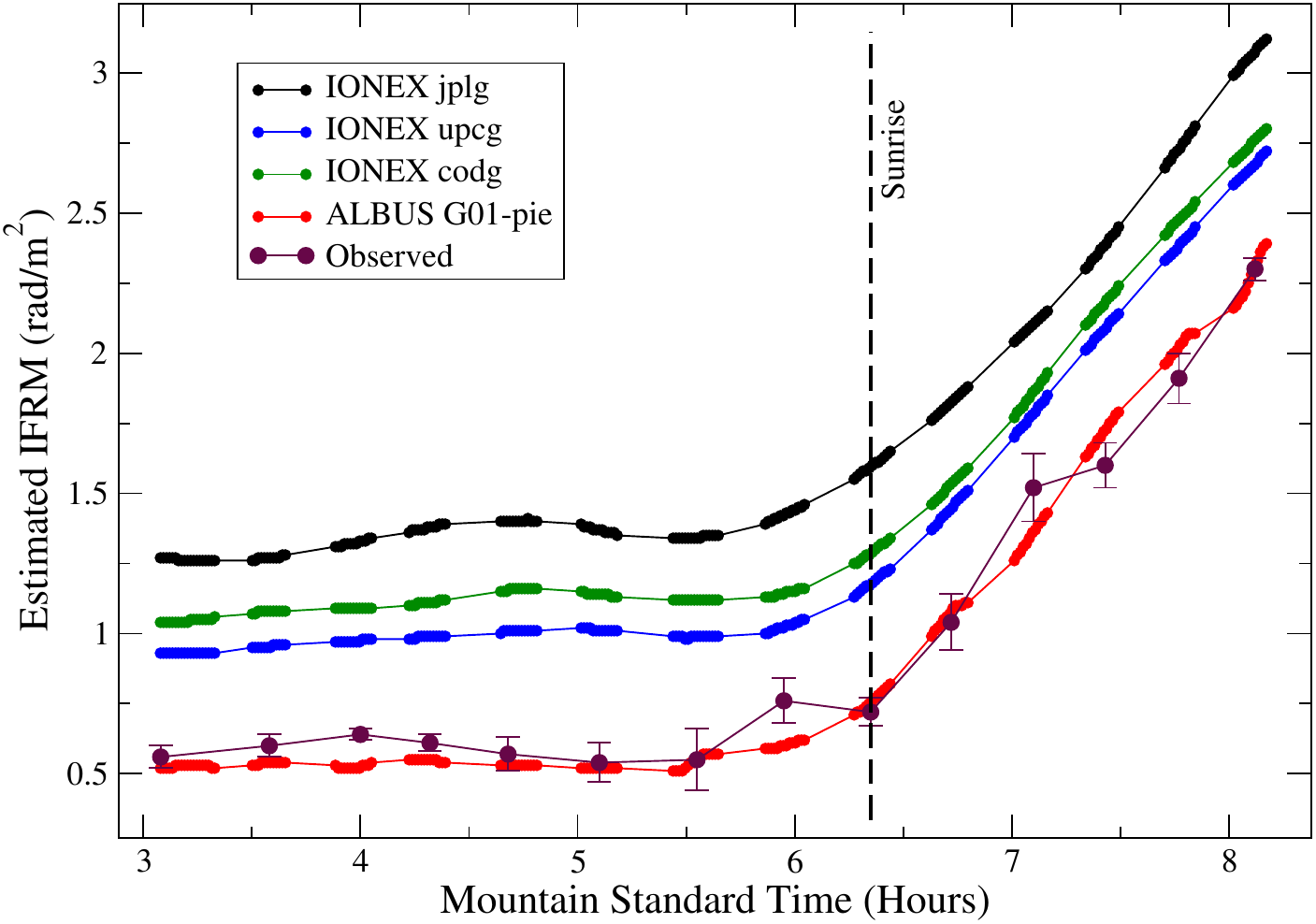}{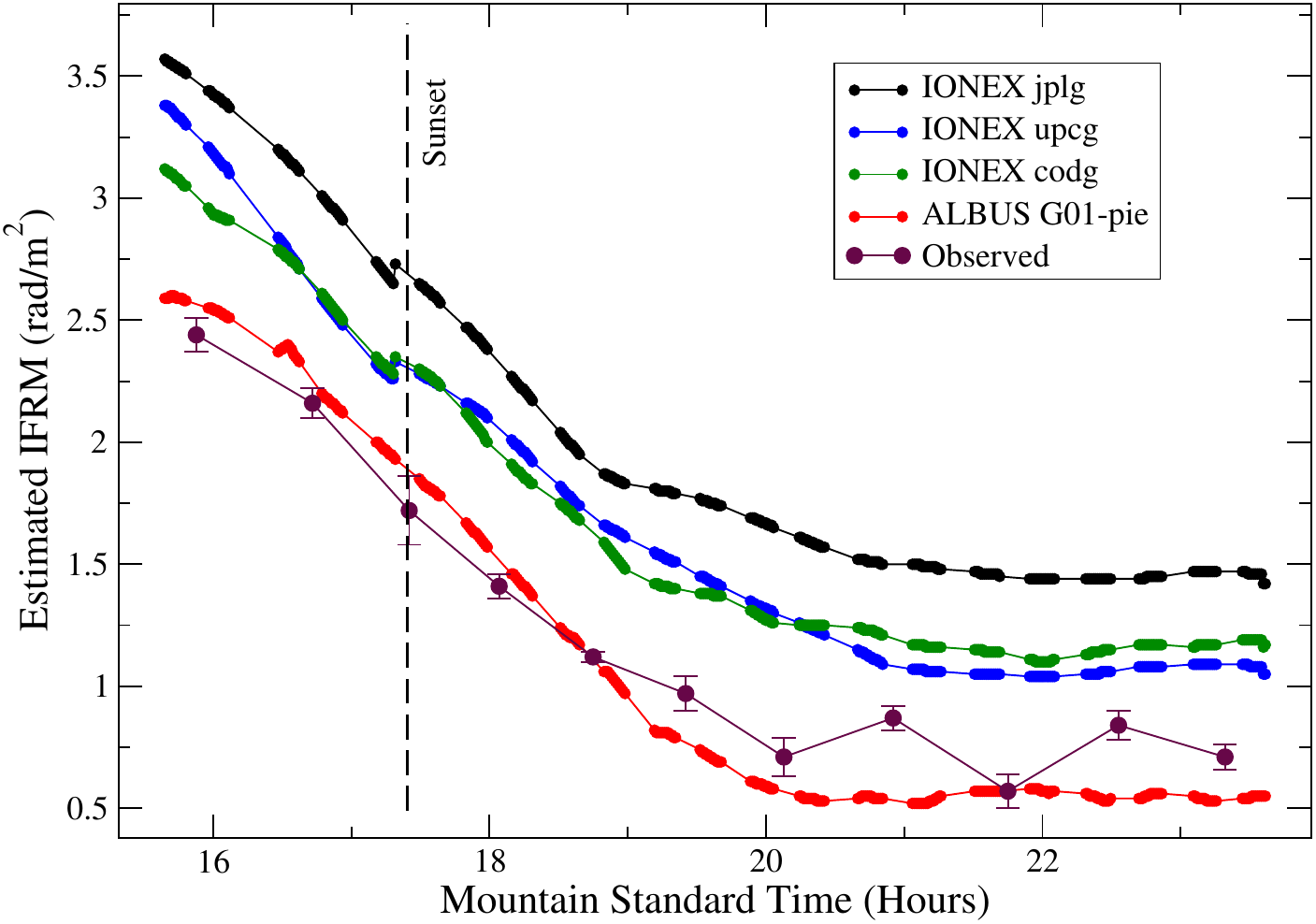}
\caption{The lunar IFRM estimates from three global and two local
  models, for the `D2' observation (left) and the 'DC' observation
  (right). The three displayed global estimates are representative of
  all seven global maps we have investigated.  The vertical dashed
  line marks sunrise or sunset.  The plotted data points (in maroon,
  with error bars) are the actual IFRM values determined from the
  lunar observations.
\label{fig:diurnal}}
\end{figure*}

In this section we address the question of how well the various IFRM
estimates correct the observed lunar EVPAs.  In
Fig.~\ref{fig:diurnal}, we show three global and one ALBUS lunar
IFRM estimates for the `D2' and `DC' observations.  The three global
estimates are based on the `jplg', `upcg', and `codg' global maps, and
are chosen to show the range of estimates from all seven that we
investigated.  The displayed ALBUS estimate is the G01-pie1
(single GNSS station using the `pie1' GNSS station).

It is immediately evident from the figure that the IFRM estimates
based on global VTEC maps are significantly higher than the local
ALBUS estimates.  Importantly, the {\em change} in IFRM from night to
day for all models is nearly the same.  Evidently, the
globally-derived estimates have an IFRM offset with respect to the
local estimates.
There is enough sensitivity in the lunar observations for the `D2' and
`DC' observations that we can measure the lunar IFRM as a function of
time. These values are shown as individual data points, with errors
determined from the dispersion in the EVPA values.  For both days, the
ALBUS G01 estimates are a far better match to the measured values than the
global estimates.

The conclusions noted above hold equally for all eight VLA lunar
P-band observations: The global IFRM estimates are always higher
than the local ALBUS estimates, and higher than the observed IFRM
values.  This IFRM offset is nearly constant over the duration of a
single observation, but differs significantly between observations.

Time-sequenced determinations of the lunar IFRM were only practical
with the `D2' and 'DC' observations, as only these two spanned the
day-night transition with a significant change in IFRM.  However,
since the major difference between the global and local estimates is a
constant, so that all IFRM estimates are equally effective in removing
the {\em change} in the day to night IFRM, lunar polarimetric imaging
utilizing all the data from a given observation is very effective in
investigating which estimate is best.  These integrated images have
excellent sensitivity, providing accurate estimates of the residual
IFRM following correction by any individual IFRM estimate.

\begin{figure}{}
  \gridline{\fig{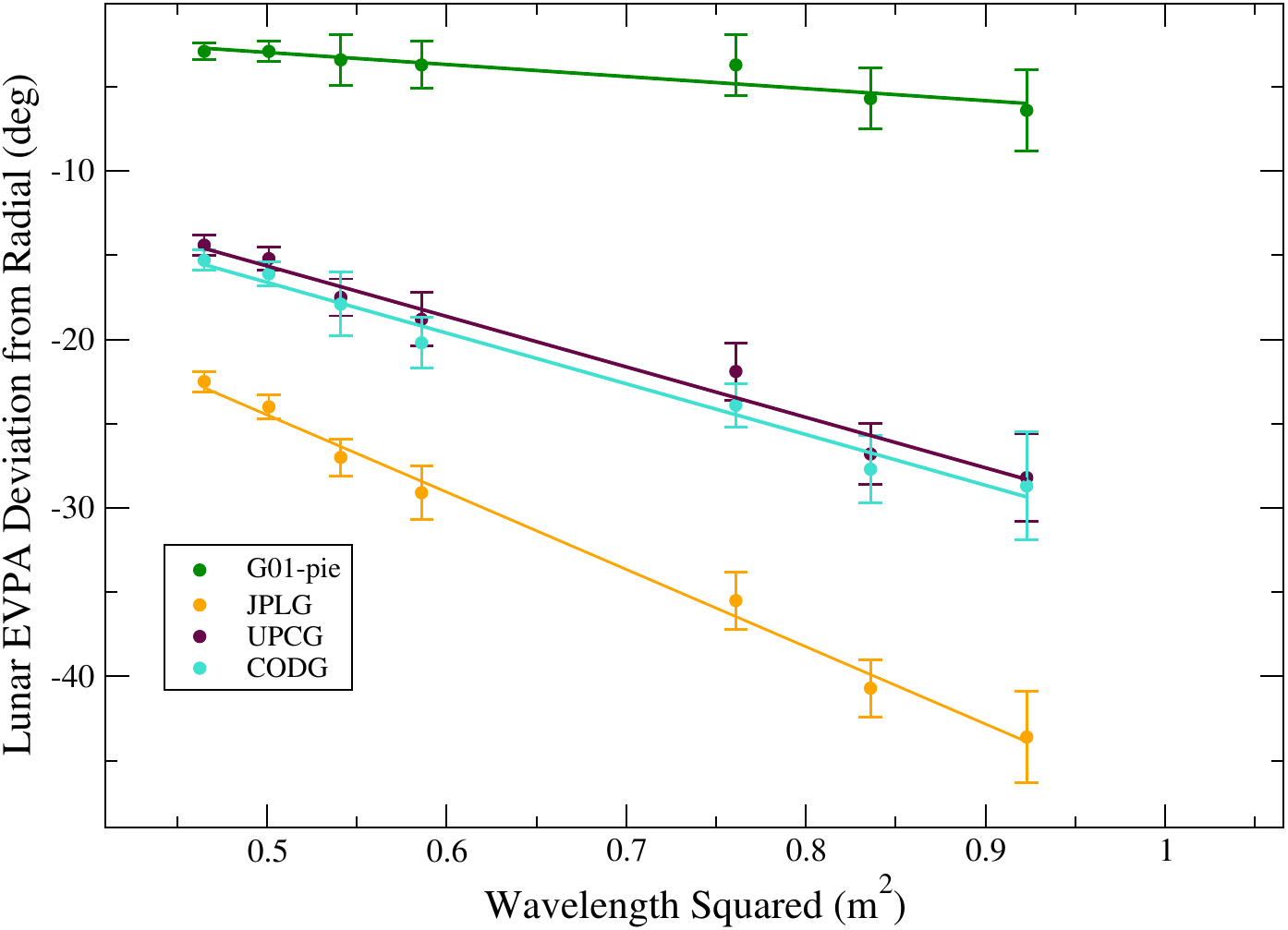}{0.33\textwidth}{D1 -- 10May2017}
            \fig{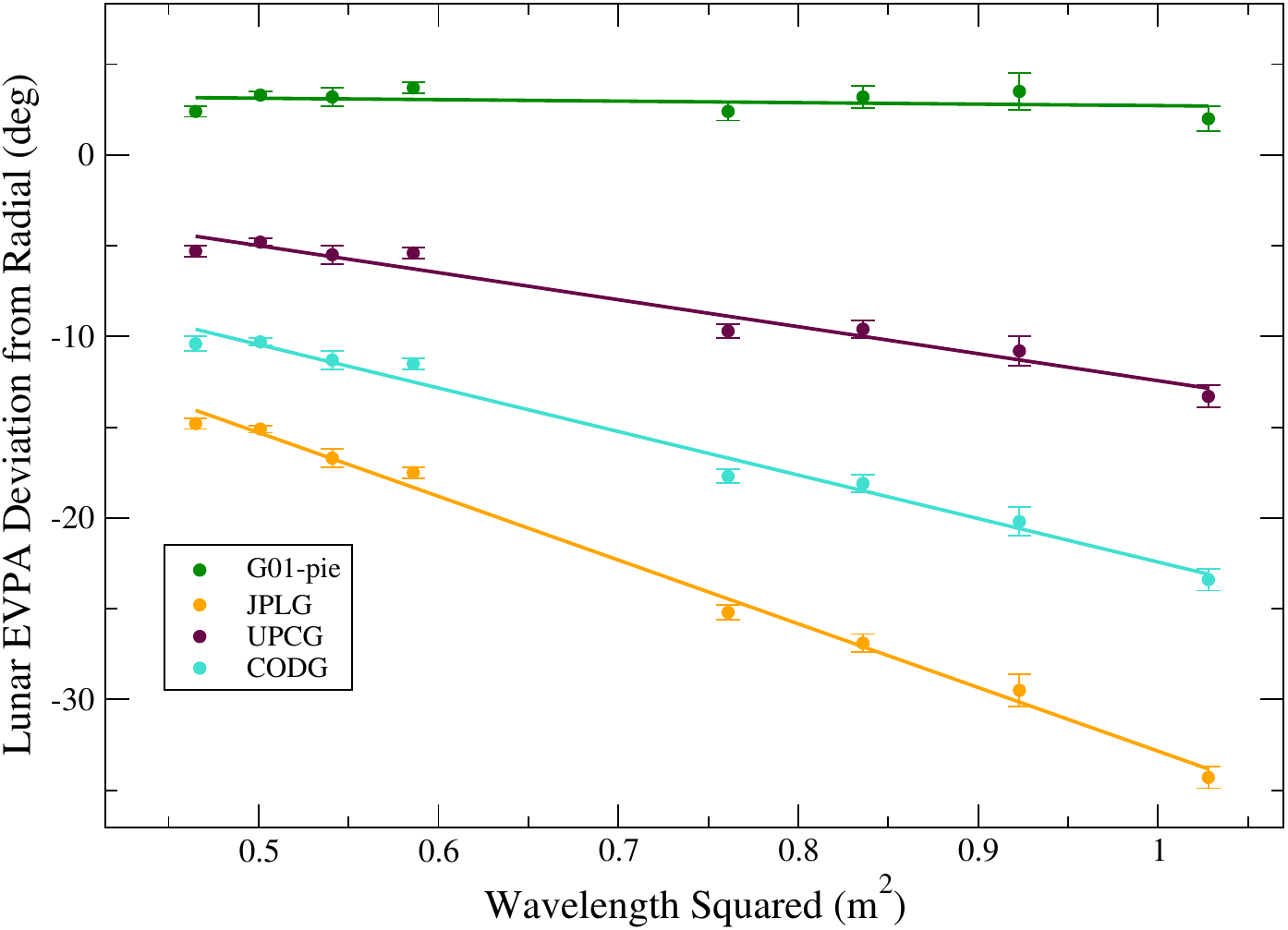}{0.33\textwidth}{C2 -- 30Dec2018}
            \fig{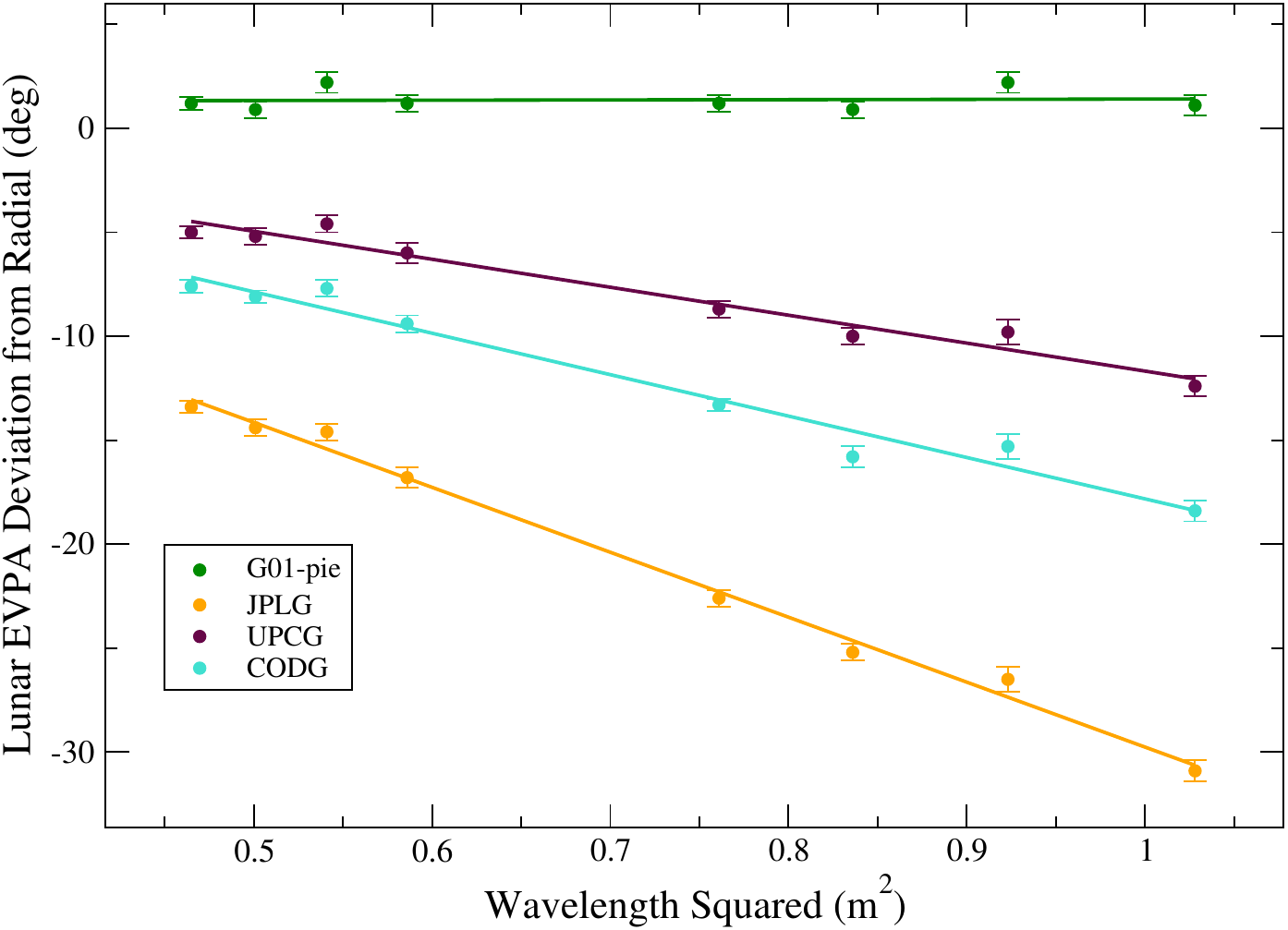}{0.33\textwidth}{C3 -- 13Dec2022}}
  \gridline{\fig{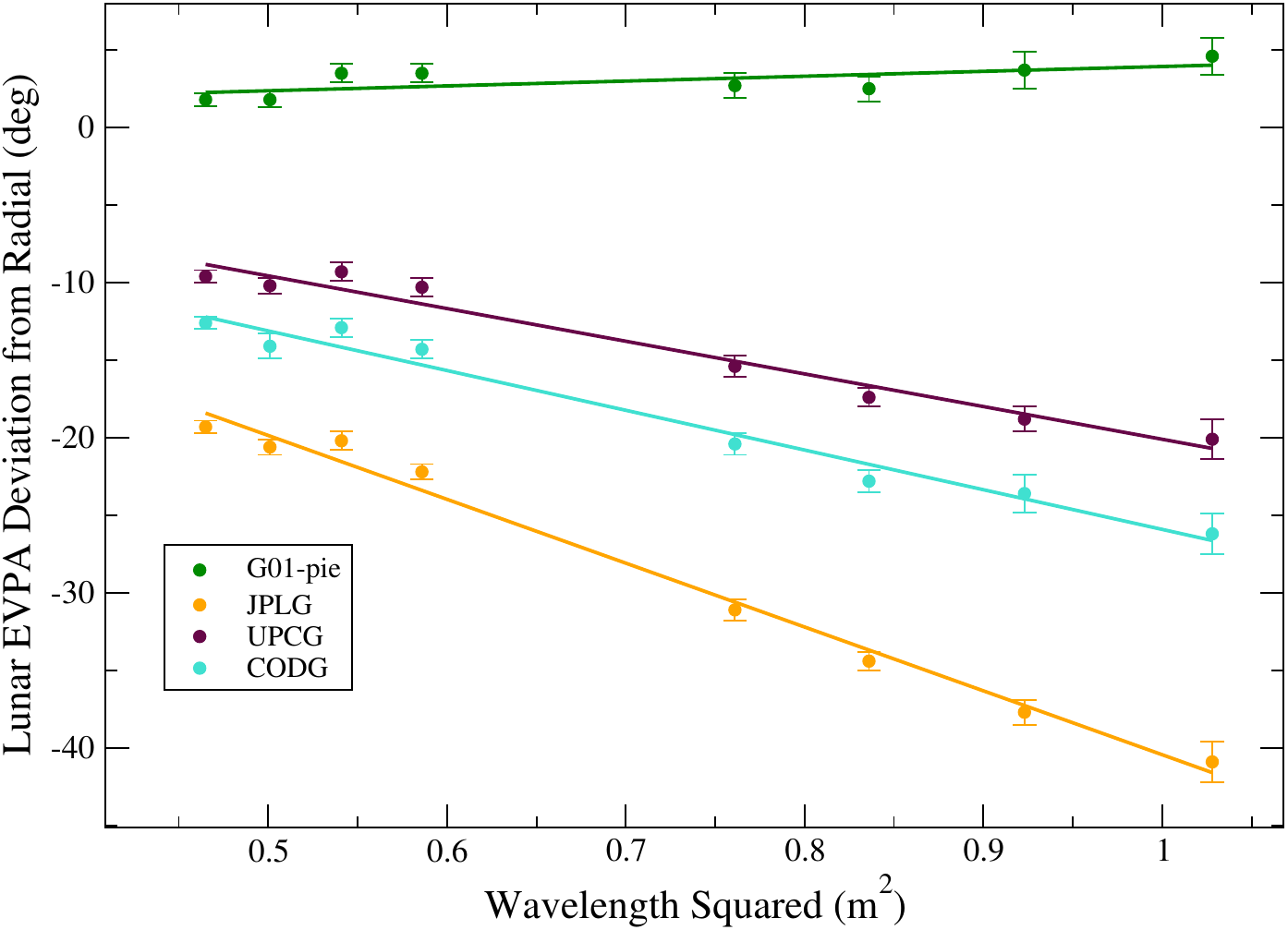}{0.33\textwidth}{D2 -- 04Nov2023}
            \fig{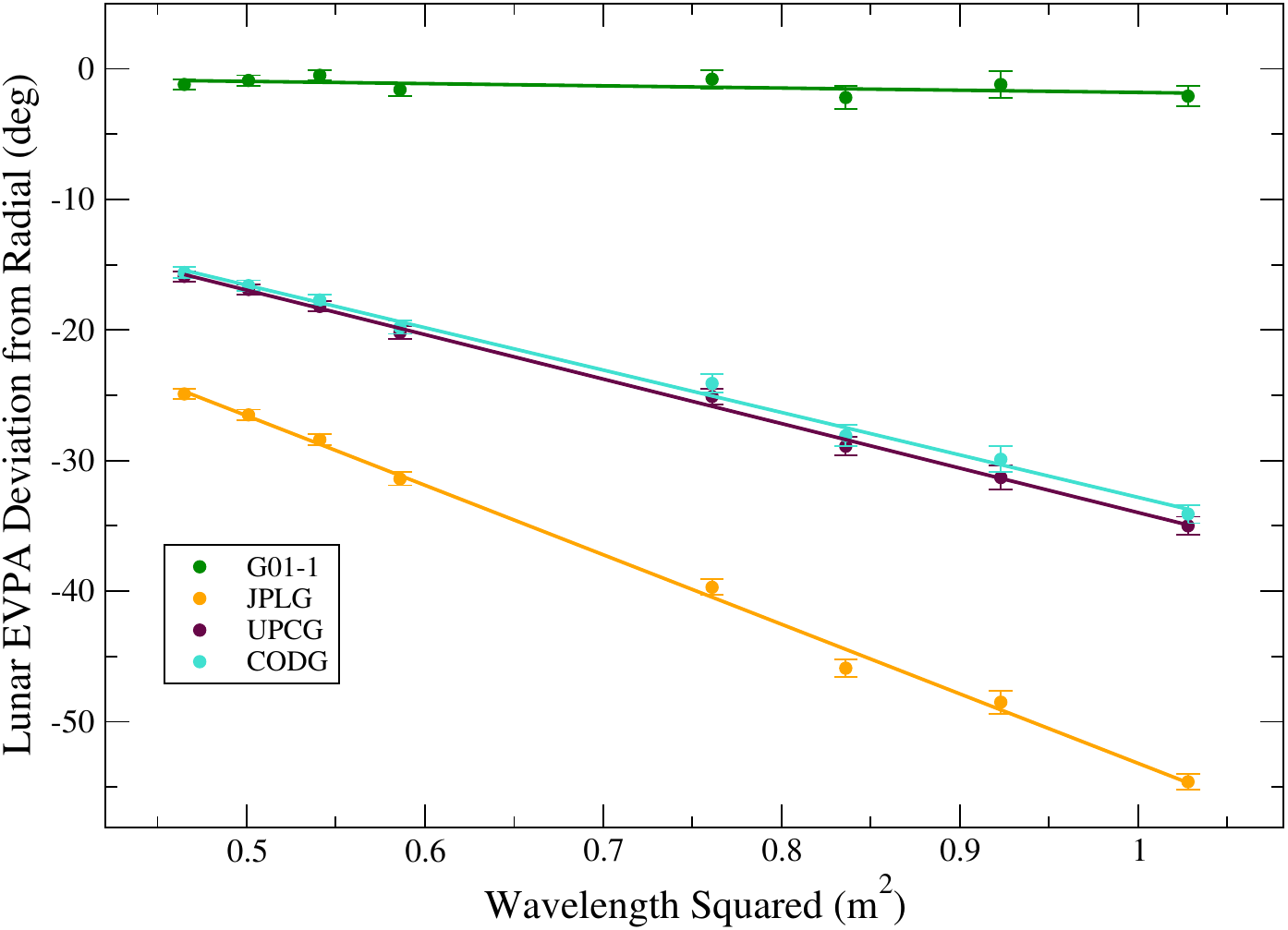}{0.33\textwidth}{DC -- 20Jan2024}
            \fig{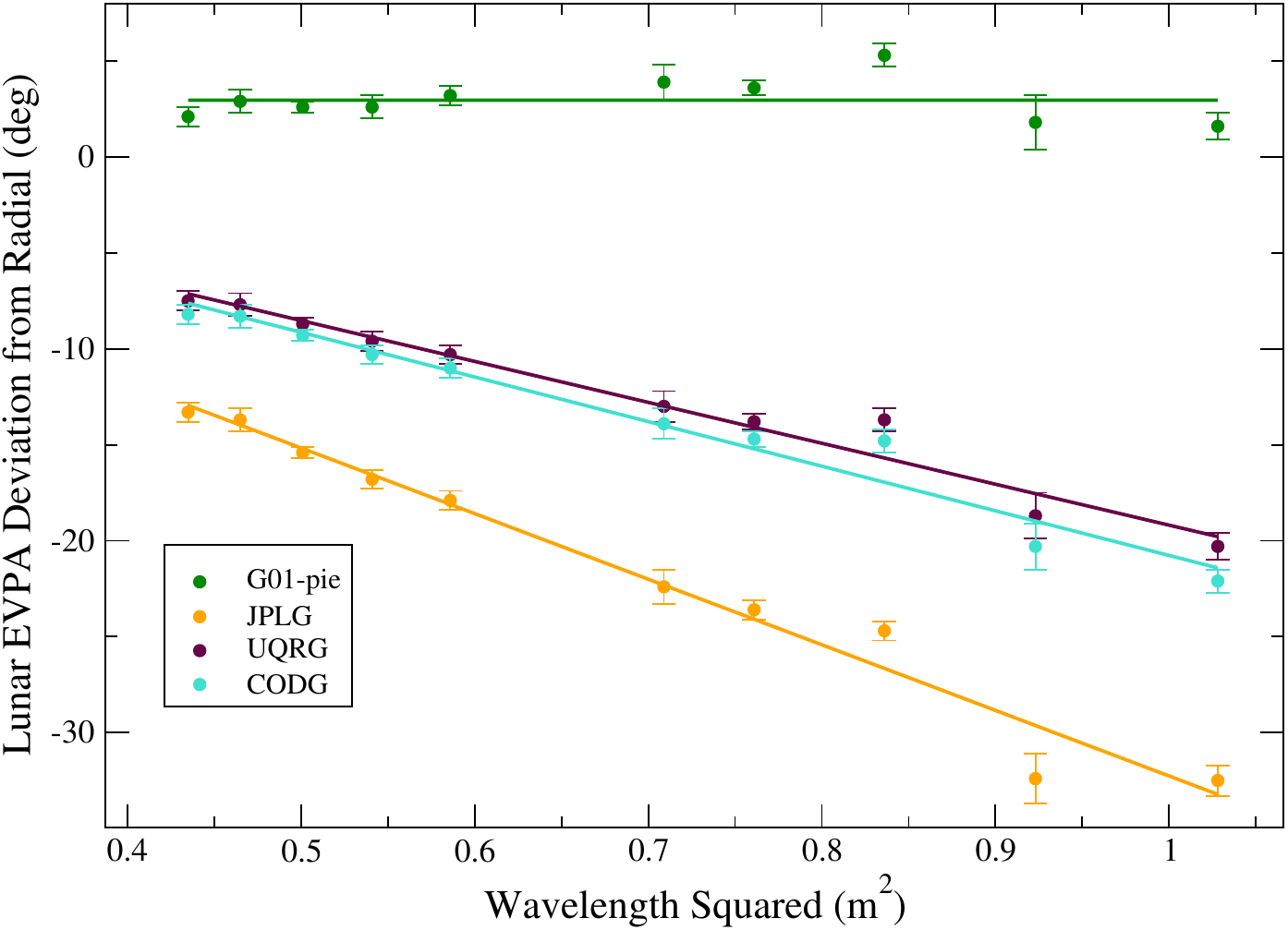}{0.33\textwidth}{CB -- 19Aug2025}}
  \caption{The lunar EVPA measurements following correction by four
    IFRM estimates, the G01-pie1 estimate from ALBUS and three from global VTEC
    maps: G01-pie1 in green, global-jplg in yellow, global-upcg in maroon,
    and global-codg in turquoise. The global estimates always
    significantly overestimate the correction, the ALBUS G01 model is
    always the best.}
  \label{fig:AllCor}
\end{figure}

The results of doing this are shown in Fig~\ref{fig:AllCor} for six of the
eight VLA observations. Here we show the post-IFRM corrected EVPA
offsets as a function of $\lambda^2$ for the lunar observations, along with
a linear least-squares fit.
Color-coding in the figure is the same in each panel: G01-pie1 (single
nearby station) in green, global-jplg in yellow, global-upcg in maroon,
and global-codg in turquoise.  The three global estimates displayed were
chosen to span the range of estimates -- all other global VTEC map-based
estimates are similar except for their offsets.

The excellent linear fits to these data demonstrate that the residuals
are due to an overestimation of the IFRM, and are not due to an offset
in EVPA.  The three estimates based on global VTEC maps always
significantly overcorrect the IFRM (resulting in a negative slope),
while the IFRM estimate based on the G01 ALBUS model results in EVPA
values with close to zero slope.  For all data shown, a linear fit of
the residual EVPAs against $\lambda^2$ was done, the results of which
are shown by the straight lines.  The slopes are the residual IFRMs.

The parameters of these linear fits to the post-correction EVPAs are
shown in Table~\ref{tab:EVPAFits}.  
`G01-pie1' is the G01 model, using only the nearest GNSS station `pie1'
(the VLBA Pietown location).  

\begin{deluxetable*}{cD@{$\pm$}DD@{$\pm$}DD@{$\pm$}DD@{$\pm$}DD@{$\pm$}DD@{$\pm$}D}
  \label{tab:EVPAFits} \tablecaption{Fit Parameters for all P-band
    Lunar Observations}
  \tablehead{
    \colhead{Code} & \multicolumn8c{G01-pie1} & \multicolumn8c{global-jplg} & \multicolumn8c{global-upcg}\\
    \colhead{} & \multicolumn4c{Slope} & \multicolumn4c{Int.} & \multicolumn4c{Slope} & \multicolumn4c{Int.} & \multicolumn4c{Slope} & \multicolumn4c{Int.}\\
    \colhead{} & \multicolumn4c{$\rmm$} & \multicolumn4c{Deg} & \multicolumn4c{$\rmm$} & \multicolumn4c{Deg} & \multicolumn4c{$\rmm$} & \multicolumn4c{Deg}
  }
  \decimals
  \startdata
  D1 & -0.12 & .02 & -0.6 & 0.9 & -.80 & .03 & -1.5 & 1.2 & -.52 & .04 & -0.7 & 1.5 \\
  C1 &   .   &  .  &   .  &  .  & -.65 & .02 & -2.2 & 0.9 & -.25 & .02 & -2.0 & 0.9 \\
  C2 &  0.01 & .02 &  3.5 & 0.8 & -.61 & .02 &  2.3 & 0.8 & -.26 & .02 &  2.4 & 0.9 \\
  C3 &  0.00 & .02 &  1.3 & 1.0 & -.54 & .02 &  1.5 & 0.8 & -.23 & .02 &  1.8 & 0.8 \\
  D2 &  0.05 & .02 &  0.8 & 1.0 & -.71 & .03 &  0.7 & 1.3 & -.37 & .03 &  1.0 & 1.1 \\
  DC & -0.03 & .02 & -0.1 & 0.7 & -.93 & .02 &  0.0 & 1.0 & -.60 & .01 &  0.1 & 0.5 \\
  D3 & -0.10 & .04 &  1.2 & 1.6 & -.68 & .04 &  1.8 & 1.7 & -.23 & .07 &  2.4 & 2.8 \\
  CB &  0.00 & .03 &  3.0 & 1.3 & -.60 & .03 &  1.9 & 1.4 & -.37 & .02 &  2.1 & 1.0 \\
  \enddata
  \tablecomments{The G01-pie1 fit for the `C1' observation is missing as the GNSS
    station data for this observation were corrupted.}
  \end{deluxetable*}
Key points to note are:
\begin{itemize}
\item Application of all IFRM estimates results in close to zero
  intercepts.  Each estimate, for a given day, gives the same
  zero-wavelength intercept, to within the errors.  The small offsets
  seen (for example, the four intercept estimates for the D2
  observation have an average offset of about 1 degree) are
  attributed to the misorientation of the reference antenna receiver
  feeds -- these offsets are well within the error tolerance of
  orienting the dipoles.
\item The observed residual slopes for the three global estimates vary
  considerably between observations. The `jplg' residuals vary from
  $-0.54$ to $-0.93\,\rmm$, the others somewhat less.  For our VLA
  observations, the `jplg'-based estimates were always the largest.
  Closer inspection of the residuals over time shows that these
  offsets vary only slightly over the length of an observation.  The
  fact that the post-correction residuals are exactly linear as a
  function of $\lambda^2$ shows that these estimates have an offset in
  estimated IFRM, which can only be due to an overestimate in the VTEC
  maps as the magnetic field models are the same for both local and
  global estimates..
\item The ALBUS IFRM estimates are much closer to the actual IFRM than
  any of those derived from the global maps.  Of the various models
  available, the G01 model -- utilizing a single, nearby,
  well-calibrated GNSS station -- is always the closest to observed
  values.
\item The scatter in the ALBUS estimates about the known zero slope is
  less than $0.1\,\rmm$.  This represents the evident accuracy of
  these corrections.
\end{itemize}

\begin{figure*}[ht!]
 \epsscale{1.1}
\plottwo{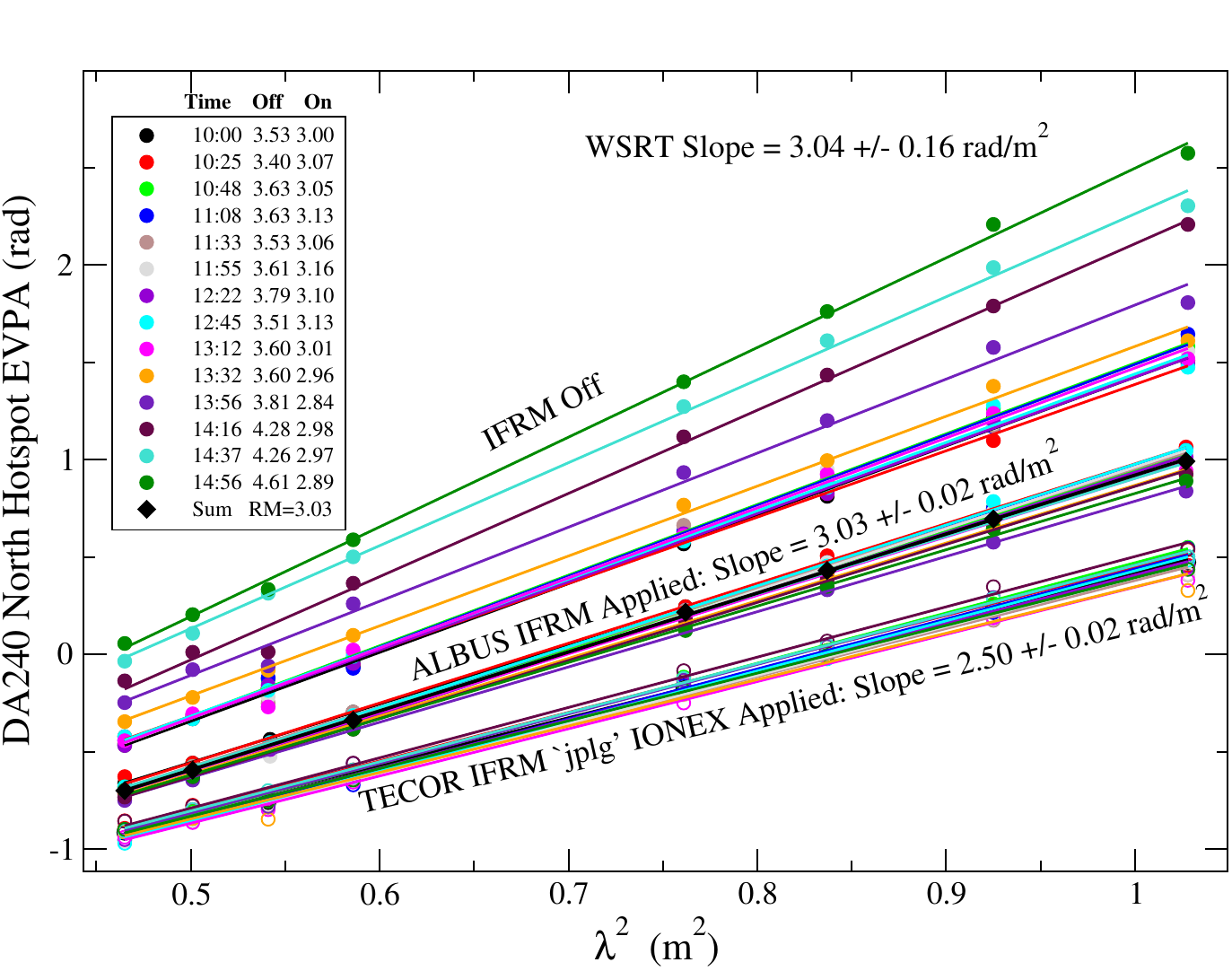}{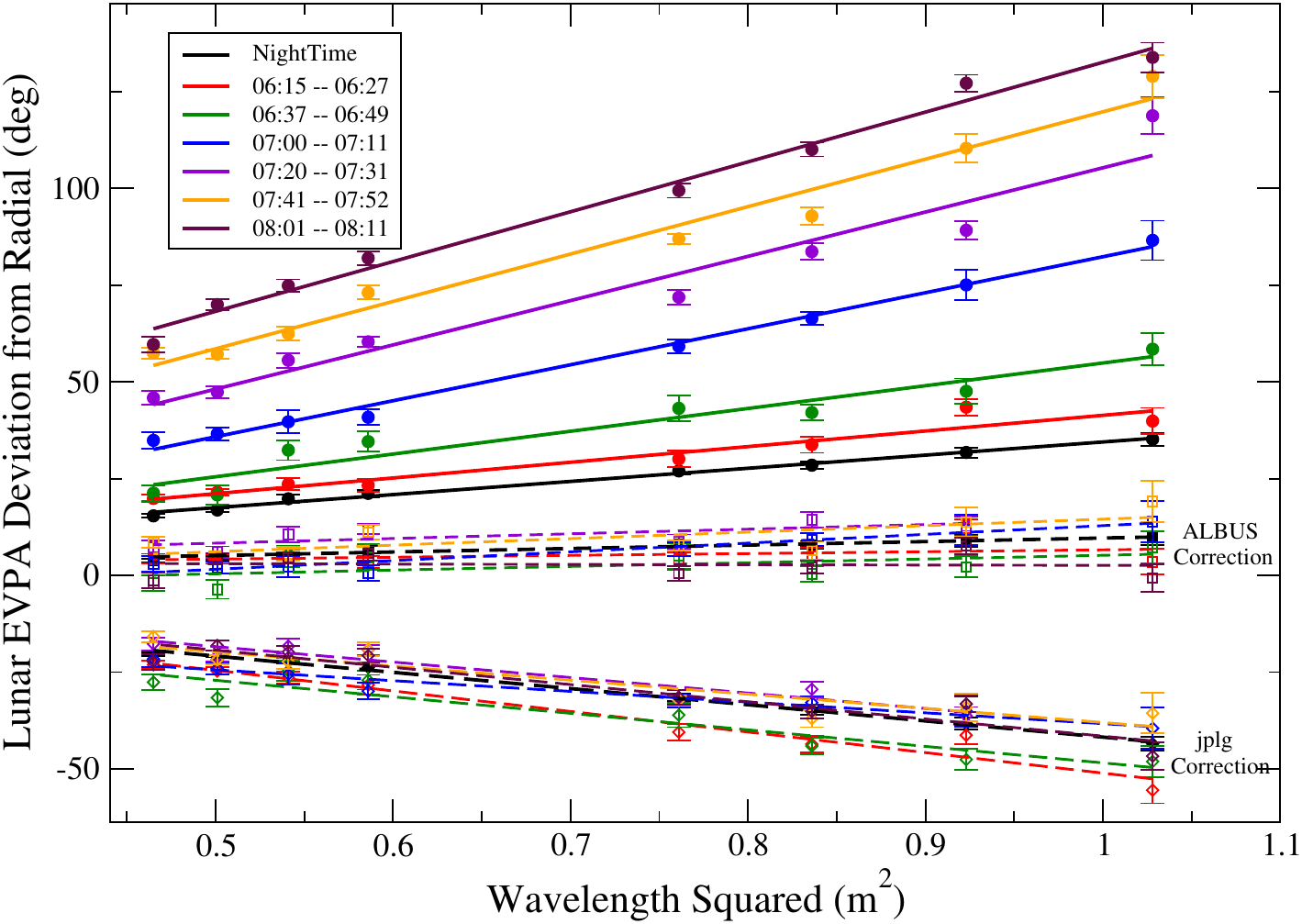}
\caption{{\bf(Left)} The observed and corrected EVPAs for DA240 for
  the `D2' observation by time. Shown are the uncorrected, and
  IFRM-corrected data for two estimates. {\bf (Right)} Same, for the
  Moon. The colored solid lines show the fits to each observation
  taken after dawn.  The black solid line is the fit for the
  night-time observations.  The dashed lines show the fits following
  ALBUS corrections (shorter dashes) and `jplg' global (longer dashes)
\label{fig:RMTime}}
\end{figure*}

Another way to show the effectiveness of the IFRM corrections is to
plot the uncorrected and corrected EVPA values for each scan within a
single observation.  As noted above, this is only practical for the
`D2' and `DC' observations.  In Figure~\ref{fig:RMTime} we show this
for the highly polarized NE hotspot in DA240 and for the Moon for each
scan for the D2 observation.
The steeply sloping solid lines, whose slopes are the observed RM,
show the RM increasing as a function of time, as the rising Sun
increases the ionization of the ionosphere.  In the left panel, the
results of applying the ALBUS model and the global `jplg' estimates to
DA240 are shown.  The resulting post-correction EVPAs are very
consistent over the scans, showing that the increase in IFRM has been
effectively removed, but the average corrected slopes are quite
different for different IFRM corrections -- a result of the offset in
IFRM between estimates shown in Figure~\ref{fig:diurnal}.  Deciding
which is correct requires knowledge of the true RM for DA240 -- which
we don't have, although there is an early result from the WSRT array
\citep{Brentjans2007} which agrees with the ALBUS result .  The right
panel shows the same results for the Moon.  The SNR is lower, but the
lunar observations have the great advantage that we know the correct
answer in advance -- the corrected slope should be zero (as the Moon
and the cis-lunar environment have no IFRM of their own), and the
zero-wavelength intercept should be close to zero, indicating the EVPA
is purely radial.  The right panel shows that the ALBUS corrections do
indeed have close to zero slope and zero intercept.

\section{MeerKAT Results} \label{Sec:MeerKAT}

MeerKAT is an outstandingly capable instrument for imaging low surface
brightness structures.  Its centrally condensed array design ($\sim$40 of its 64
antennas are within 500 m of the array center) provides very high surface
brightness sensitivity on the angular scales of interest for lunar
polarimetry.  In addition, considerable care has been taken in
preventing antenna- and site-generated RFI from causing spurious
correlations.  Due to its remote site, the external RFI in the UHF
band is minimal (in comparison to the VLA's P-band environment), with
less than 3\% of the spectrum having to be removed due to RFI.

We have six lunar observations taken with MeerKAT, listed in
Table~\ref{tab:MoonData}, but only the last three are of use for
judging the accuracy of IFRM corrections, as the first three were
taken overnight during a period of very low and steady IFRM.  Two
recent observations, taken August 18 and 19, 2025 were arranged to
span sunrise, while the latest observation, taken October 30, spanned
sunset.

MeerKAT's array design provides us with outstanding image sensitivity
on the arcminute scales needed for this work -- for each 8-minute
lunar observation, the EVPA estimates for each of the 16 spectral
windows are determined with typical 1-$\sigma$ errors of 0.2 degrees.
The resulting error in the determination of the IFRM for each scan is
typically 0.2 $\rmm$.

\subsection{IFRM Estimation}

\begin{figure}{}
  \gridline{\fig{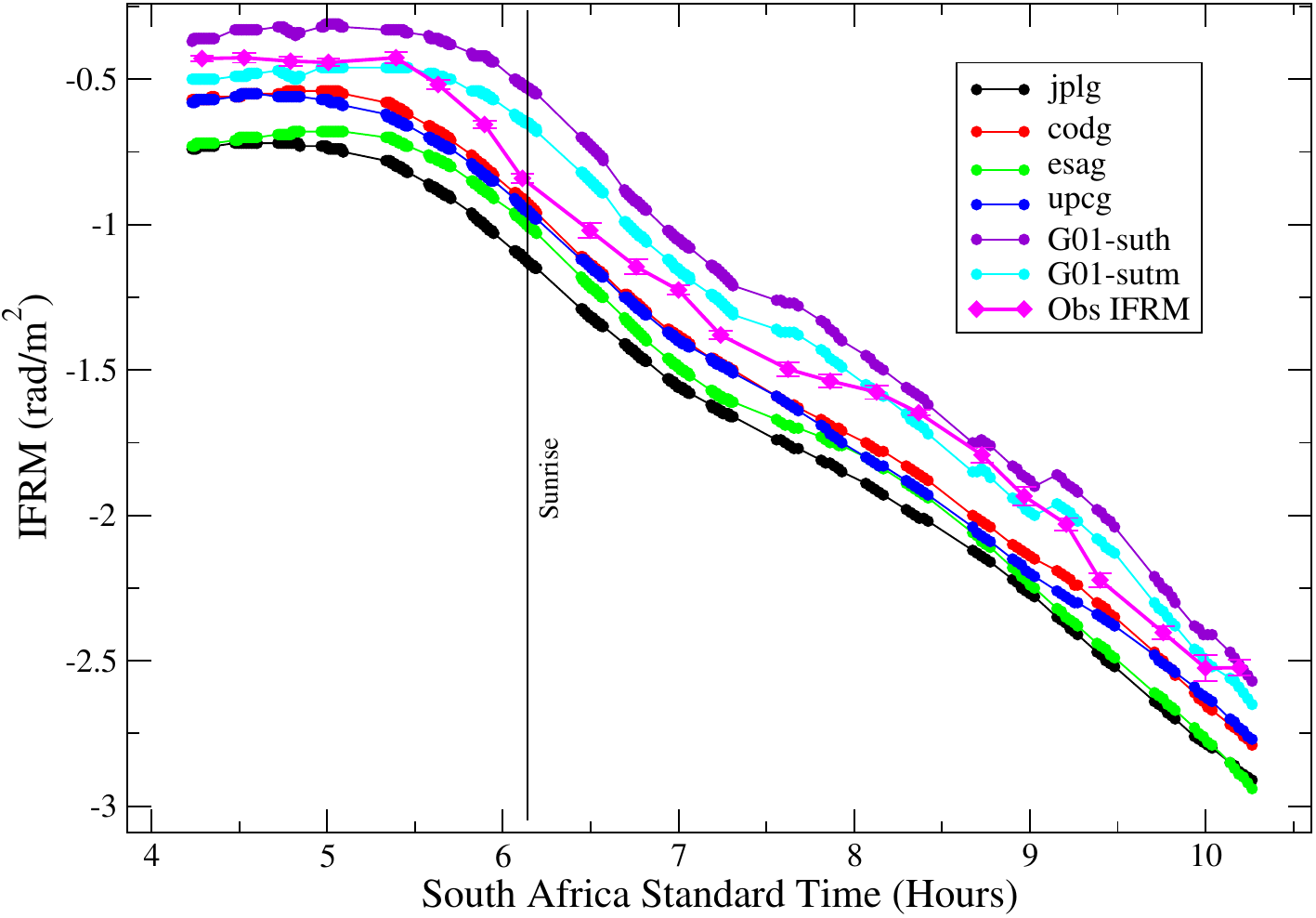}{0.33\textwidth}{Aug 18 2025}
            \fig{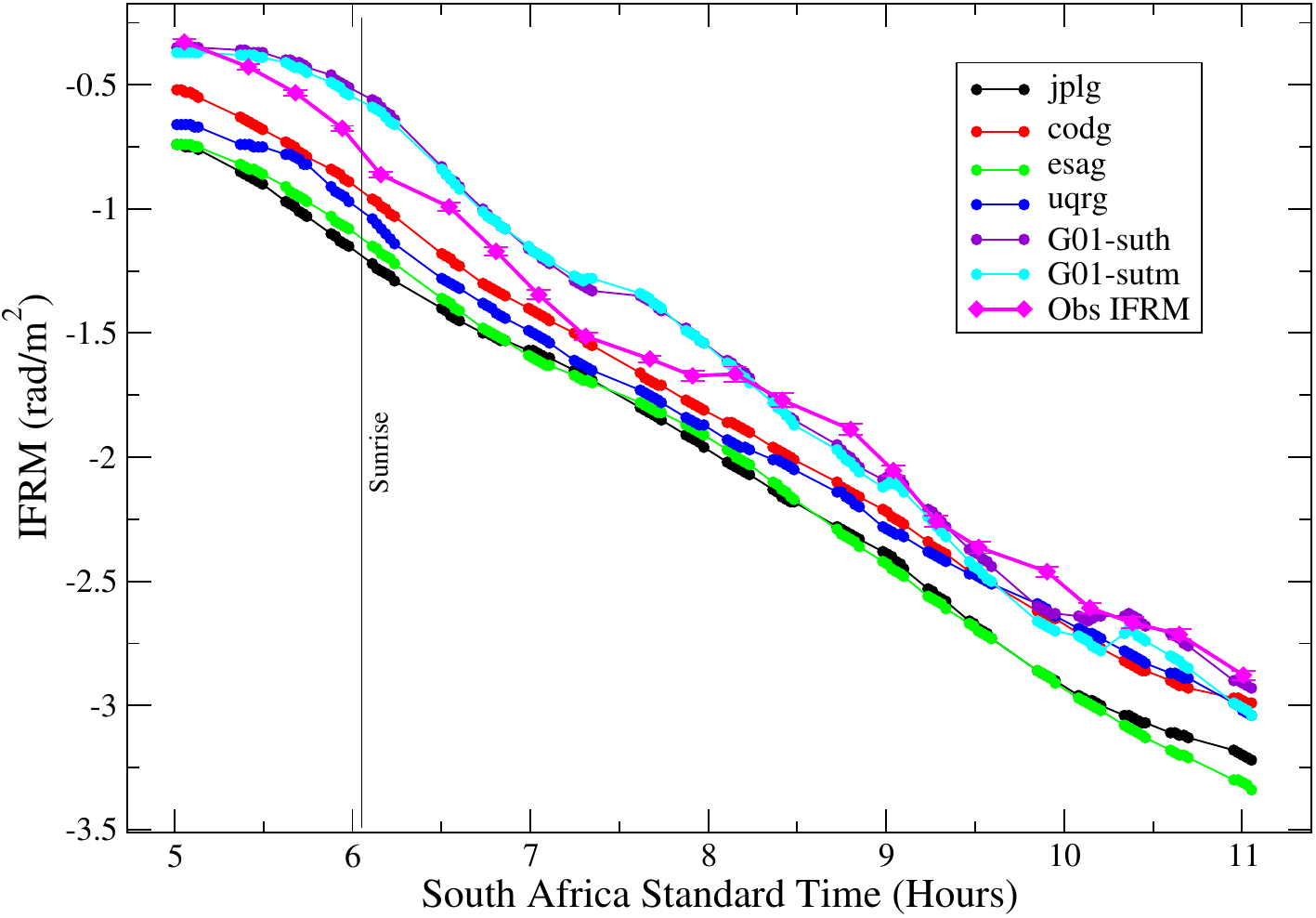}{0.33\textwidth}{Aug 19 2025}
            \fig{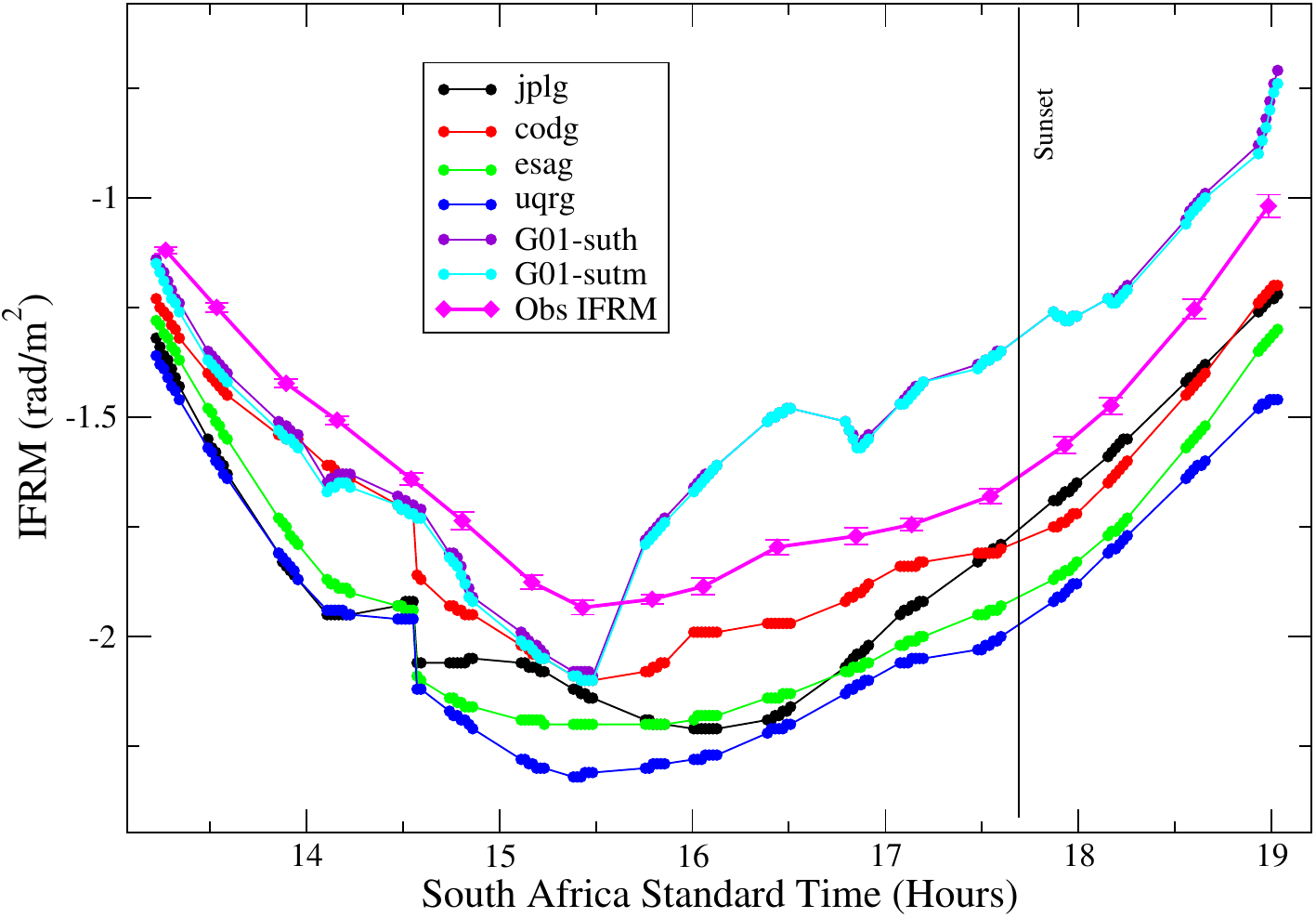}{0.33\textwidth}{Oct 30 2025}}
  \caption{The estimated and observed IFRM values for the three
    MeerKAT observations.  For all three, the estimates based on
    global VTEC maps overestimate the IFRM at all times.  The
    single-station ALBUS estimates are the closest to the observed
    values, except for the Oct 30 data. }
  \label{fig:Meer-IFRM}
\end{figure}

As with the VLA observations, we have measured the actual IFRM using
the lunar data for each scan for the three latest MeerKAT UHF band
observations.  These are plotted in Fig~\ref{fig:Meer-IFRM}, along
with estimates generated by TECOR using the global VTEC estimates, and
those generated by ALBUS, using the closest calibrated GNSS station,
at Sutherland.

As with the VLA, the global estimates are always too large (negative
for this latitude), although the offset is not as great as with the
VLA.  We show two ALBUS estimates -- both using the well calibrated
data from two of the Sutherland sites -- `suth' and `sutm'.  For both
of these, receiver bias values are available and have been used by
ALBUS.  The resulting estimates are a good fit to the data for the two
August observations, but are not as good for the October 30 data,
where it appears that a sudden change at about 15:30 SAST caused the
IFRM estimates to jump upwards by nearly 0.5 $\rmm$.  The origin of
this is unknown to us.

\begin{figure}{bht}
  \gridline{\fig{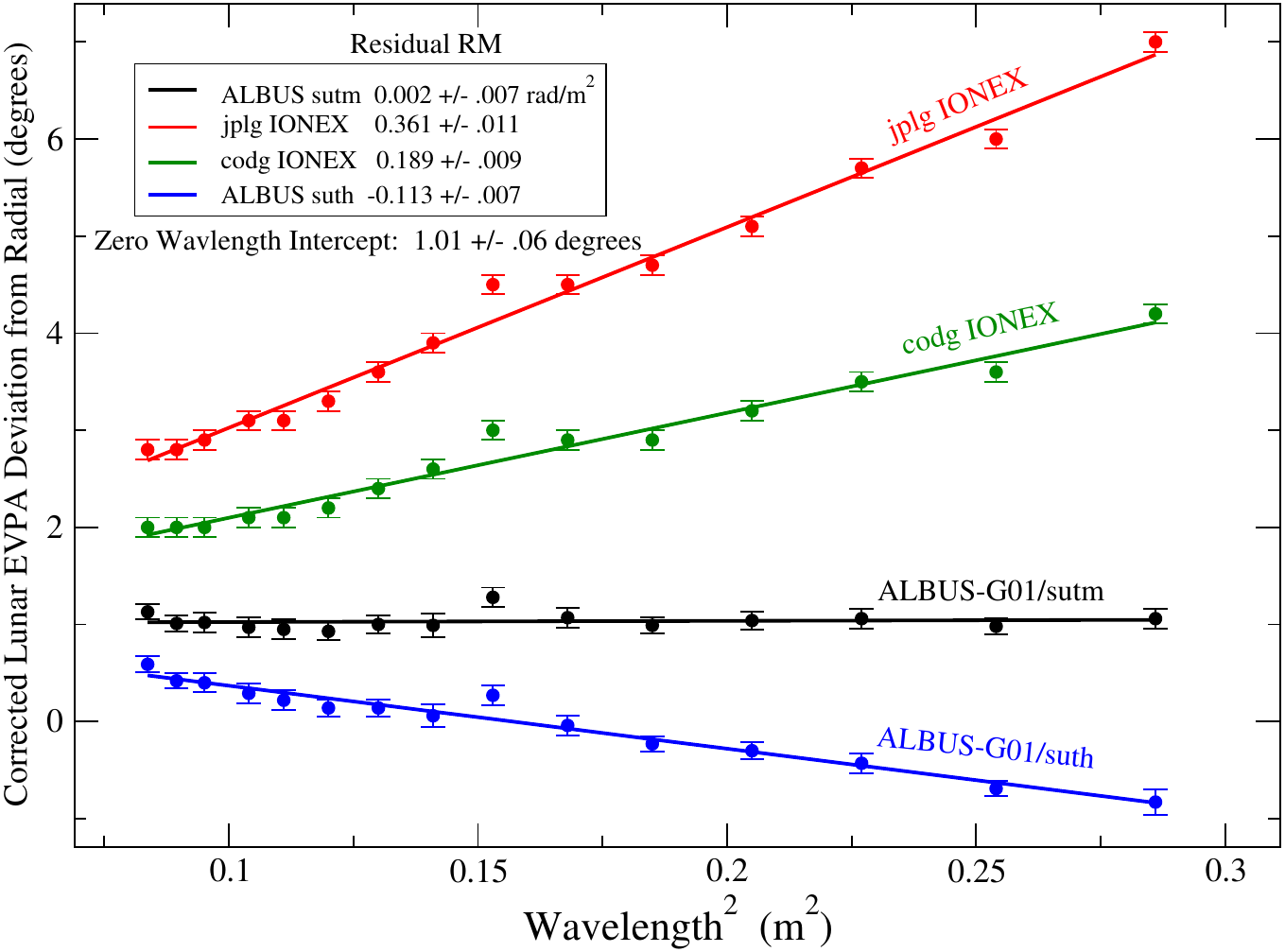}{0.33\textwidth}{Aug 18 2025}
            \fig{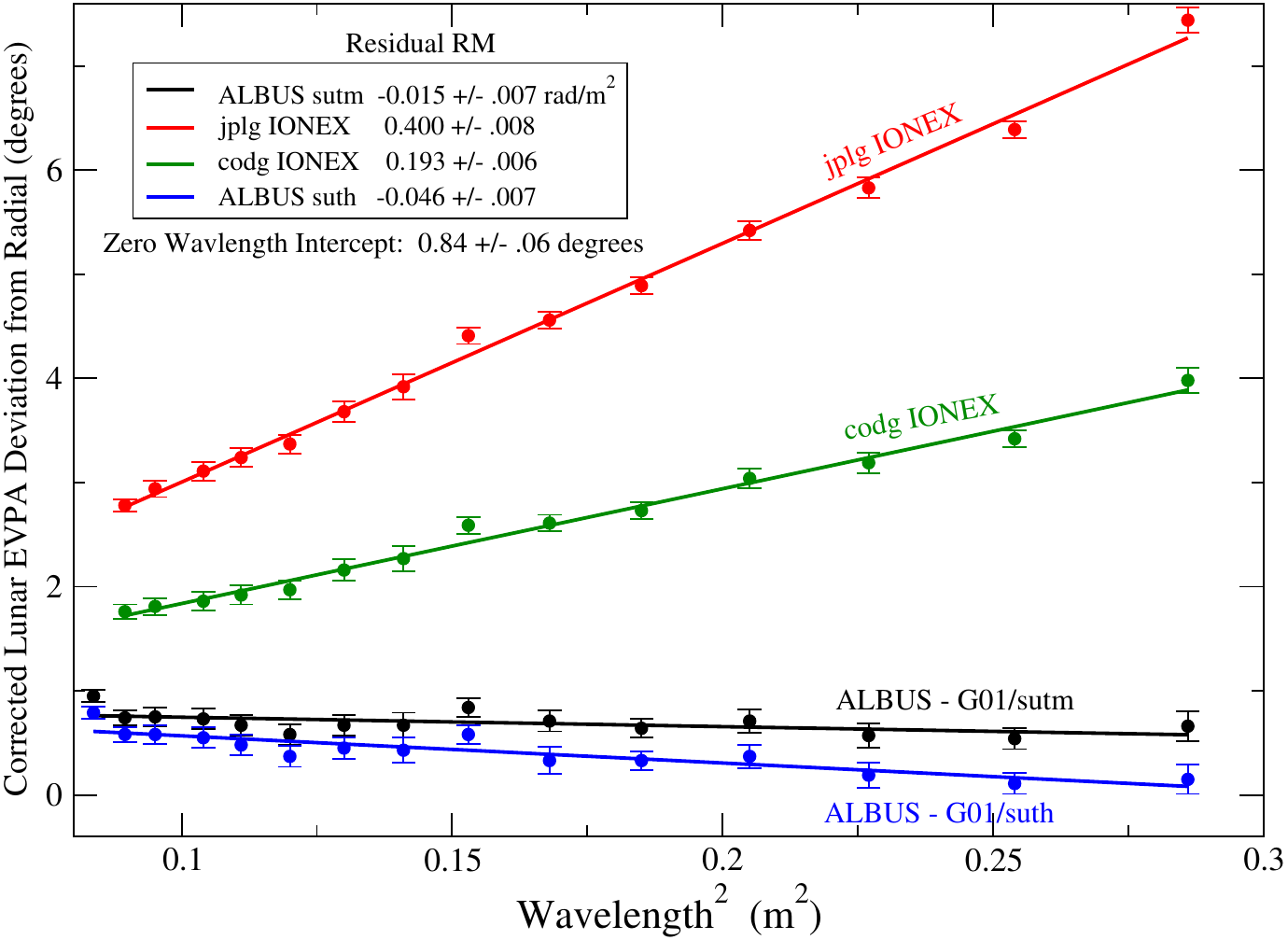}{0.33\textwidth}{Aug 19 2025}
            \fig{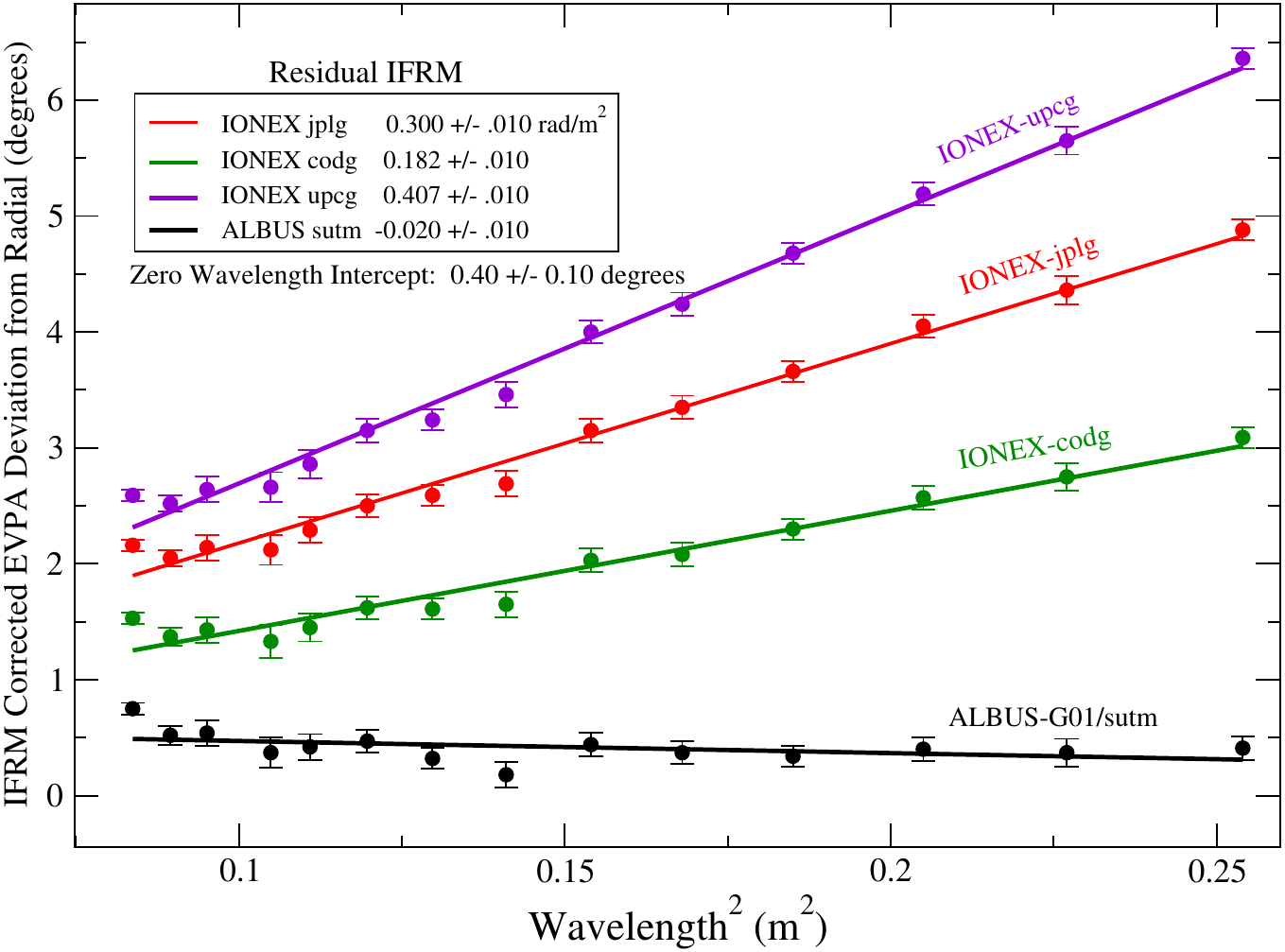}{0.33\textwidth}{Oct 30 2025}}
  \caption{Showing the integrated residual EVPA values following IFRM
    corrections by various estimates.  As with the VLA observations,
    the global VTEC-based estimates overcorrect the EVPA values.
    The ALBUS estimates based on a single local well-calibrated
    receiver again give the most accurate corrections. }
  \label{fig:Meer-LunarObs}
\end{figure}

Time-integrated observations of the Moon with various corrections,
resulting in estimates of the average error incurred with each
estimation over the length of each observation, are shown in
Fig~\ref{fig:Meer-LunarObs}.
Notable here is that residual IFRM values are best for the G01
correction using the well-calibrated receiver 'sutm' -- with a
residual of less than 0.02 $\rmm$.  The 'suth' data results are very
similar, but are slightly less accurate, while the `global' estimates
always overcorrect the rotation, although the overcorrection is
significantly less than for the VLA observations -- typically $-0.2$
to $-0.4$ $\rmm$.

\begin{figure*}[ht!]
 \epsscale{1.1}
\plottwo{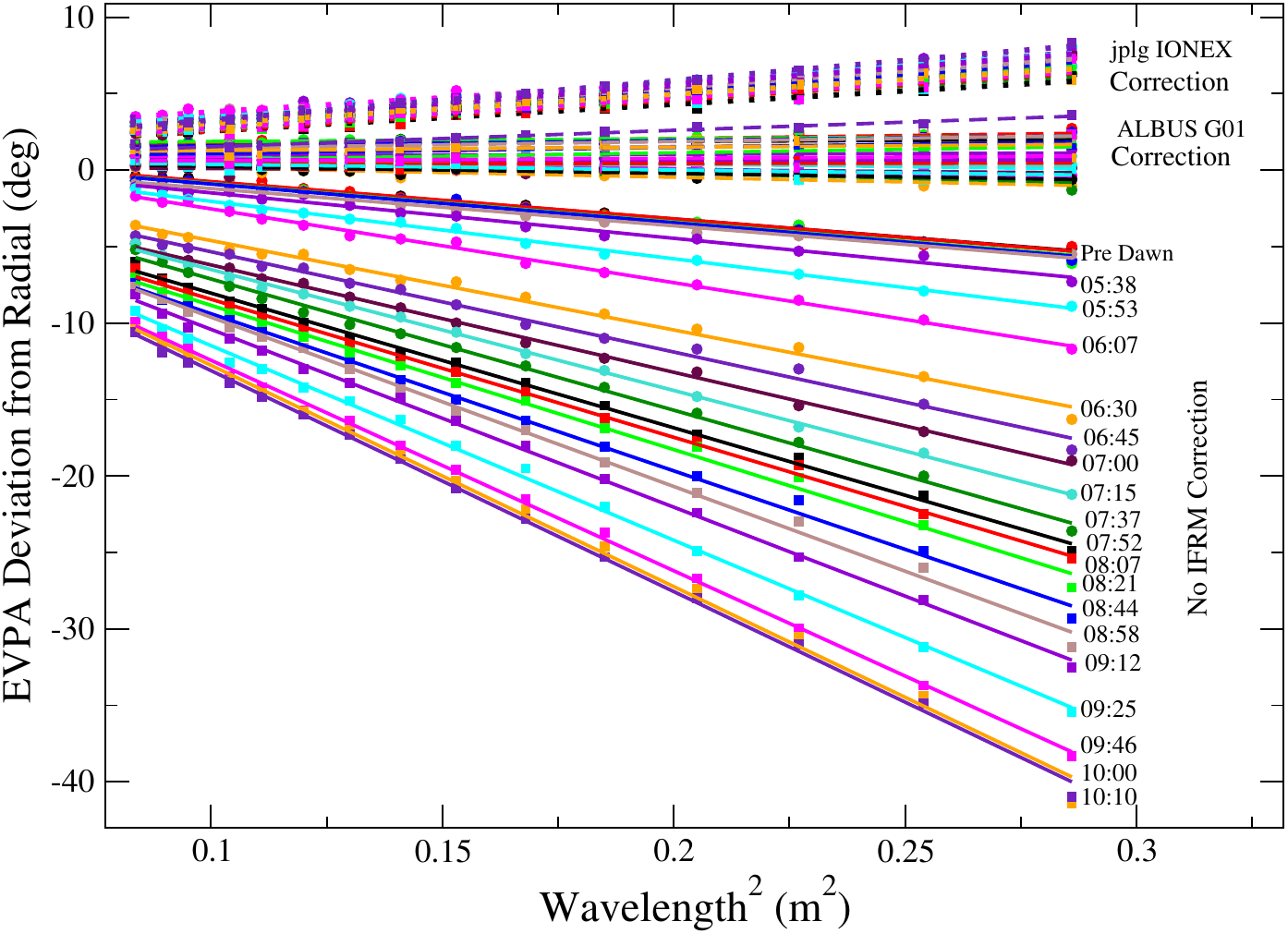}{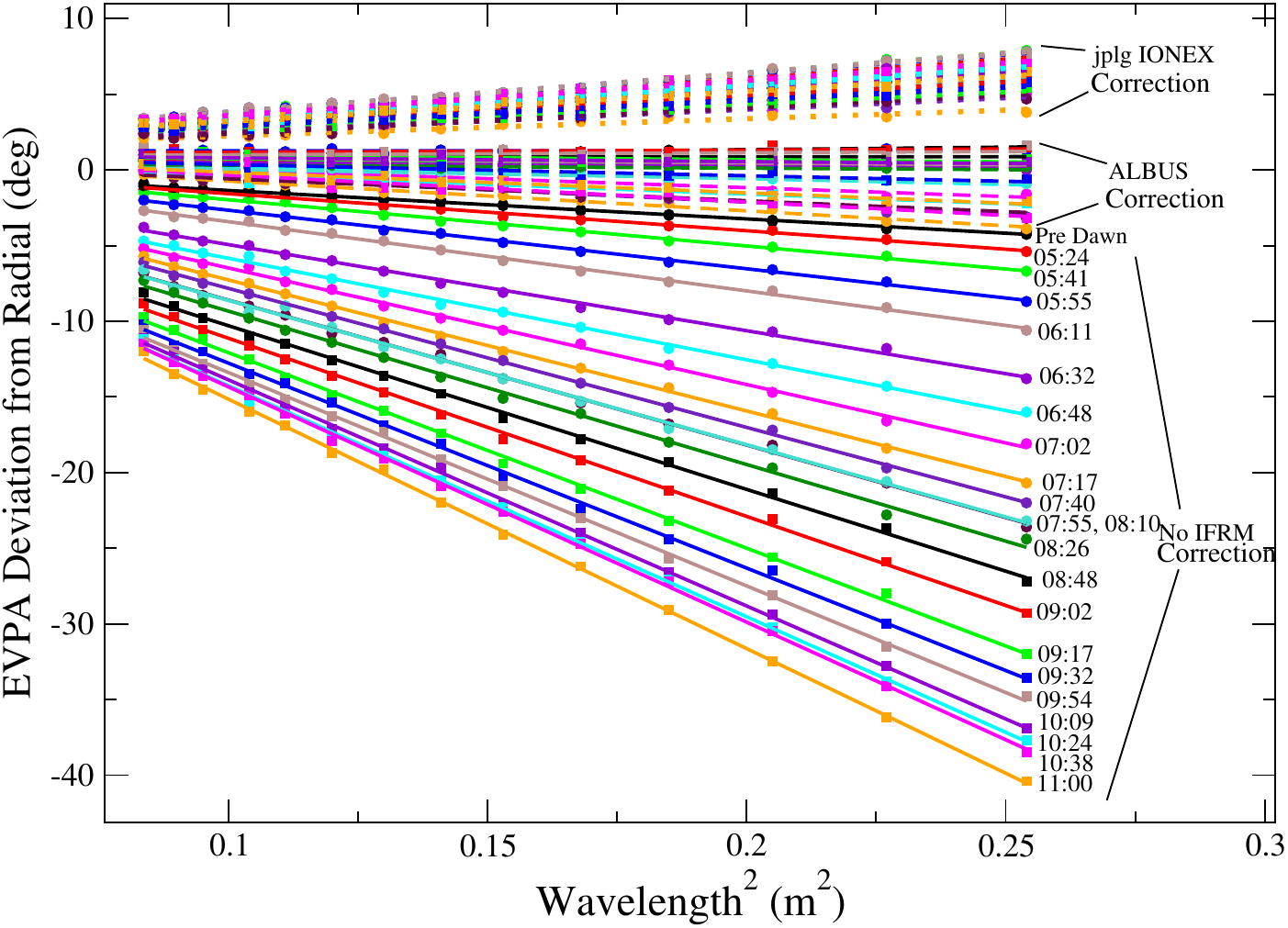}
\caption{Showing the uncorrected observed lunar EVPA offsets, and the
  values after correction by the `jplg' global estimate and the ALBUS
  estimate for two MeerKAT observations -- 18Aug2025 of the left, and
  19Aug2025 on the right.}
\label{fig:MeerIFRM-Time}
\end{figure*}

We show in Figure~\ref{fig:MeerIFRM-Time} the detailed before and
after EVPA deviations for each 8-minute scan for the two August 2025
observations.  This plot shows the raw EVPA offsets with no
corrections with solid lines, and the resulting values following
correction by the `jplg' global estimation (short dashes), and the
local ALBUS correction (long dashes).

\section{Polarization Characteristics for 3C286 and 3C138} \label{sec:3C286}

\begin{figure*}[ht!]
 \epsscale{1.1}
\plottwo{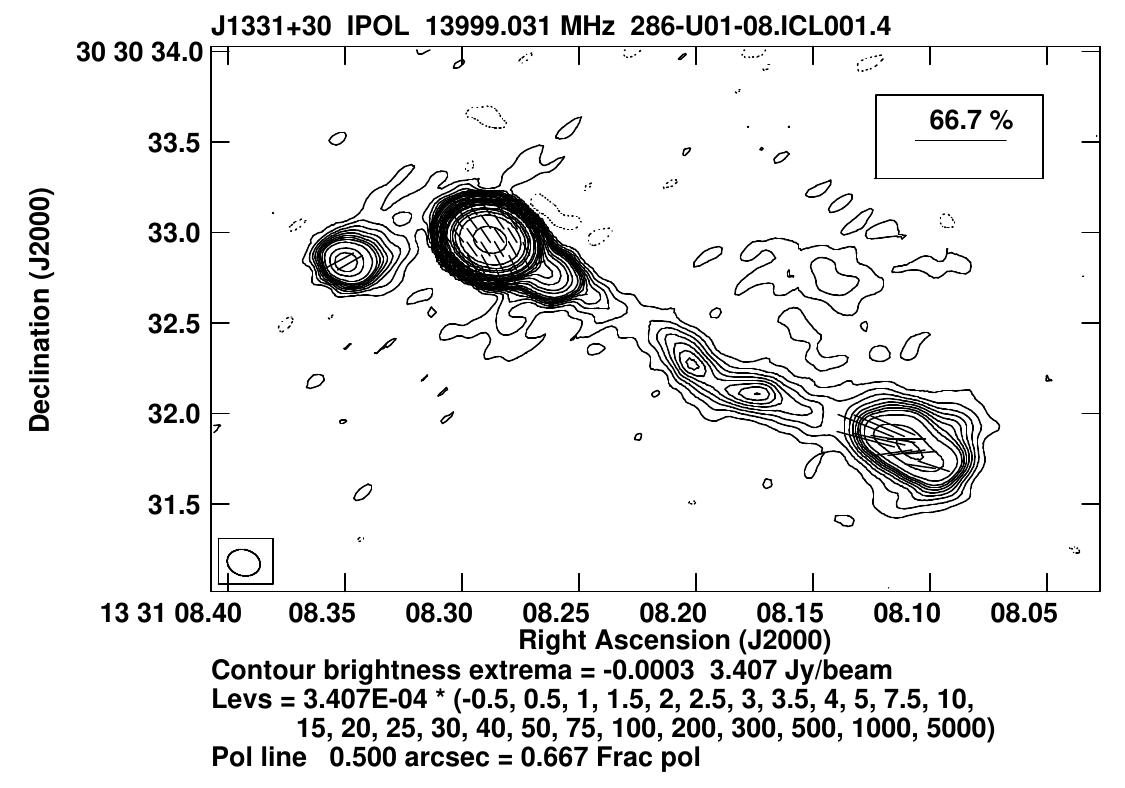}{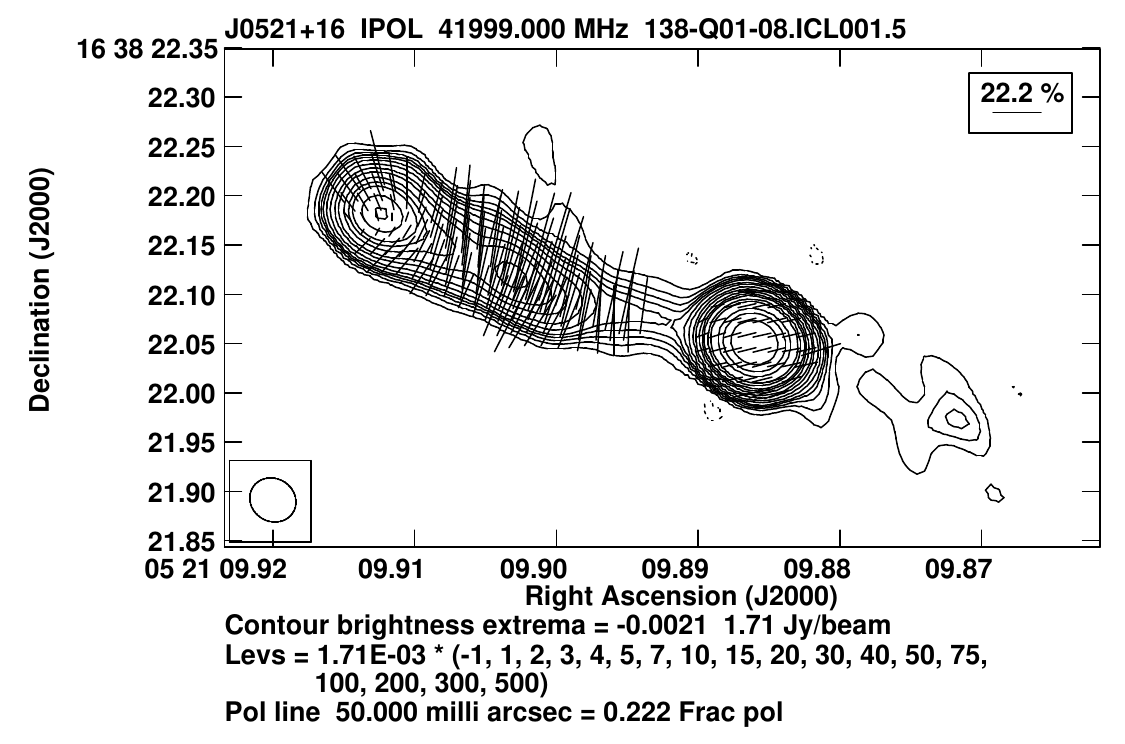}
\caption{Maps of the polarized calibrators 3C286 (left) at 14 GHz with
  150 mas resolution, and 3C138 (right) at 42 GHz with 45 mas
  resolution.  For 3C286, both the Stokes I and polarized flux
  density are dominated by the nuclear emission -- the Stokes I
  extended emission accounts for only 166 mJy compared to 3.58 Jy for
  the nuclear core.  For 3C138, the situation is quite different --
  the Stokes I extended emission is much stronger, accounting for 571
  mJy compared to 1.710 Jy for the nucleus, while the polarized
  emission at this band is about equal between the jet and the
  nucleus.
\label{fig:PolMaps}}
\end{figure*}

A secondary purpose of this work is to determine the polarization
characteristics for the standard calibrators 3C286 and 3C138 over the
entire frequency range that its polarized emission can be used for
correction of the cross-hand phases for circularly polarized receiver
systems.  Both sources are significantly resolved to the VLA, so care
must be taken in utilizing their structures for calibration purposes.
Maps of each, utilizing recent JVLA A-configuration observations, are
shown in Fig~\ref{fig:PolMaps}.

For both the VLA and MeerKAT, determining the intrinsic polarization
structures requres correction for any systematic offsets in the
determination of the EVPA.  For circularly polarized systems, such an
offset will be caused by an incorrect phase offset between the RCP and
LCP data -- mostly likely due to an incorrect assumption of the EVPA
of the polarized calibrator.  For linear systems, an offset will occur
due to misorientation of the receiver in the antenna used as the phase
reference during the calibration process.  Other origins of offsets
can also be imagined.

The removal of these offsets can be analyzed in the following way.
Let $\chi$ represent the EVPA for a given position within the
polarized image.  The observed EVPA, $\chi_{obs}$, is related to the
true EVPA, $\chi_{true}$, by
\begin{equation}
  \chi_{obs} = \chi_{true} + \mathrm{RM}\lambda^2 + \delta\chi
\end{equation}
where RM is the actual rotation measure at a given time for the
source, and $\delta\chi$ is any instrumental offset, such as
mis-oriented receivers for a linear system, or incorrect cross-hand
phase for a circular system.  After correction by a model rotation
measure $\mathrm{RM}_m$, the corrected EVPA will be given by
\begin{equation}
  \chi_{obs,corr} = \chi_{true} + \Delta\mathrm{RM}\lambda^2 + \delta\chi
\end{equation}
where $\Delta\mathrm{RM} = \mathrm{RM} - \mathrm{RM_m}$ represents the
difference between the true and estimated rotation measure along the
line of sight to the source.  For a lunar or planetary observation, we
interpret $\chi_{true}$ as the deviation from radial, so $\chi_{true} = 0$, giving
\begin{equation}
  \chi_{obs,corr,moon} = \Delta\mathrm{RM}\lambda^2 + \delta\chi
\end{equation}
This is the formulation which forms the basis of the analysis done in
previous sections.

For a subsequent determination of a calibrator EVPA (where
$\chi_{true}\neq 0$), we subtract the planetary residual corrected EVPA
determination from the calibrator determination, and presuming the
offset $\delta\chi$ is common to both sources, we find
\begin{equation}
  \chi_{true,src} = \chi_{corr,src} + (\Delta\mathrm{RM}_{src} -
  \Delta\mathrm{RM}_{moon}).
\end{equation}
EVPA offsets which are common to both planetary and calibrator
observations are removed in this subtraction process.  Furthermore, if
the IFRM determinations for both target and planet were in error by a
constant amount due to a constant offset in the TEC calculation, such
offsets will also be removed.

In words, the polarization EVPA values for a target source require a
determination of the Stokes Q and U values for both the source and for
the planet (Mars, Venus, Moon), with the best IFRM estimates applied.
Under the assumption that the offset in the EVPA on the calibration
source is constant in time, this offset is subtracted from the EVPA
value determined for the target source to give the best estimate of
the source EVPA.

\subsection{3C286}

As noted in the introduction, high frequency VLA polarimetric
observations by \citet{PB13} showed that the EVPA of 3C286 increases
from 33 degrees at 10 GHz to approximately 36 degrees at 45 GHz.
Subsequent observations by high frequency telescopes have confirmed
and extended this trend \citep[][and references therin]{2025Kam}.  The
\citet{PB13} work did not include measurements below 8 GHz, and it was
not until early MeerKAT polarimetric observations of 3C286 indicated
the EVPA continues to decline from 33 degrees at frequencies less than
6 GHz that further VLA observations were made.

\begin{figure*}[ht!]
 \epsscale{1.1}
\plottwo{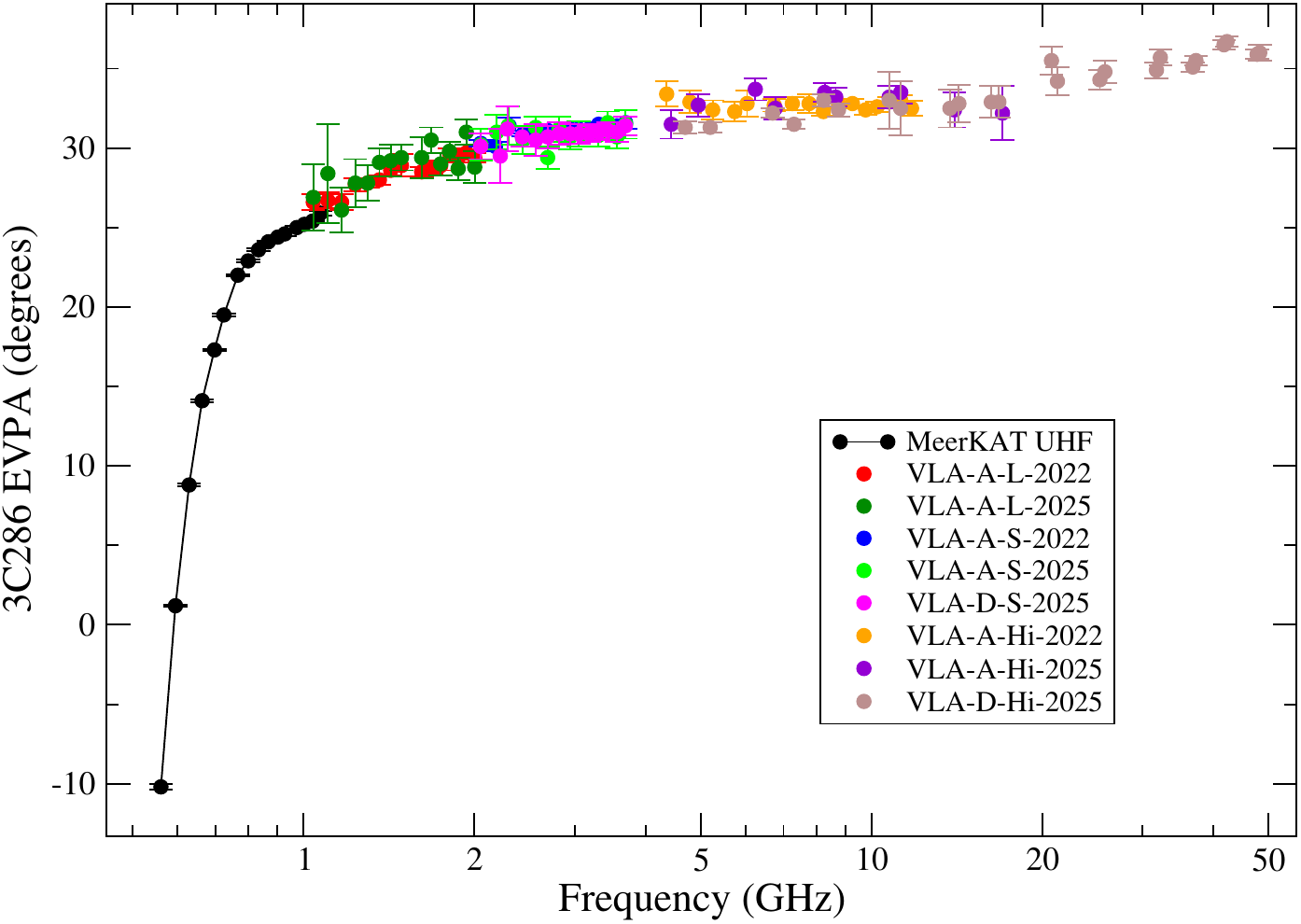}{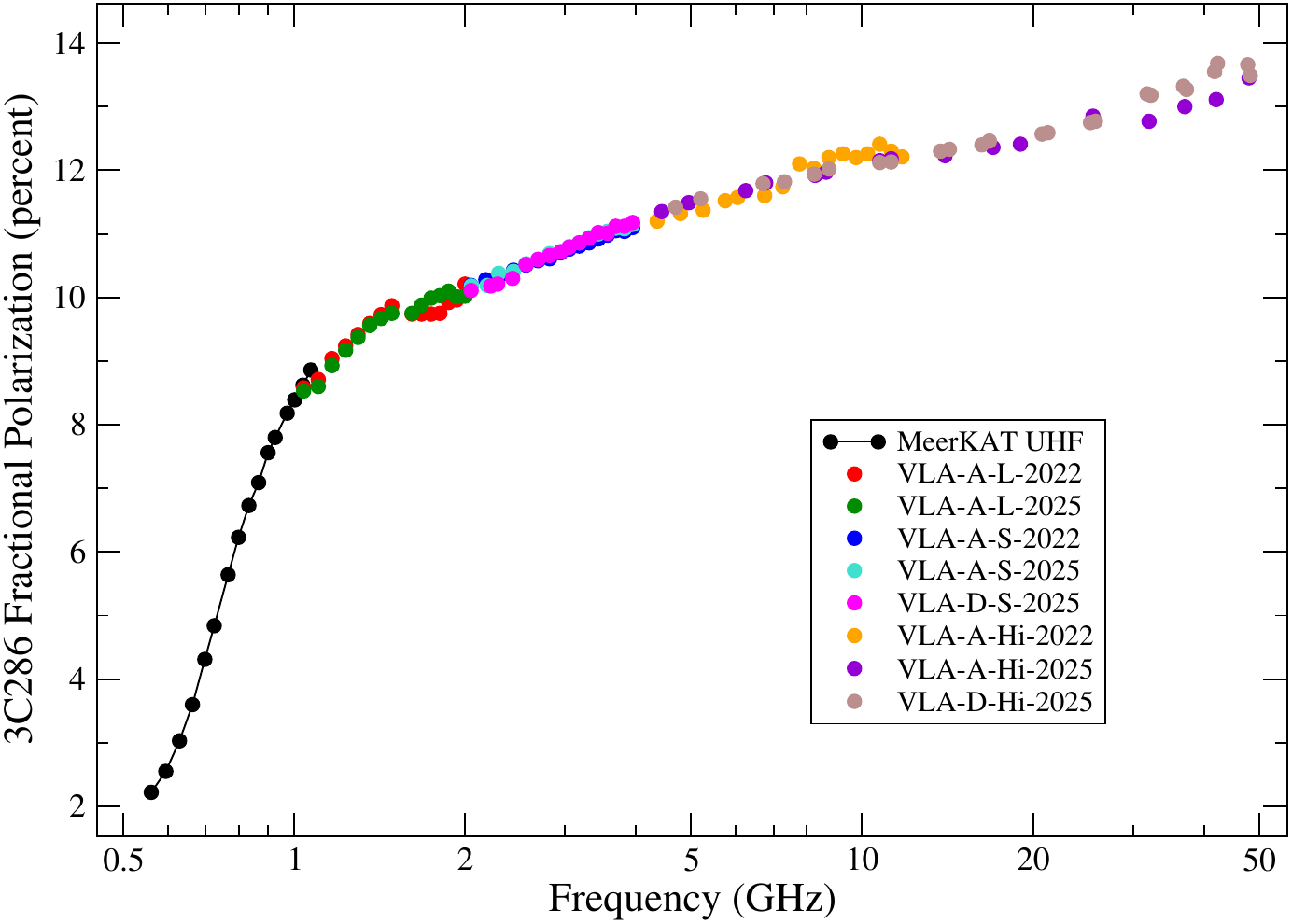}
\caption{Showing the intrinsic EVPA (left) and polarization fraction
  (right) of 3C286, from VLA and MeerKAT data. These data have been corrected
  with the ALBUS G01 IFRM estimates.  There is excellent agreement
  between VLA bands, and between MeerKAT and VLA low frequency measurements.
\label{fig:3C286Pol}}
\end{figure*}

As part of our investigation of IFRM corrections, we have taken more
observations with the VLA with the intention of using the same
methods described here and in \citet{PB13} to determine the intrinsic
EVPAs of standard calibrators, mostly notably 3C286, over the full
frequency range that its polarization can be detected.  For the VLA's
L-band and the MeerKAT array's UHF band, ionopheric corrections are
essential.  The results of our work, over all frequencies, and with
both arrays, with ALBUS G01 corrections applied, are shown in
Figure~\ref{fig:3C286Pol}.  As 3C286 and 3C138 are significantly
resolved by the VLA in most configurations, these results make use of
integrated measures -- the I, Q, and U brightnesses are integrated
over the source extents to provide the utilized flux densities.
Not all the available data were utilized in this figure.  For the VLA,
only those observations whose EVPAs are referenced to emission from
Venus or Mars were used, as the large angular extent of the Moon
causes potential polarization EVPA offsets due to the polarized
structure of the VLA's primary beam.  The angular extents of Venus and
Mars are small enough that beam polarization contamination is not a
limiting issue.

For MeerKAT, we have not used the L and S band observations due to its
strongly polarized primary beam structure beyond the half-power point.
Furthermore, for the 3C286 polarization measurements, only the 2021
data were used, as the IFRM for that date was very low -- less than
$-0.1$ $\rmm$ -- while for the October 2025 observation, the IFRM
estimates were an unrealistically high $-6$ $\rmm$, leading to a
significant overcorrection of the Faraday rotation for this source.
This overestimate is likely due to the very low elevation of 3C286 as
seen by MeerKAT combined with an apparently erroneous gradient
determination towards the north.

The left panel shows the derived intrinsic EVPA for 3C286 as a
function of frequency. There is a very satisfactory agreement between
the VLA and MeerKAT results.  The EVPA slowly declines from $\sim$ 36
degrees at 48 GHz to a value near 25 degrees at 1 GHz. The
fractional linear polarization similarly declines from nearly 14\% at
48 GHz to about 8.5\% at 1 GHz.  Below this frequency, there is a very
rapid depolarization and EVPA rotation, such that at the lowest
MeerKAT UHF lowest frequency of 540 MHz, the source is polarized at
only 2\%, with an EVPA near $-10$ degrees.

To assist accurate polarimetry for all radio telescopes, we provide
analytic expressions for the intrinsic EVPA for 3C286.  For this
purpose, we have smoothed and sampled the data shown in the left panel
of Figure~\ref{fig:3C286Pol}, then fitted a polynomial as a function
of $\log(\nu_G)$, where $\nu_G$ is the frequency in GHz.  Simpler
expressions are obtained by dividing the frequency span into two
ranges: For frequencies between 0.5 and 1.0 GHz
  \begin{equation}
    EVPA = 26.0 + 57.0x + 615x^2 + 3790x^3
  \end{equation}
  and for frequencies between 1.0 and 50 GHz
  \begin{equation}
    EVPA = 26.1 + 17.1x -16.1x^2 + 5.75x^3
  \end{equation}
  where $x = \log(\nu_G)$.  These expressions fit the smoothed data to
  better than 0.3 degrees.

\subsection{3C138}

\begin{figure*}[ht!]
 \epsscale{1.1}
\plottwo{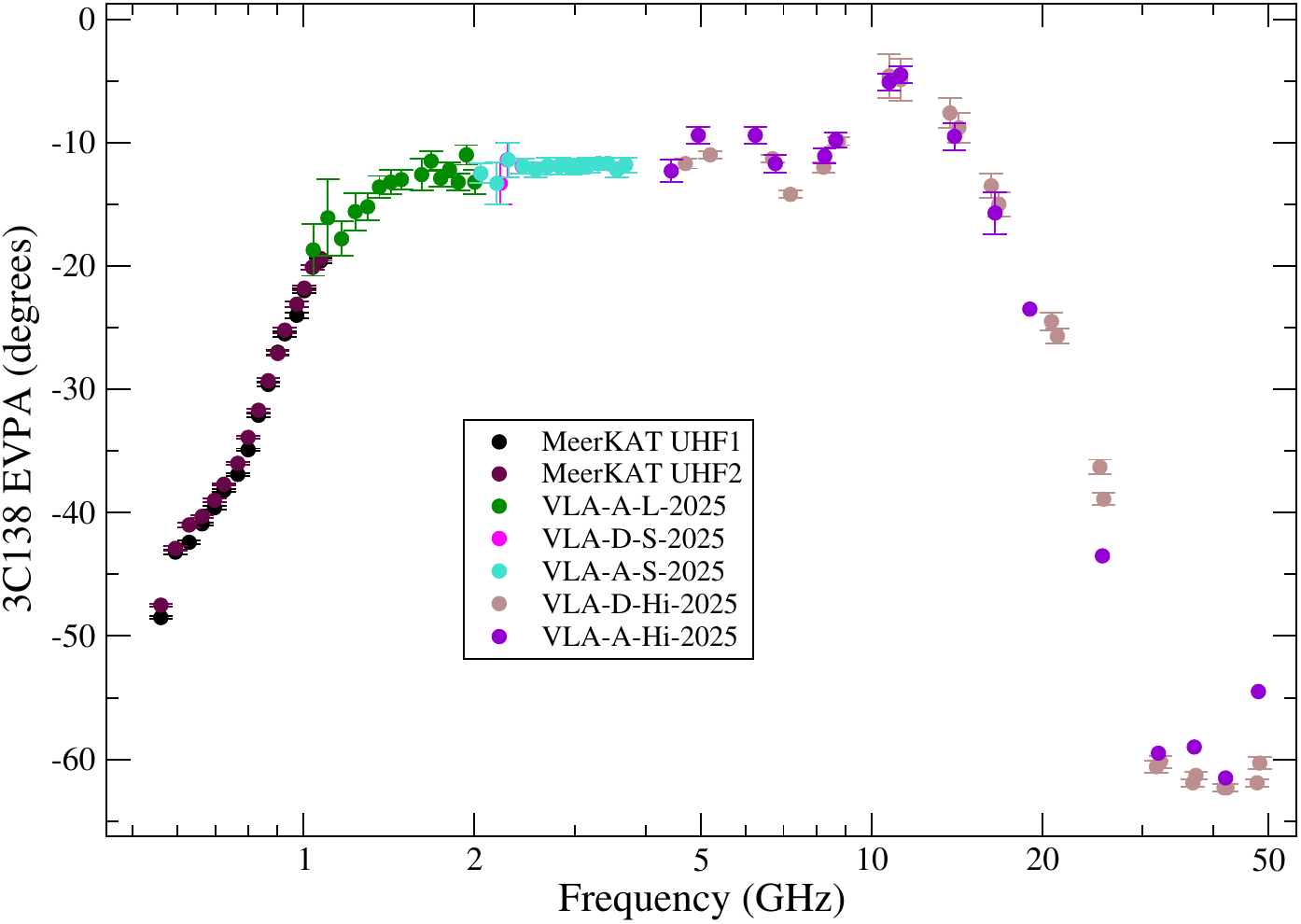}{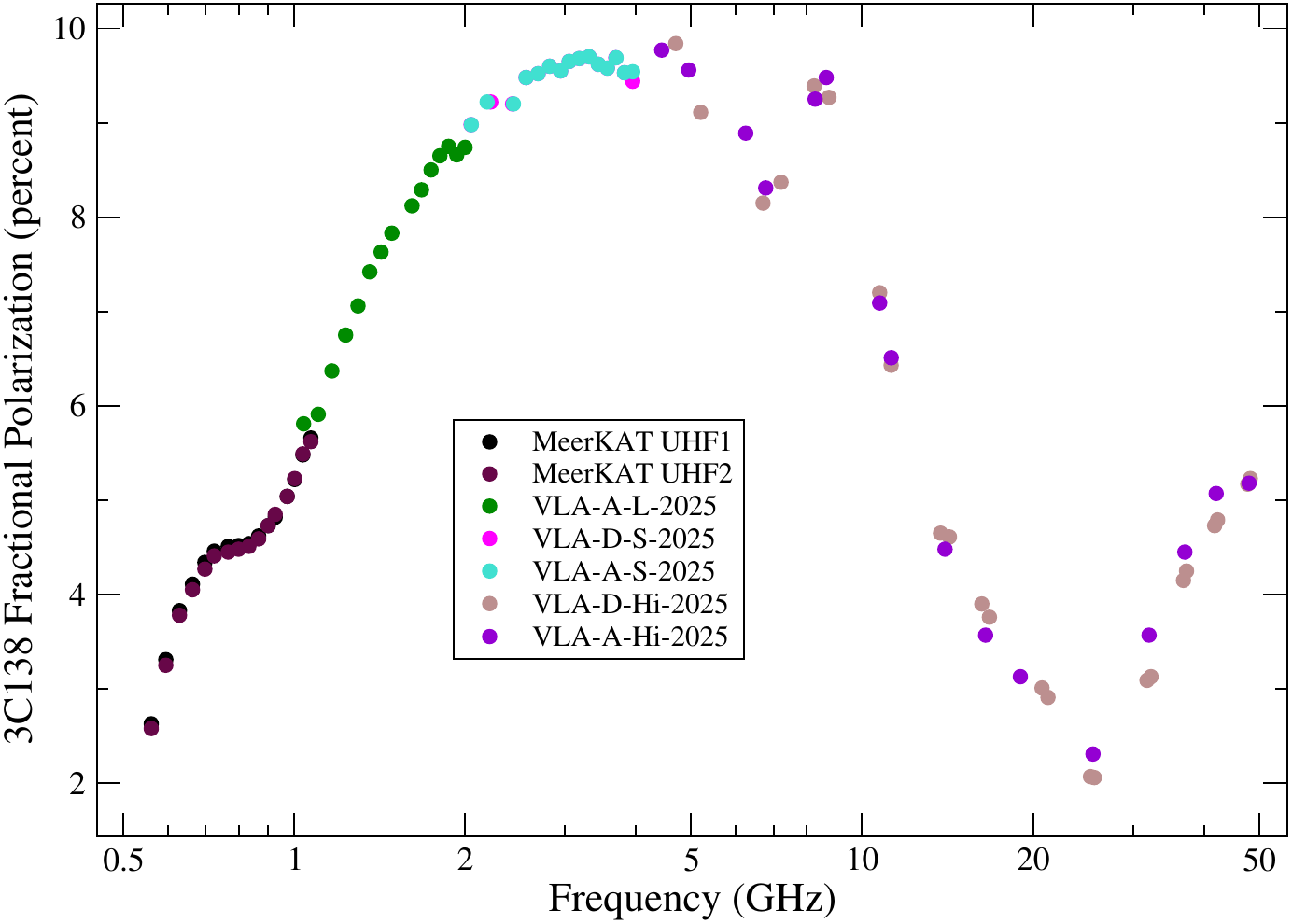}
\caption{Showing the intrinsic EVPA (left) and polarization fraction
  (right) for 3C138 from VLA and MeerKAT data. These data have been corrected
  with ALBUS G01 IFRM estimates.  The oscillatory behaviour seen above
  4 GHz is due to a beating between the nuclear polarized emission and
  that from the sub-arcsecond jet.
\label{fig:3C138Pol}}
\end{figure*}

This highly polarized source is not normally recommended for
polarization calibration, as it is highly variable, with strong flares
beginning in 2002 and extending through the decade \citep{PB13a}, and
again in 2025.  These flares -- which can increase the nuclear flux
density at 45 GHz by a factor of four or more -- change both the
source flux density and polarization, especially at higher
frequencies.  The variability does not seem to affect the low
frequency characteristics by much, as we note the new polarization
results below 4 GHz are unchanged from those shown by \citet{PB13}.
We present here in Fig~\ref{fig:3C138Pol} the results provided by the
recent VLA and MeerKAT observations

The oscillatory behavior in both the EVPA and the fractional
polarization are due to beating between the polarized emission from
the extended jet and the nucleus.  These oscillations maximize between
10 and 50 GHz as the polarized fluxes from these two components are
about equal, but with very different EVPAs.  This behavior is not seen
below 5 GHz as the inverted spectrum nucleus is too weak to
significantly contribute to the total flux density, nor above 50 GHz
as the steep-spectrum polarized jet is too weak.

Noting that these oscillations are not seen at frequencies lower than
$\sim$ 4 GHz, and also that the derived polarization properties from
these recent data are in good agreement with those shown by
\citet{PB13} at low frequencies, we conclude that this source remains
a useful polarization calibrator for low frequency work.

 As for 3C286, we provide analytic expressions for the intrinsic EVPA
 for 3C138, although for this source, due to its variability, the
 expressions are only valid below 4 GHz.  For this purpose, we have
 smoothed and resampled the data shown in the left panel of
 Figure~\ref{fig:3C138Pol}, then fitted a polynomial as a function of
 $\log(\nu_G)$, where $\nu_G$ is the frequency in GHz.  Simpler
 expressions are again obtained by dividing the frequency span into
 two ranges: For frequencies between 0.5 and 1.0 GHz
  \begin{equation}
    EVPA = -21.9 +71.1x -1435x^2 -11110x^3 - 23190x^4
  \end{equation}
  and for frequencies between 1.0 and 4.0 GHz
  \begin{equation}
    EVPA = -22.0 + 95.4x - 340x^2 + 534x^3 - 308x^4
  \end{equation}
  where $x = \log(\nu_G)$.  These expressions fit the observed data to
  better than 0.3 degrees. 

\section{Summary} \label{Discussion}

Extensive low-frequency observations of the Moon with the VLA and
MeerKAT have provided highly sensitive imaging of the lunar
polarization properties, allowing accurate and detailed comparison of
the IFRM derived from observations to estimates made from GNSS-based
estimates of the STEC.  We have compared the observed
IFRM-induced variations in EVPA from the known intrinsic radial
orientation for the Moon with model estimates generated from published
VTEC maps provided by various agencies, and with regional estimates
generated by the ALBUS software.  From the comparisons we can conclude
that for the VLA and MeerKAT:
\begin{itemize}

\item All IFRM estimates -- both global and regional -- show similar changes
  in IFRM between night and day.  These changes closely match the
  observed changes seen in the lunar and polarized calibrator observations.

\item All IFRM estimates generated from global VTEC maps (as implemented by
  TECOR, RMextract and spinifex) overpredict the IFRM for the VLA
  observations by 0.5 to 1.0 $\rmm$ for the VLA, and by $\sim$ $-0.3$
  $\rmm$ for MeerKAT, except when the IFRM values are unusually low.
  These offsets vary between observations, but is roughly constant for
  any single observation.  The thin-shell models using the `jplg' global
  VTEC maps have the highest overcorrection for all ten VLA
  observations.

\item The estimates provided by the ALBUS program are very close to
  the measured values.  Of the ten ALBUS models available, the G01
  model -- using only the closest station with known timing bias
  values, and a simple 2-D (thin-shell) ionospheric model -- provides
  the most accurate and reliable STEC estimates, nearly always within
  0.1 $\rmm$ of the actual value.  The closest GNSS receiver to the VLA
  is located at the VLBA Pietown antenna (`pie1'), 53 km from the VLA,
  which as part of the IGS service has a well calibrated receiver
  offset.  None of the other GNSS stations within 300 km of the VLA
  have known offsets.  For MeerKAT, the nearest well-calibrated
  stations are at Sutherland, nearly 200 km away.  Despite this
  distance, use of the station `sutm' alone provides the most accurate
  estimates of the IFRM.
\end{itemize}
    
The origin of the IFRM offsets in the thin-shell models using global
VTEC maps is unknown. It must arise from an overestimation of the
STEC, or useage of too large a value of the terrestrial magnetic
field, or a combination of both.  Because the ALBUS G01 model,
utilizing a well-calibrated local GNSS station provides an excellent
estimate of the IFRM, we believe the more likely explanation is an
over-estimation of the VTEC by the global maps.  There are references
in the geophysical literature to this same conclusion
\citep{L21,W21,Z21}.  These papers suggest the overestimation of the
VTEC may be due to unmodelled latitudinal gradients.  There are also
suggestions that the thin-shell model, used in all the implementations
utilizing these global VTEC maps, is responsible.  {\cite{Porayko19},
  and \cite{Porayko23}, noting the same IFRM overestimates via LOFAR
  pulsar observations, discuss the possibility that the thin-shell
  model, combined with an incorrect value of the height at which the
  electrons are assumed to be, is responsible.

The simple 2-D ALBUS regional model, utilizing a well calibrated
nearby receiver, provides the best estimates of the IFRM, suggesting
that the problem with the global estimates of VTEC lies with the
extensive smoothing required to produce a global map using typically
200-km separated ground stations.  This observation leads us to
suggest that much more accurate estimations, in both time and
direction, are within reach of ALBUS, or an
  equivalent program which makes use of local GNSS stations to
  estimate the STEC, if we can obtain accurate bias values for the
$\sim$ two dozen stations within 300 km of the VLA and MeerKAT sites.
Future work should aim at obtaining, or deriving, these bias values,
and investigating which of the ten ALBUS models provides the most
accurate and reliable IFRM estimates.  However, as noted in
Appendix~\ref{sec:ALBUS}, the internal workings of ALBUS remain a
mystery, and it is likely that a significant effort will be required
to understand and improve this program.

\appendix

\section{Comparing RMExtract, spinifex, and TECOR} \label{sec:RMExtractVsTECOR}

There are two alternative programs known to us which provide IFRM
estimates -- {\tt ionFR} \citep{S13} and {\tt RMextract} \citep{M18}.
Both of these programs work similarly to TECOR, i.e., using the same
model of the Earth's magnetic field, thin-shell approximation, and
global VTEC maps.
ionFR\footnote{\href{https://github.com/csobey/ionFR}{https://github.com/csobey/ionFR}}
is no longer maintained.  RMextract has recently been superseded by {\tt
spinifex}\footnote{\href{https://git.astron.nl/RD/spinifex}{https://git.astron.nl/RD/spinifex}}\citep{spinifex2025},
which is under development for use with LOFAR and SKA-LO observations,
both of which operate at wavelengths longer than 1 meter.

To compare the RMextract, spinifex, and TECOR estimates, we show in
Fig~\ref{fig:ModelCompare} the lunar IFRM estimates for the D2
observation using the `jplg' and `uqrg' VTEC maps.
The uqrg maps are generated by the Polytechnic University of
Catalonia, and differ from the upcg maps by being estimated on a
15-minute cadence, rather than the two hours of upcg and jplg.
\begin{figure*}[ht!]
 \epsscale{1.1}
\plottwo{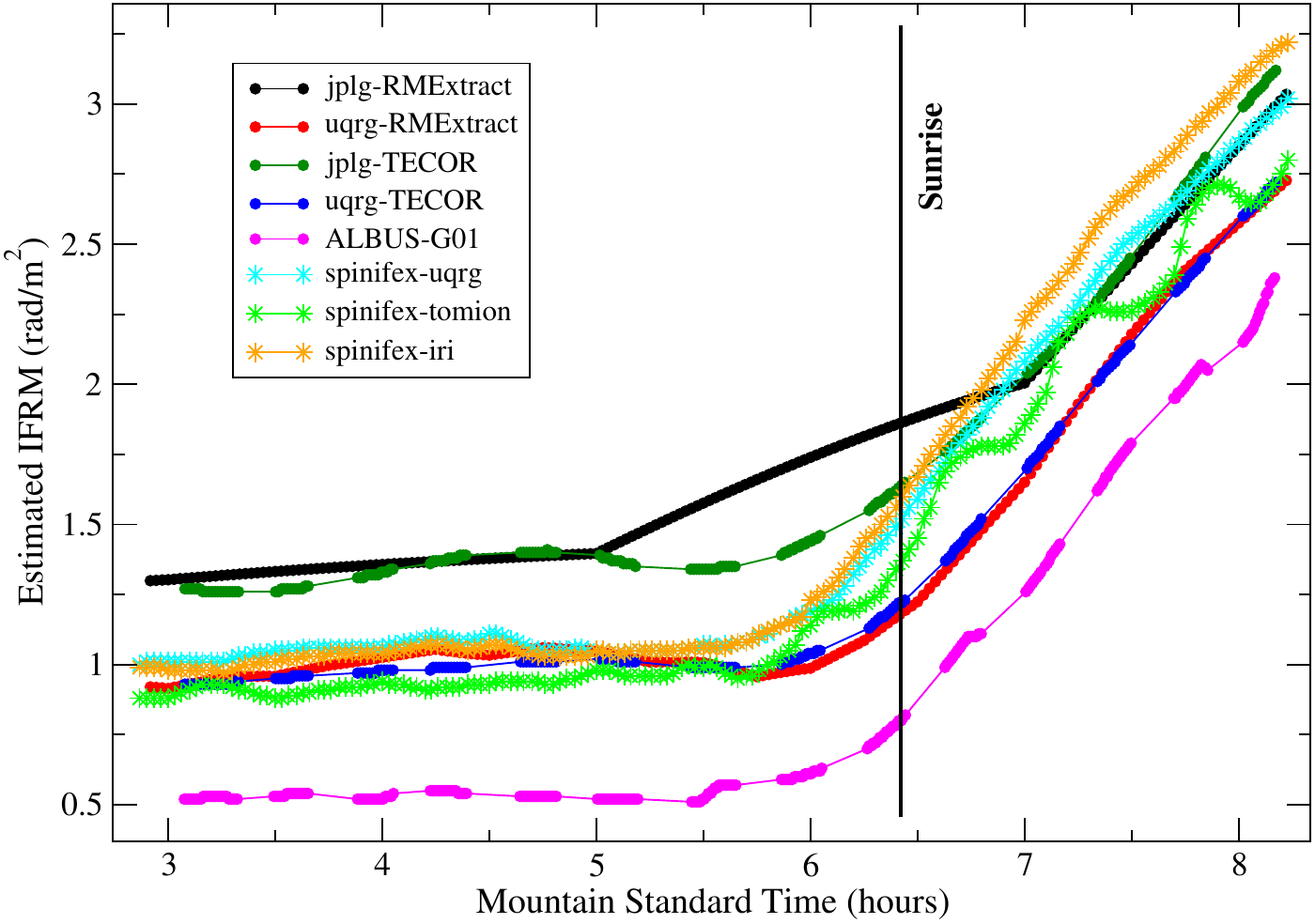}{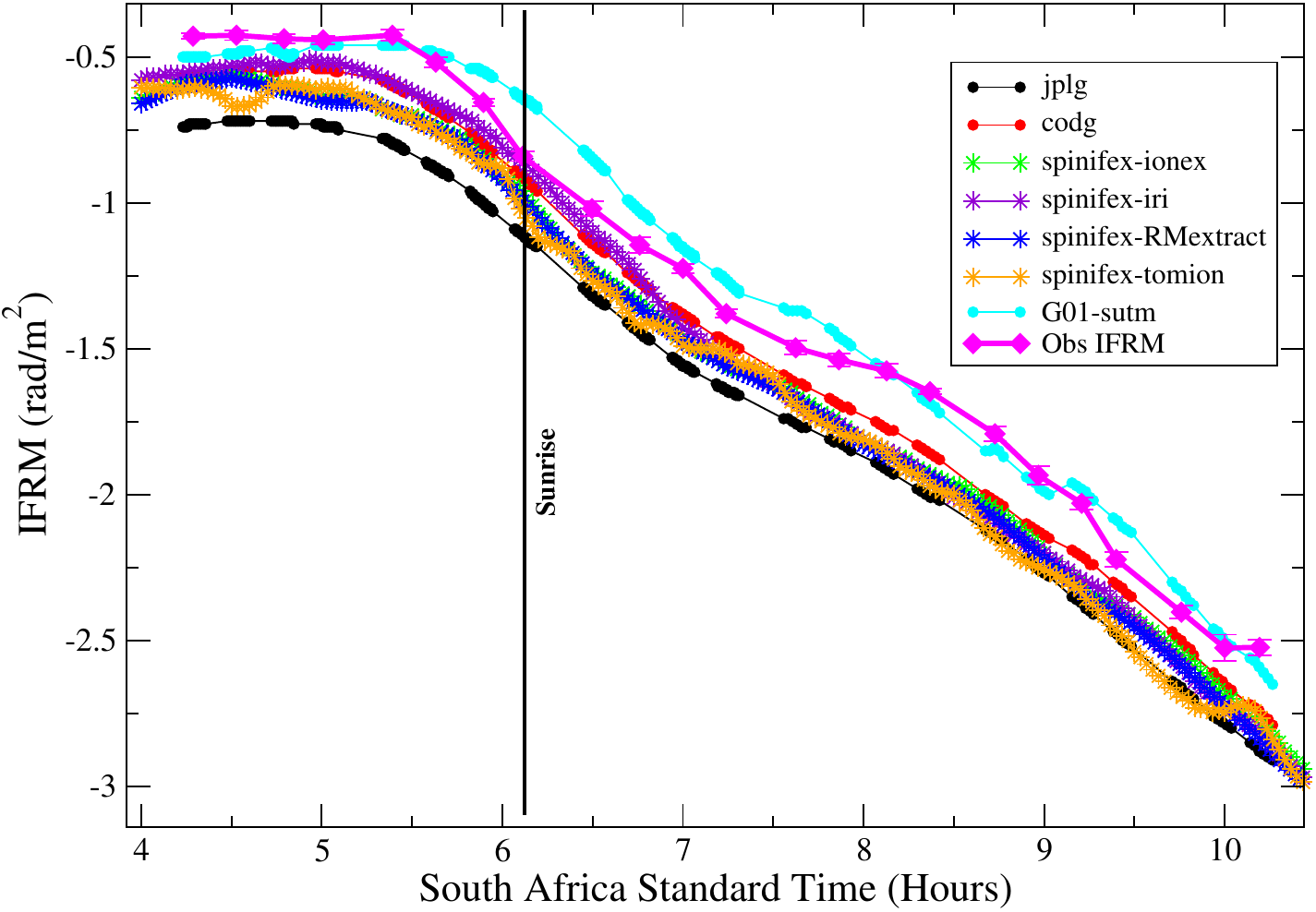}
\caption{Comparing the IFRM estimates provided by TECOR, RMextract,
  ALBUS and spinifex for the D2 observation.  (Left) For the VLA `D2'
  observation on 04 November 2023. The ALBUS values, in magenta, are
  an excellent match to the observed IFRM values. All models based on
  global VTEC estimates overcorrect the IFRM by 0.5 to 1.0 $\rmm$.
  (Right) For the MeerKAT observation on 18 August 2025.  The observed
  IFRM is in magenta, the G01-sutm estimate in blue.  All models based
  on VTEC maps again overestimate the IFRM, although by less than for
  the VLA -- about $-0.3$ $\rmm$.
\label{fig:ModelCompare}}
\end{figure*}

The VLA results for the `D2' observation are shown in the left panel.
TECOR, using the jplg VTEC map, provides the IFRM estimates shown in
green.  The black trace shows the RMextract solutions using the same
jplg VTEC map.  It will be noted there is agreement between them once
every two hours, at the times at which the map values are defined.
The difference between the two is in how they interpolate between map
times.  TECOR rotates the map following the sun's movement,
effectively assuming the ionospheric profile is fixed to the subsolar
point.  It appears that RMextract simply interpolates between the
fixed-time values, ignoring solar movement.

ASTRON recommends the use of the `uqrg' VTEC maps for IFRM estimates.
There is also some evidence in the ionospheric literature that it is the
most accurate of the global maps (see, e.g., \cite{W21} and
\cite{Z21}.)
The TECOR and RMextract estimates using this map are shown in the blue
(TECOR) and red (RMextract) traces.  The issue of interpolation is
largely avoided here because of the 15-minute cadence of the uqrg
maps.  Close inspection of the red trace shows the inflection points
every 15 minutes, corresponding to the times at which the maps are
updated.  The agreement with TECOR is now very good.  The ASTRON
recommendation is likely based on both this, and on the observation
that the IFRM offset is less for the uqrg maps than for the jplg
maps.

For comparison, the ALBUS model prediction, which matches closely the
observed IFRM values, is shown in magenta, significantly less than
either IONEX model, as already noted.  Also shown are three estimates
by the successor to RMextract -- spinifex.

The MeerKAT results for 18 August 2025 are shown in the right panel.
For this instrument, the differences between models are much less than
for the VLA -- typically less than 0.3 $\rmm$.  But the same trends as
for the VLA are evident -- all estimates based on global VTEC maps
overcorrect (here, in the negative sense) the IFRM, while the G01-sutm
estimate is the closest to the observed values -- despite the nearly
200 km separation of that station from the MeerKAT array.

\section{More Details on ALBUS} \label{sec:ALBUS}

ALBUS generates an estimate of the IFRM for a given time, location and
direction by utilizing available nearby GNSS station data.  The
current version of ALBUS utilizes only GPS data, although the program
can potentially utilize the data from at least two other satellite
constellations.  The version utilized in this paper allows
specification of the desired stations to be polled, as well as a
maximum distance for the stations from the desired location.

Following retrieval of these datasets, ALBUS polls the `codg' IGS data
to obtain the satellite and receiver timing bias values.  As noted
below, knowledge of these values is critical in obtaining accurate
IFRM estimates.  If no station bias values are found, the uncorrected
data are used.

Following data retrieval and calibration, the observed differential
times between two frequencies within L-band are converted to total
electron column density for each of the satellites within view.  From
these are derived a time-dependent map of the TEC for the entire
hemisphere, from which the STEC for the source of interest is derived.
ALBUS offers ten different ionospheric models, briefly described in
Table~ \ref{tab:ALBUSMods}, including both 2- and 3-dimensional
distributions of the electron density.
\begin{deluxetable}{cc}
\label{tab:ALBUSMods}
\tablewidth{0pt} \tablecaption{Description of ALBUS Ionosphere Models}
\tablehead{ \colhead{Model} & \colhead{Description}}
\startdata
G00 & Single station, nearest satellite\\
G01 & Single station, simple 2D, all satellites\\
G02 & Multiple stations, simple 2D, all satellites\\
G03 & Multiple stations, 3D model (same long, lat for all h)\\
G04 & Multiple stations, 3D model (different long, lat for all h)\\
G05 & Multiple stations, 3D model (spherical harmonics with different
h layers)\\
G06 & Multiple stations, 2D model with time dependence\\
G07 & Multiple stations, 2D model with gradient least squares\\
G08 & Multiple stations, 2D model with time, gradient least squares\\
G09 & Multiple stations, 2D model with spherical, gradient least squares\\
  \enddata
\end{deluxetable}

An important aspect of utilizing GNSS data is the issue of satellite
and receiver biases.  Each transmitter, and each receiver, has its own
timing offset.  These biases are of the same magnitude as the actual
dispersive timing differences used in the calculation of the STEC (see
\cite{E01} for examples) -- hence it is of considerable importance
that these be removed from the raw data.  In
Figure~\ref{fig:ALBUSBias} we show in the left panel the estimated
STEC (using the G01 model and selecting the single-station option) for
the three individual GNSS stations closest to the VLA: `pie1' is
located at the VLBA site near Pietown NM , 52 km from the VLA at
azimuth $-62$, `sc01' is on top of Socorro Peak, 60 km from the VLA at
azimuth 91, and `p107' is near Grants NM, 120 km from the VLA at
azimuth $-12$.  Of these three, only the `pie1' station has a known
receiver bias, which has been applied in the calculations.  The biases
for the other two stations are not known, and the values in the plot
reflect this lack.  The right-hand panel shows the resulting estimates
of the IFRM.  As noted in this paper, the blue trace (`pie1') closely
matches the sobserved IFRM determined from the lunar observations.

The large differences between estimates emphasizes the importance of
including bias values.  Note that the `sco1' data estimates a negative
electron column density, and a negative IFRM -- both are physical
impossiblities.  
\begin{figure}[]
  \plottwo{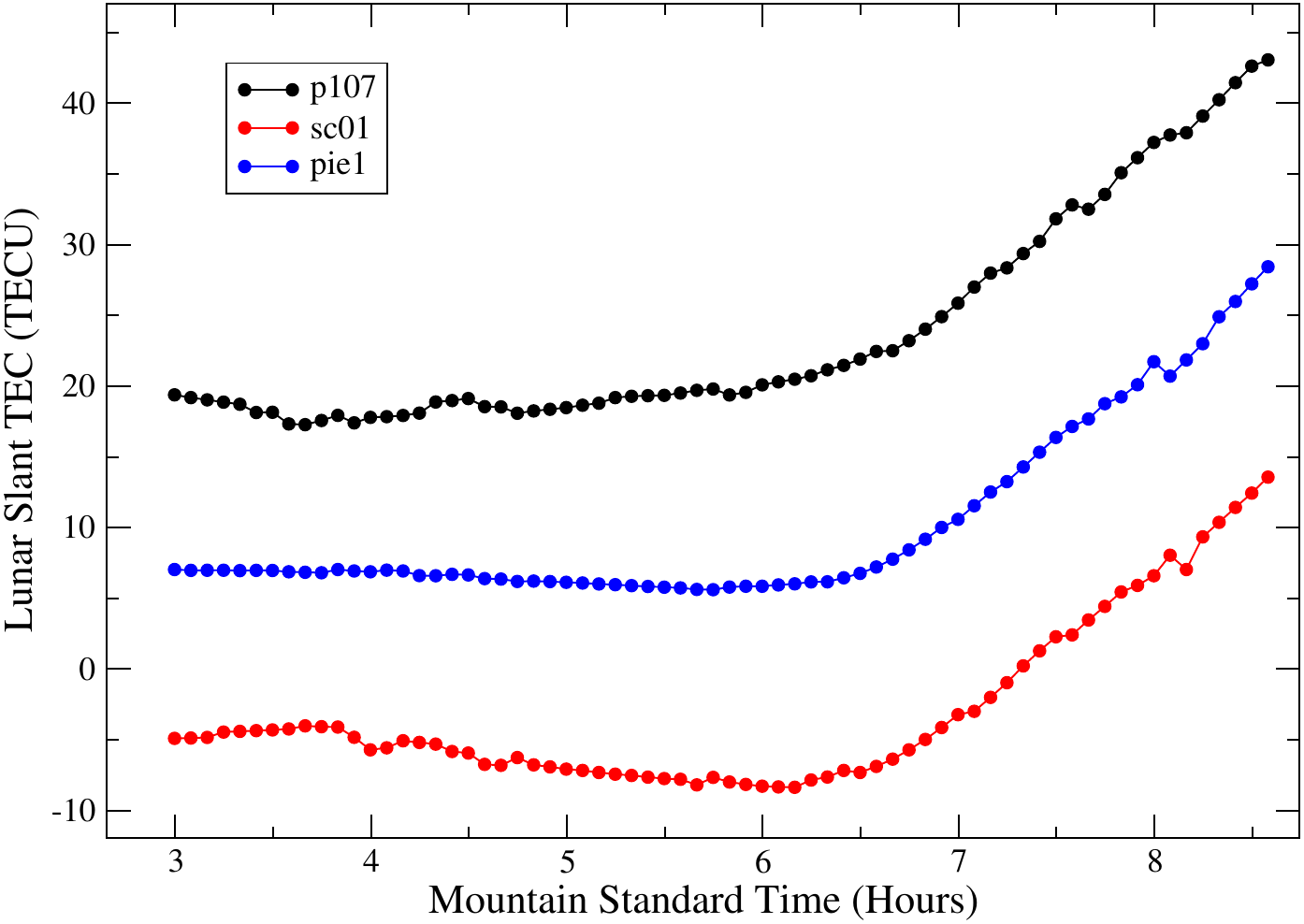}{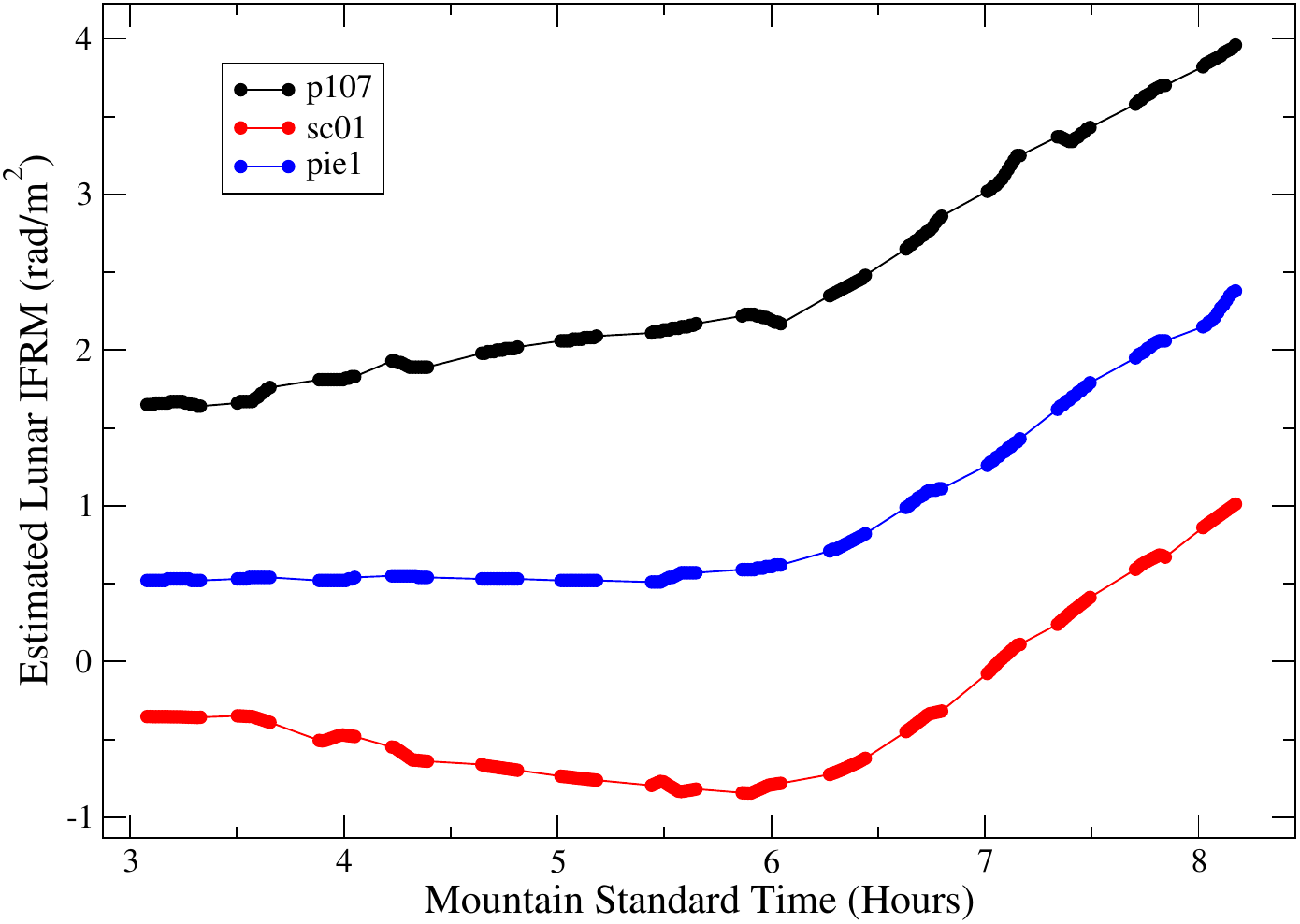}
  \caption{Showing STEC (left) and IFRM (right) estimates using data
    from the closest three GNSS stations to the VLA for the VLA D2
    observation.  Only the `pie1' station (blue traces) has known
    receiver bias values.  Use of the data from the other two stations
    results in improbable, or impossible, estimates of the STEC and
    IFRM.}
  \label{fig:ALBUSBias}
\end{figure}

As part of this work, we investigated the characteristics of each of
these models, using the `D2' observations.  The estimated IFRM values
of all ten models for the `D2' data are shown in the left panel of
Fig~\ref{fig:ALBUSPlots}. For this example, we have used a GNSS
station search of 165 km, since use of a single station causes the
`3-d' models G03, G04, and G05 to generate impossible values of the
IFRM.  From this plot we see that the G02 and G06, G03, G04, and G05,
and the G07 and G08 estimates are identical.  The G00, G07, G08, and
G09 estimates are unstable, and we do not consider these further.
\begin{figure}{}
  \gridline{\fig{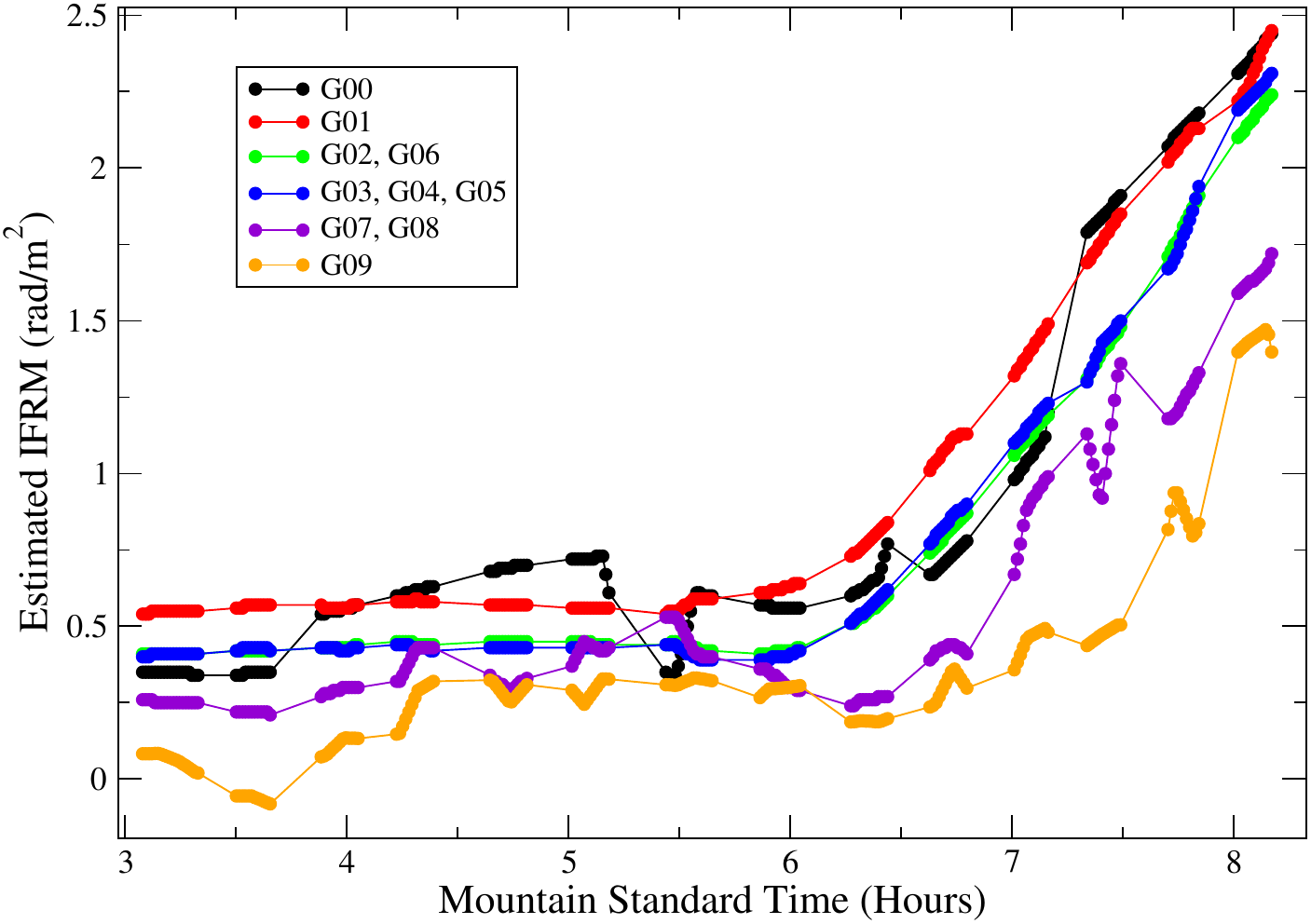}{0.33\textwidth}{}
            \fig{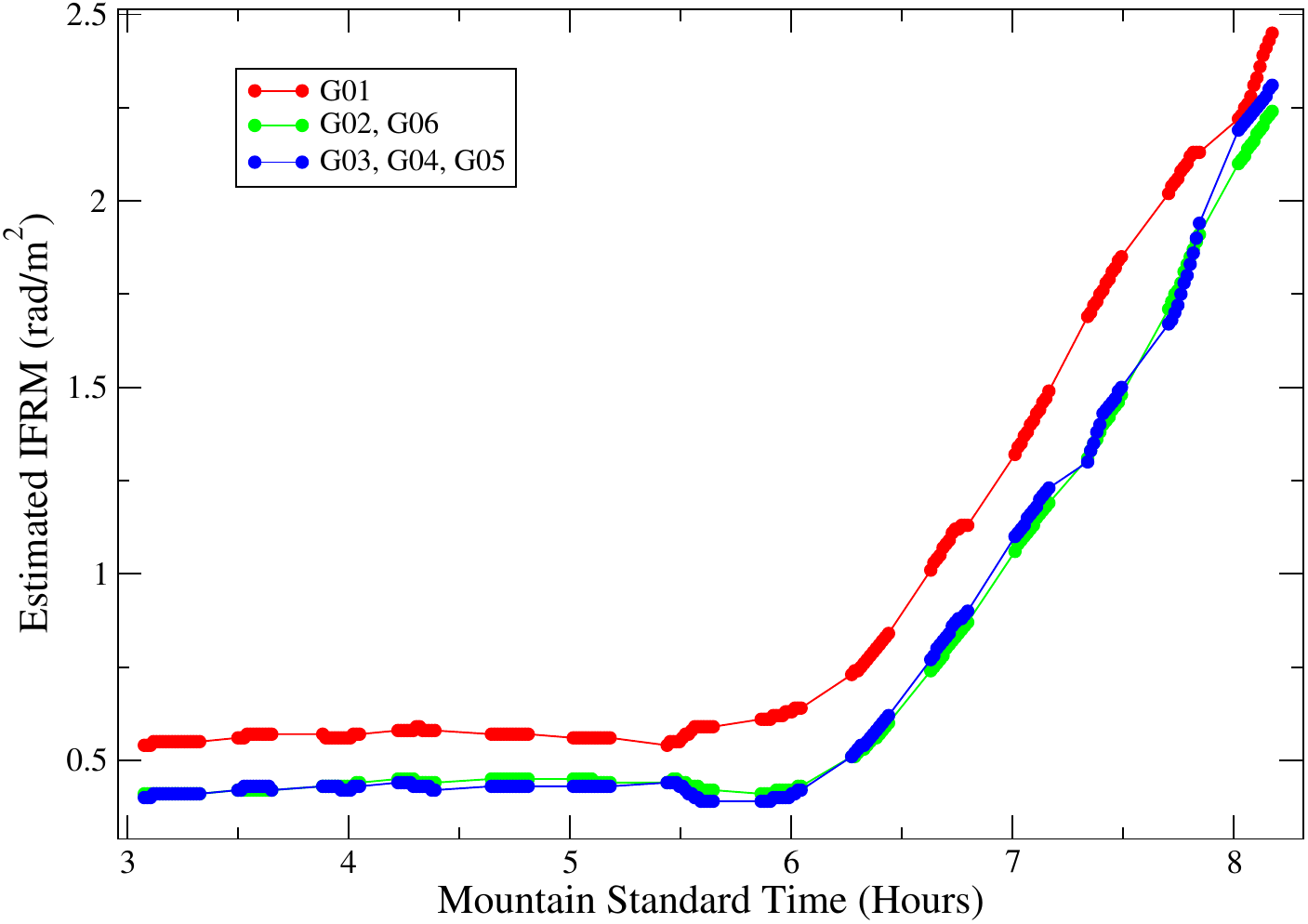}{0.33\textwidth}{}
            \fig{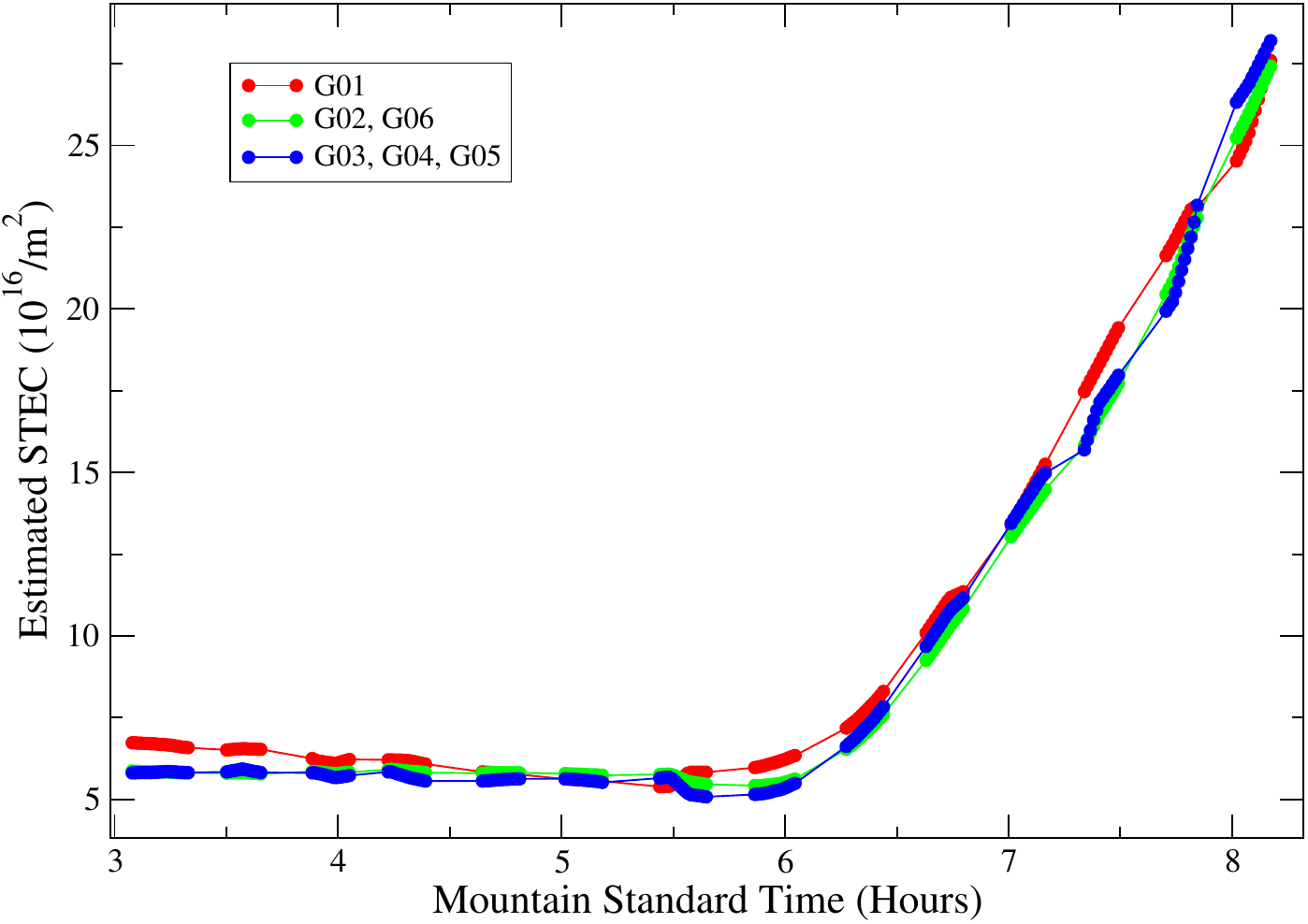}{0.33\textwidth}{}}
  \caption{Showing the ALBUS estimates of IFRM and STEC for the `D2'
    observation. (Left) The estimated IFRM for all ten models.
    (Center) The estimated IFRM for the six stable models.  Note that
    the estimate for the G01 model is higher than the others, and
    closely matches the observed lunar values.  (Right) The STEC for
    the six stable models.  Note that although all stable models have
    similar values of STEC, the ALBUS code estimates different values
    of IFRM.}
  \label{fig:ALBUSPlots}
\end{figure}
The stable G01 through G06 IFRM estimates are shown in the center panel of
the figure.  Note that the G01 estimate is higher than the others,
which are all very similar.  As noted in this paper, the G01 estimate
closely matches the observed IFRM values determined from the lunar
observations, while the others, although similar, are significantly
lower.  In the right-hand panel, we show the estimated STEC values
used to generate the IFRM values shown in the middle panel.
Surprisingly, all STEC values are very similar -- the clear higher
value in IFRM from the G01 model is not reflected in the STEC values.
We do not understand how this can be -- the conversion from STEC to
IFRM involves integration of the product of the STEC distribution with
the line-of-sight values of the magnetic field.  Presuming the program
uses the same magnetic field model for this computation, the estimated
IFRM values should reflect the STEC values -- especially for the G01
and G02 models, both of which presume a 2-d electron distribution --
the thin-shell model.

This apparent discrepancy illustrates a problem with the ALBUS program
-- we don't understand in detail what its internal computations are
doing.  The program is large and complex, with a Python layer on top of
lower level C and C++ software. See \cite{W22} for details.  Other tests
that we have performed show characteristics that are not understood.
What we can state with certainty is that the the simple G01 model, using
a single nearby station with known receiver bias values, provides the
best estimates of the IFRM.  But we do not understand why a simple 2-D
(thin-shell) model gives better results than a full 3-D model, which
physically we would expect to give better results.  Further progress
will require a detailed examination of the code, with a goal of
implementing the more sophisticated models using high-quality GNSS data
with known receiver biases.

\begin{acknowledgments}
  The MeerKAT telescope is operated by the South African Radio
  Astronomy Observatory, which is a facility of the National Research
  Foundation, an agency of the Department of Science and Innovation.

  The National Radio Astronomy Observatory and Green Bank Observatory
  are facilities of the U.S. National Science Foundation operated
  under cooperative agreement by Associated Universities, Inc.

  This work has made use of the `MPIfR S-band receiver system',
  designed, constructed, and maintained by funding of the MPI f\"{u}r
  Radioastronomy and the Max-Planck Society.

  We thank Joe Helmboldt and Anthea Coster for helpful advice on
  ionospheric physics, and for providing ionosonde and MADRIGAL data
  for comparison to the global VTEC estimates.  We thank the anonymous
  referee for useful comments on the original text.
\end{acknowledgments}

\begin{contribution}

  RAP set up the observations, calibrated the data, analyzed the
  results, and led the paper writing.  BJB provided essential
  information on ionospheric modelling and on the use of planetary
  bodies for polarimetric calibration.  ET developed the containers
  for running ALBUS within the AIPS environment.  EWG implemented many
  changes within AIPS to enable the imaging, and implemented the
  container needed to run ALBUS within the AIPS environment.  BVH
  facilitated the MeerKAT lunar observations, and provided advice for
  accessing trignet data.  AGW provided essential advice on the ALBUS
  program, and implemented many modifications to speed up its
  execution and to improve its capabilities and user interface.  All
  authors contributed to the writing and editing of this paper.

\end{contribution}

%

\facilities{NRAO (VLA), SARAO (MeerKAT)}
\software{AIPS \citep{G03}, ALBUS \citep{W22}, RMextract \citep{M18},
  spinifex \citep{spinifex2025}}

\bibliography{IFRMCor17Feb}{}
\bibliographystyle{aasjournalv7}



\end{document}